\documentclass[twocolumn,superscriptaddress,showpacs,nofootinbib]{revtex4}

\usepackage{bm,color}

\usepackage{graphicx}
\usepackage{amsmath}
\usepackage{amssymb}
\usepackage{amsfonts}
\usepackage{bm}
\usepackage{color}

\def\be{\begin{equation}}
\def\ee{\end{equation}}
\def\bea{\begin{eqnarray}}
\def\eea{\end{eqnarray}}

\begin{document}

\title{Cosmological models based on a complex scalar field with a power-law
potential associated with a polytropic equation
of state}

\author{Pierre-Henri Chavanis}
\email{chavanis@irsamc.ups-tlse.fr}
\affiliation{Laboratoire de Physique Th\'eorique, Universit\'e de Toulouse,
CNRS, UPS, France}

\begin{abstract}

We construct cosmological models based on a complex scalar field with 
a power-law potential 
$V=\frac{K}{\gamma-1}(\frac{m}{\hbar})^{2\gamma}|\varphi|^{2\gamma}$ associated
with a polytropic
equation
of state $P=K\rho^{\gamma}$ (the potential associated with an isothermal
equation of state $P=\rho k_B T/m$ is
$V=\frac{m k_B T}{\hbar^2}|\varphi|^2 [\ln(m^2|\varphi|^2/\rho_*\hbar^2)-1]$ and
the potential associated with a logotropic
equation of state $P=A\ln(\rho/\rho_P)$ is
$V=-A[\ln(m^2|\varphi|^2/\hbar^2\rho_P)+1]$). We consider a fast oscillation
regime of ``spintessence'' where the equations of the problem can be simplified.
We study all possible cases with arbitrary (positive and
negative) values of the polytropic constant and polytropic index.
The $\Lambda$CDM model,
the Chaplygin gas model and the Bose-Einstein condensate model are recovered as
particular cases of our study corresponding to a constant potential
($\gamma=0$), an inverse square-law potential ($\gamma=-1$), and a quartic
potential ($\gamma=2$). We also derive the
two-fluid representation of the Chaplygin gas model.

\end{abstract}



\maketitle


\section{Introduction}

Scalar fields (SFs) have been invoked in different domains of particle physics, 
quantum field theory, astrophysics, and cosmology. In particle physics and
string theory, they arise in a natural manner as bosonic spin-$0$ particles
described by the Klein-Gordon (KG) equation \cite{kolb,zee}. Examples include
the
Higgs particle, the inflaton, the dilaton field of superstring theory, pseudo
Nambu-Goldstone bosons,  tachyons etc. SFs also arise in the Kaluza-Klein and
Brans-Dicke theories \cite{b1m} as well as in quantum gravity,
supergravity, and superstring models (e.g., hidden sector fields and moduli). In
cosmology, SFs were first introduced to
explain the phase of inflation in the primordial universe assumed to be driven
by vacuum energy  \cite{linde}. The inflaton, which has its origin in the
quantum fluctuations of the vacuum, is usually associated with a
nonequilibrium phase transition. In canonical SF models based on  a Lagrangian
of the form ${\cal L}=X-V(\varphi)$, where
$X=\frac{1}{2}\partial_{\mu}\varphi\partial^{\mu}\varphi$ is the kinetic term
and
$V(\varphi)$ is the potential term, the physical information
on the system is contained in the potential $V(\varphi)$ of the SF. In
k-inflation models
based on a Lagrangian of the form ${\cal L}={\cal L}(X)$ \cite{adm,gm}, the 
physical information on the system is encapsulated in the nonstandard kinetic
term.\footnote{General Lagrangians of the form ${\cal
L}(X,\varphi)$  have
also been considered in Refs. \cite{adm,gm}.}
After the discovery of the accelerating expansion
of the universe \cite{novae1,novae2,novae3,novae4}, SFs have been used  to
describe dark energy (DE). Different
forms of SFs were introduced such as quintessence
\cite{ratra,fhsw,cds}, $k$-essence
\cite{chiba,ams}, phantom (ghost) fields \cite{caldwell,ckw},  tachyons
\cite{sen,gibbons,frolov,paddytachyon,feinstein,pc}, quintom
\cite{gpzz,ssa,amani} etc.
They usually involve a mass of the
order
of the current Hubble scale ($m\sim H_0\hbar/c^2\sim 10^{-33}\,{\rm eV/c^2}$).
It has also
been proposed that dark matter (DM) and superfluid stars
may
be described in terms of a SF which represents the wave function of a
self-gravitating Bose-Einstein condensate (BEC). This leads to the concept of
boson stars \cite{kaup,rb,colpi} and BEC stars
\cite{chavharko} that could describe DM stars or  the superfluid core of neutron
stars.  DM halos may also be interpreted as giant self-gravitating  BECs. In
particular, the  fuzzy dark matter (FDM) model is based on the assumption that
DM is made of extremely light scalar particles (ultralight axions) with mass
$m\sim
10^{-22}\,{\rm eV/c^2}$ \cite{hu,hui}. At large scales, FDM  behaves as CDM but
at
small scales ($\lesssim 1\, {\rm kpc}$), the wave (quantum) properties of the
bosons manifest themselves and may solve the problems of CDM such as the
cusp-core problem \cite{moore}, the missing satellite problem
\cite{satellites1,satellites2,satellites3}, and the ``too big to fail'' problem
\cite{boylan}.

In cosmology, most models attempting to describe the evolution 
of the universe by a SF consider a real SF. However, complex SFs should be
considered as well because they can be given a physical justification in
relation to the Higgs mechanism in particle physics or to the phenomenon  of
Bose-Einstein condensation in ultra cold gases. In addition, complex SFs are
potentially more relevant than real SFs because they can form stable DM halos
while DM halos made of real SFs are either dynamically unstable \cite{jsreal} or
oscillating
\cite{oscillaton}.\footnote{These dynamical instabilities occur in the
relativistic regime for boson stars. Although DM halos can usually be described
by
nonrelativistic equations, there are certain situations in which relativity may
be important. In addition, bosons described by a
complex SF with a global $U(1)$ symmetry associated with a conserved charge
(Noether theorem) can form BECs even in the early universe while this is more
difficult for a system of bosons described by a real SF like the QCD axion.}
This is basically due to the fact that the charge of a complex SF is conserved
while real SFs have no conserved charge.
Therefore, a complex SF seems more promising  than a real SF to account both for
the cosmic evolution of the universe (cosmological background) and the formation
of large-scale structures (galaxies and DM halos).

The cosmological evolution of a spatially homogeneous complex SF with 
a quartic self-interaction potential (possibly representing the wave
function of a relativistic BEC) was recently studied by
Li {\it et al.}
\cite{shapiro} and
Su\'arez and Chavanis \cite{abrilphas}.\footnote{Li {\it et
al.}
\cite{shapiro} focused on a repulsive quartic self-interaction while Su\'arez
and
Chavanis \cite{abrilphas} considered both repulsive and attractive
quartic self-interactions. They also developed a general formalism that is valid
for an
arbitrary potential of interaction $V(|\varphi|^2)$.} These authors considered a
slow oscillation regime where the SF rolls down the potential without
oscillating and a fast oscillation regime of ``spintessence''
\cite{spintessence} where the direction  (phase) of the SF rotates rapidly in
the complex plane while its modulus
changes
slowly (adiabatically). In the slow oscillation
regime, there is an
interesting situation of ``kination'' where the kinetic term dominates the
potential term and the SF behaves as stiff matter
\cite{kination}. In
the fast oscillation
regime, the equations of the problem can be averaged over the oscillations of
the SF and simplified. Using a hydrodynamic representation of the KG equation,
Su\'arez and Chavanis \cite{abrilphas} showed that the fast oscillation regime
of spintessence is
equivalent to the Thomas-Fermi (TF) approximation where the quantum potential
can be neglected (this amounts to taking $\hbar=0$ in the hydrodynamic
equations). They investigated the domain of validity of this regime in detail.

A complex SF with a repulsive $\frac{\lambda}{4\hbar
c}|\varphi|^4$ ($\lambda>0$)
self-interaction undergoes, in the
fast oscillation regime, a radiationlike\footnote{The radiationlike era is due
to the $|\varphi|^4$ self-interaction of the SF. It is different from the
standard
radiation era due to photons or ultrarelativistic particles like neutrinos.}
era (with an equation of state $P\sim\epsilon/3$) followed by a matterlike  era
($P\simeq 0$). However, the fast oscillation regime is not valid at very early
times. It is preceded, in the primordial universe, by a stiff matter era
($P=\epsilon$) which is valid in the slow oscillation regime. Therefore,  a
complex SF with a repulsive $|\varphi|^4$
self-interaction undergoes successively a stiff matter era, a radiationlike era,
and a DM era.\footnote{It is
well-known that a noninteracting real SF undergoes a stiff matter era followed
by an inflation era and, finally (in the fast oscillation regime), by a matter
era  
\cite{belinsky,belinsky2,belinsky3,piran}. This evolution (stiff $\rightarrow$
inflation $\rightarrow$ matter) was also found by  Scialom and Jetzer
\cite{sj,js} for a self-interacting complex SF. These authors showed that the
inflation era occuring in the slow-roll regime is sufficiently long provided
that the bosonic charge $Q$ is close to zero (in that case, the phase of the SF
remains approximately constant and thus inflation is essentially driven by
one component of the SF as if it were real). On the other hand, for a
massive complex SF, inflation is less effective when the
self-interaction $\lambda$ increases. This is because the potential becomes less
flat and, as a consequence, the slow-roll approximation needed for inflation to
occur is no longer well satisfied. Therefore, inflation takes place only if the
charge and the self-interaction of the SF are sufficiently
small. Otherwise, the background passes directly from the stiff matter era to
the oscillatory phase \cite{shapiro,abrilphas} (there may, however, be an
inflation era prior to the stiff matter era \cite{stiff}). 
Arbey {\it et al.} \cite{arbeycosmo} also considered (independently from
\cite{sj,js}) a self-interacting complex SF. For $\lambda=0$ they found a stiff
matter era, followed by an epoch where the SF is constant (like the inflation
era reported previously but occuring during the standard radiation era), and a
DM era. For
$\lambda\neq 0$ they found, in the fast oscillation regime, a dark radiation era
followed by a DM era. For a complex SF in the
fast oscillation regime, only the phase $\theta$ of the SF changes
(spintessence). The modulus $|\varphi|$ of the SF evolves slowly without
oscillating. By contrast, for a real SF in the fast oscillation
regime, $\varphi(t)$ oscillates rapidly by taking positive and negative
values. Arbey {\it et al.} \cite{arbeycosmo} considered a fast
oscillation
regime different from spintessence where the
complex SF behaves as an effective oscillating real SF (axion). In
the present paper, when considering the fast oscillation regime of a complex SF,
we shall systematically assume that it corresponds to the spintessence
regime.} It is also possible to account  for the
present
acceleration of the universe (DE) by introducing a cosmological
constant $\Lambda$ or by shifting the origin of the SF potential to a small
positive value equal to the cosmological energy density
$\epsilon_\Lambda=\rho_\Lambda c^2$ so that
$V_{\rm tot}=\rho_\Lambda
c^2+\frac{m^2c^2}{2\hbar^2}|\varphi|^2+\frac{\lambda}{4\hbar c}|\varphi|^4$
(see Appendix E of
\cite{graal}).  The radiationlike era only exists for sufficiently large
values of the
self-interaction parameter \cite{shapiro,abrilphas}. Self-interacting SFs may be
more relevant than noninteracting SFs. Indeed, it is found in
\cite{shapiro,abrilphas} that the FDM model which assumes that the bosons are
noninteracting is not consistent with cosmological observations.

A complex SF with an attractive
$\frac{\lambda}{4\hbar c}|\varphi|^4$
($\lambda<0$) self-interaction can display, in the
fast oscillation regime, two types of behaviors which are associated
with two branches of solutions. These solutions start at a
nonzero scale factor with a finite energy density. In this sense, the SF emerges
``suddenly'' in the universe. Initially, there is a very short
cosmic stringlike
era ($P\sim -\epsilon/3$).  On the normal (nonrelativistic)
branch, the energy density decreases to zero as $a^{-3}$ and the SF behaves as
pressureless DM
($P\simeq 0$). On the peculiar (ultrarelativistic) branch, the
energy density slowly decreases and tends to a constant at late times giving
rise to a
de Sitter evolution. In that case, the  SF behaves as DE ($P\sim
-\epsilon$).\footnote{This intriguing behaviour was first reported by Su\'arez
and
Chavanis \cite{abrilphas} and 
further discussed by
Carvente {\it et al.} \cite{carvente}.} On
the normal branch, the fast oscillation regime  is not valid at early times so
that quantum mechanics ($\hbar\neq 0$) must be taken into account. In the very
early universe, a
complex SF with an attractive self-interaction may undergo an inflation era. If
the
self-interaction is sufficiently small, the inflation era is followed by a stiff
matter era. On the peculiar branch, if the self-interaction is
sufficiently
large, the fast oscillation regime is valid initially but it
ceases to be valid
at late times so that quantum mechanics ($\hbar\neq 0$) must be taken into
account ultimately. As
a result, the de Sitter regime may stop and the SF may eventually enter in a
stiff matter era implying that the late universe passes from a phase of
acceleration
to a phase of deceleration.

In the present paper, we study how 
the evolution of the universe depends on the form of the SF potential. Many
types of potentials have been studied in the past in the case of a real SF. The
novelty of our approach is to consider the cosmic evolution of a complex SF.
Furthermore, we
focus on the fast oscillation regime where the SF experiences the process of
spintessence.  The general equations of the problem, valid for an arbitrary
self-interaction potential, are given in \cite{abrilphas}. In the present paper,
we consider
algebraic potentials of the form $V=
\frac{K}{\gamma-1}(\frac{m}{\hbar})^{2\gamma}|\varphi|^{2\gamma}$ where $K$
and $\gamma$ can
be
positive or negative.  These power-law potentials are associated with
the polytropic equation of
state $P=K\rho^{\gamma}$ where $\rho=\frac{m^2}{\hbar^2}|\varphi|^2$ is the
pseudo rest-mass density. The
$\Lambda$CDM model is recovered for a constant potential
$V=\rho_{\Lambda}c^2$ ($\gamma=0$ and
$K=-\rho_{\Lambda}c^2$) corresponding  to a constant
equation of state
$P=-\rho_{\Lambda}c^2$. The Chaplygin gas is recovered for an inverse square law
potential
$V=\frac{A}{2}(\frac{\hbar}{m})^2\frac{1}{|\varphi|^2}$
($\gamma=-1$ and
$K=-A$)  corresponding to the equation of state
$P=-A/\rho$. The standard BEC model is
recovered for a quartic potential $V=\frac{2\pi
a_s\hbar^2}{m^3}(\frac{m}{\hbar})^4|\varphi|^4$ ($\gamma=2$ and
$K={2\pi
a_s\hbar^2}/{m^3}$)  corresponding to
the equation of state
$P=\frac{2\pi
a_s\hbar^2}{m^3}\rho^2$. We also consider a potential of the form 
$V=\frac{m k_B T}{\hbar^2}|\varphi|^2 [\ln(m^2|\varphi|^2/\rho_*\hbar^2)-1]$
associated with the isothermal equation of state $P=\rho k_B T/m$.
This is the limit of the polytropic equation of state when
$\gamma\rightarrow 1$. In a companion paper \cite{graal}, we specifically study
a
potential
of the form $V=-A[\ln(m^2|\varphi|^2/\hbar^2\rho_P)+1]$ associated with the
logotropic equation of state $P=A\ln(\rho/\rho_P)$.
This is the limit of the polytropic equation of state when
$\gamma\rightarrow 0$ and $K\rightarrow +\infty$ with $A=K\gamma$ constant
(see Sec. 3 of \cite{epjp} and Appendix A of \cite{graal} for a precise
statement). The logotropic 
model gives a good agreement with the cosmological observations of our
universe and is able to
account for the universal value of the surface density
$\Sigma_0^{\rm
obs}=141_{-52}^{+83}\, M_{\odot}/{\rm pc}^2$ of DM
halos without free parameter.  In the present paper, we
do not necessarily try to
construct a realistic model of universe. We consider an arbitrary power-law
potential $V\sim K|\varphi|^{2\gamma}$  and describe the different types of
evolutions that it produces depending on the value of $\gamma$ and on the sign
of $K$. Our work therefore complements the study of the $|\varphi|^4$ potential
performed in
\cite{shapiro,abrilphas}.

The paper is organized as follows. In Sec. \ref{sec_csf} we consider a
spatially homogeneous SF in an expanding universe. Following our previous
paper \cite{abrilphas} we establish the 
general equations of the problem in the fast oscillation (or TF) regime for an
arbitrary
potential of interaction $V(|\varphi|^2)$ and show that the SF behaves as a
barotropic fluid described by an equation of state $P(\rho)$ determined by the
potential. In Sec. \ref{sec_dmde} we show that the energy density $\epsilon$ can
be written as the sum of a rest-mass energy density $\rho_m c^2$ and an internal
energy density $u$ which play respectively the roles of DM and DE. We relate the
internal energy density $u$ to the SF potential $V$. We also introduce a
two-fluid
model associated with the single dark fluid (or SF) model. In Sec. \ref{sec_pwl}
we consider
power-law potentials associated with polytropic and isothermal equations of
state. In Sec. \ref{sec_kp} we describe all possible cosmological evolutions
corresponding to a positive polytropic constant $K>0$.
In Sec. \ref{sec_kn} we describe all possible cosmological evolutions
corresponding to a negative polytropic constant $K<0$. A summary of our main
results is provided in the conclusion. Complements about the general formalism
are given in the Appendices.

\section{Theory of a complex SF}
\label{sec_csf}

In this section,
we recall the basic
equations governing the cosmological evolution of a spatially homogeneous
complex SF with an arbitrary  self-interaction potential in a
Friedmann-Lema\^itre-Robertson-Walker (FLRW) universe. We also recall how these
equations can be simplified in the fast oscillation regime (equivalent to the
classical or TF approximation) that is considered in the following
sections. We refer to
our previous papers \cite{abrilphas,action,graal} and references therein for a
more
detailed
discussion.

\subsection{Spatially homogeneous SF}
\label{sec_hsf}

Let us consider a complex  SF $\varphi$ with a
self-interaction potential $V(|\varphi|^2)$ described  by the KG
equation. For a spatially homogeneous SF $\varphi(t)$ evolving in an
expanding background, the KG equation takes the form\footnote{See
Appendix 
\ref{sec_ra} for the general expression of the KG equation valid for
a
spatially inhomogeneous
SF.}
\begin{eqnarray}
\frac{1}{c^2}\frac{d^2\varphi}{dt^2}+\frac{3H}{c^2}\frac{d\varphi}{dt}+\frac
{m^2
c^2}{\hbar^2}\varphi
+2\frac{dV}{d|\varphi|^2}\varphi=0,
\label{hsf1}
\end{eqnarray}
where $H=\dot a/a$ is the Hubble parameter and $a(t)$ is the scale factor. The
second term in Eq. (\ref{hsf1}) is the Hubble drag. The
rest-mass term (third term) can be written as $\varphi/\lambda_C^2$ where
$\lambda_C=\hbar/mc$ is the Compton wavelength ($m$
is the mass of the SF). 
The total potential including the rest-mass term and
the self-interaction term writes
\begin{equation}
V_{\rm tot}(|\varphi|^2)=\frac{m^2c^2}{2\hbar^2}|\varphi|^2+V(|\varphi|^2).
\label{hsf1b}
\end{equation}
The energy density
$\epsilon(t)$ and the pressure $P(t)$ of the SF are given by (see, e.g.,
\cite{action} for details)
\begin{equation}
\epsilon=\frac{1}{2c^2}\left |\frac{d\varphi}{d
t}\right|^2+\frac{m^2c^2}{2\hbar^2}|\varphi|^2+V(|\varphi|^2),
\label{hsf2}
\end{equation}
\begin{equation}
P=\frac{1}{2c^2}\left |\frac{d\varphi}{d
t}\right|^2-\frac{m^2c^2}{2\hbar^2}|\varphi|^2-V(|\varphi|^2).
\label{hsf3}
\end{equation}
The equation of state parameter is defined by $w=P/\epsilon$.

The Friedmann equations determining the evolution of the  homogeneous
background are
\begin{equation}
\frac{d\epsilon}{dt}+3H(\epsilon+P)=0
\label{hsf4}
\end{equation}
and 
\begin{equation}
\label{fe1}
H^2=\frac{8\pi
G}{3c^2}\epsilon.
\end{equation}
Equation (\ref{fe1}) is valid in a  flat universe  ($k=0$)  without cosmological
constant ($\Lambda=0$).  
The Friedmann equations can be derived from the Einstein field equations by
using the FLRW metric. The  energy conservation  equation (\ref{hsf4})
results from the identity $D_{\nu}T^{\mu\nu}=0$, where $T^{\mu\nu}$ is the
energy-momentum tensor. It can
also be obtained from the KG equation (\ref{hsf1}) by using Eqs. (\ref{hsf2})
and (\ref{hsf3}) (see Appendix G of \cite{action}). Inversely,
the KG equation can be derived from Eqs. 
(\ref{hsf2})-(\ref{hsf4}). Once the SF
potential $V(|\varphi|^2)$
is given, the Klein-Gordon-Friedmann (KGF) equations
provide a complete set of
equations that can in principle be solved to obtain the evolution of the
universe assuming that the energy density is entirely due to the SF (for
simplicity we 
do not consider here the effect of other species like standard radiation and
baryonic matter).

{\it Remark:} From Eq. (\ref{hsf4}) we see that the energy density 
decreases with the scale factor when $w>-1$ and increases with the scale factor
when $w<-1$. In the second case, the universe has a phantom behavior (we will
see that this regime is not allowed for a complex SF in the fast oscillation
regime). On the other hand, when $k=\Lambda=0$, the deceleration parameter
$q=-\ddot a a/{\dot a}^2$ is given by $q=(1+3w)/2$. Therefore, the universe is
decelerating when $w>-1/3$ and accelerating when $w<-1/3$.

\subsection{Charge of the SF}
\label{sec_ext}

Writing the complex SF as
\begin{eqnarray}
\varphi=|\varphi|e^{i\theta},
\label{ext2}
\end{eqnarray}
where $|\varphi|$ is the modulus of the SF and $\theta$ is its phase (angle),
inserting this decomposition into the KG equation (\ref{hsf1}), and separating
the
real and imaginary parts, we obtain the following pair of equations
\begin{eqnarray}
\frac{1}{c^2}\left (2\frac{d|\varphi|}{dt}\frac{d\theta}{dt}+|\varphi|
\frac{d^2\theta}{dt^2}\right
)+\frac{3H}{c^2}|\varphi|\frac{d\theta}{dt}=0,
\label{ext4}
\end{eqnarray}
\begin{eqnarray}
\frac{1}{c^2}\left\lbrack \frac{d^2|\varphi|}{dt^2}-|\varphi| \left
(\frac{d\theta}{dt}\right
)^2\right\rbrack+\frac{3H}{c^2}\frac{d|\varphi|}{dt}\nonumber\\
+ \frac{m^2
c^2}{\hbar^2}|\varphi|+2\frac{dV}{d|\varphi|^2}|\varphi|=0.
\label{ext3}
\end{eqnarray}
Equation (\ref{ext4}) can be rewritten as a conservation equation
\begin{eqnarray}
\frac{d}{dt}\left (a^3|\varphi|^2\frac{d\theta}{dt}\right )=0.
\label{ext6}
\end{eqnarray}
Introducing the pulsation $\omega=-\dot\theta$, we get 
\begin{eqnarray}
\omega=\frac{Q\hbar c^2}{a^3|\varphi|^2},
\label{ext7}
\end{eqnarray}
where $Q$ is a
constant of integration which represents the charge of
the 
SF 
\cite{arbeycosmo,gh,shapiro,abrilph,abrilphas}. This conservation law
results from the identity $D_{\mu}J^{\mu}=0$ where $J^{\mu}$ is the KG current.
The conservation of charge is equivalent to the conservation of bosons provided
that anti-bosons are counted negatively \cite{landaulifshitz}. Eq. (\ref{ext3})
can be rewritten as 
\begin{equation}
\frac{d^2|\varphi|}{dt^2}- \omega^2|\varphi|+3H\frac{d|\varphi|}{dt}
+ \frac{m^2
c^4}{\hbar^2}|\varphi|+2c^2\frac{dV}{d|\varphi|^2}|\varphi|=0.
\label{ext3bn}
\end{equation}
Substituting  Eq. (\ref{ext7}) into Eq. (\ref{ext3bn}) we obtain the
differential
equation
\begin{equation}
\frac{d^2|\varphi|}{dt^2}+3H\frac{d|\varphi|}{dt}
+ \frac{m^2
c^4}{\hbar^2}|\varphi|
+2c^2\frac{dV}{d|\varphi|^2}|\varphi|-\frac{
Q^2\hbar^2c^4}{a^6|\varphi|^3}=0.
\label{suna1}
\end{equation}
This equation is exact. It determines the
evolution of the modulus of
the complex SF. It
differs from the KG equation of a real SF by the presence of the
last term and the fact that $\varphi$ is replaced by $|\varphi|$. On the other
hand, substituting Eq. (\ref{ext2}) into Eqs. (\ref{hsf2}) and (\ref{hsf3}) we
find that the energy density and the pressure are given by
\begin{equation}
\epsilon=\frac{1}{2c^2}\left (\frac{d|\varphi|}{d
t}\right )^2+\left (\frac{\omega^2}{2c^2}+\frac{m^2c^2}{2\hbar^2}\right )|\varphi|^2+V(|\varphi|^2),
\label{hsf2b}
\end{equation}
\begin{equation}
P=\frac{1}{2c^2}\left (\frac{d|\varphi|}{d
t}\right )^2+\left (\frac{\omega^2}{2c^2}-\frac{m^2c^2}{2\hbar^2}\right )|\varphi|^2-V(|\varphi|^2).
\label{hsf3b}
\end{equation}

\subsection{Fast oscillation regime and spintessence}
\label{sec_spin}

In the fast oscillation regime
$\omega=|d\theta/dt|\gg H=\dot a/a$ where the pulsation is high with respect to
the Hubble expansion rate, Eq.
(\ref{ext3bn})
reduces to\footnote{For a free field ($V=0$), the pulsation $\omega$ is
proportional
to the mass of the SF ($\omega=mc^2/\hbar$) and the fast oscillation
condition reduces to $mc^2/\hbar\gg H$.}
\begin{eqnarray}
\omega^2=\frac{m^2c^4}{\hbar^2}+2c^2\frac{dV}{d|\varphi|^2}.
\label{ext5}
\end{eqnarray}
This equation can be interpreted  as a condition of equilibrium between the centrifugal force
 $\omega^2 |\varphi|$ and the force  $c^2dV_{\rm tot}/d|\varphi|$ produced by the total SF potential (see Sec.
V.A. of \cite{abrilphas}). When this condition is satisfied, the direction
(phase) of the SF
rotates rapidly in the complex plane while its 
modulus changes slowly (adiabatically). 
This is what Boyle {\it et al.} \cite{spintessence} call
``spintessence''. There is no relation such as  Eq. (\ref{ext5}) for a real
SF. Combining Eqs. (\ref{ext7}) and
(\ref{ext5}), we obtain
\begin{equation}
\frac{Q^2\hbar^2c^4}{a^6|\varphi|^4}=\frac{m^2
c^4}{\hbar^2}
+2c^2\frac{dV}{d|\varphi|^2}.
\label{suna2}
\end{equation}
This equation relates the modulus $|\varphi|$  of the SF to the scale
factor $a$ in the fast oscillation regime. The pulsation $\omega$ of the SF is
then given by Eq. (\ref{ext7}) or (\ref{ext5}).

\subsection{Equation of state in the fast oscillation regime}
\label{sec_eosfo}

To establish the equation of state of the SF in the fast oscillation regime, we
can proceed as follows \cite{turner,ford,pv,mul,shapiro,abrilphas}.
Multiplying the KG
equation (\ref{hsf1}) by
$\varphi^*$ and averaging over a
time interval  that is much longer than the field oscillation period
$\omega^{-1}$, but much shorter than the Hubble time $H^{-1}$,  we obtain 
\begin{eqnarray}
\frac{1}{c^2}\left\langle \left |\frac{d\varphi}{dt}\right
|^2\right\rangle=\frac{m^2c^2}{\hbar^2}\langle |\varphi|^2\rangle+2\left\langle
\frac{dV}{d|\varphi|^2}|\varphi|^2\right\rangle.
\label{ext10}
\end{eqnarray}
This relation constitutes a sort of virial theorem. On the other hand, for a
spatially
homogeneous SF, the energy density and the pressure are given
by Eqs. (\ref{hsf2}) and (\ref{hsf3}). Taking the average value of the energy
density and pressure, using Eq.
(\ref{ext10}), and making the approximation
\begin{eqnarray}
\left\langle \frac{dV}{d|\varphi|^2}|\varphi|^2\right\rangle\simeq V'(\langle
|\varphi|^2\rangle)\langle |\varphi|^2\rangle,
\label{ext11}
\end{eqnarray} 
we get
\begin{eqnarray}
\langle\epsilon\rangle=\frac{m^2c^2}{\hbar^2}\langle
|\varphi|^2\rangle+V'(\langle
|\varphi|^2\rangle)\langle |\varphi|^2\rangle+V(\langle
|\varphi|^2\rangle),\quad
\label{ext12}
\end{eqnarray} 
\begin{eqnarray}
\langle P \rangle=V'(\langle
|\varphi|^2\rangle)\langle |\varphi|^2\rangle-V(\langle |\varphi|^2\rangle).
\label{ext13}
\end{eqnarray} 
The equation of state parameter is then given by
\begin{eqnarray}
w=\frac{P}{\epsilon}=\frac{V'(\langle
|\varphi|^2\rangle)\langle |\varphi|^2\rangle-V(\langle
|\varphi|^2\rangle)}{\frac{m^2c^2}{\hbar^2}\langle |\varphi|^2\rangle+V'(\langle
|\varphi|^2\rangle)\langle |\varphi|^2\rangle+V(\langle
|\varphi|^2\rangle)}.\nonumber\\
\label{ext14}
\end{eqnarray}
We note that the averages are not strictly necessary in Eqs. (\ref{ext12})-(\ref{ext14})
since the modulus of the
SF changes slowly with time. Eqs. (\ref{ext12}) and (\ref{ext13}) can also be
obtained from Eqs. (\ref{hsf2b}) and (\ref{hsf3b}) by using Eq. (\ref{ext5}) and
neglecting the term $(d|\varphi|/dt)^2$.

\subsection{Hydrodynamic variables and TF approximation}
\label{sec_ge}

Instead of working with the SF $\varphi(t)$, we can use hydrodynamic
variables.\footnote{See Appendixes \ref{sec_ra} and \ref{sec_mtt}, and our
previous
works 
\cite{abrilph,playa,abrilphas,chavmatos,action}, for a 
general hydrodynamical 
description of the SF valid for possibly inhomogeneous systems.} 
We write the SF in the de Broglie form (for a homogeneous SF)
\begin{eqnarray}
\varphi(t)=\frac{\hbar}{m}\sqrt{\rho(t)}e^{i
S_{\rm tot}(t)/\hbar},
\label{ge2}
\end{eqnarray}
where 
\begin{eqnarray}
\rho=\frac{m^2}{\hbar^2}|\varphi|^2
\label{ge1}
\end{eqnarray}
is the pseudo rest-mass density\footnote{We stress that it is
only in the nonrelativistic limit $c\rightarrow +\infty$
that $\rho$
has the interpretation of a rest-mass density (in this limit, we also have
$\epsilon\sim \rho c^2$). In the relativistic regime,
$\rho$ does not have a clear physical interpretation but it can
always be introduced as a convenient notation.}
 and $S_{\rm
tot}=(1/2)i\hbar\ln(\varphi^*/\varphi)$  is the total 
action of the SF. In terms of
the pseudo rest-mass density
the total potential (\ref{hsf1b}) can be written as
\begin{equation}
V_{\rm tot}(\rho)=\frac{1}{2}\rho c^2+V(\rho).
\label{hsf1brho}
\end{equation}
On the other hand, the total energy  of the SF (including its rest
mass  energy $mc^2$) is
\begin{eqnarray}
E_{\rm tot}(t)= -\frac{d S_{\rm tot}}{d
t}.
\label{ge3}
\end{eqnarray}

Substituting Eq. (\ref{ge2}) into the KG equation (\ref{hsf1}) and taking the
imaginary part, we obtain the conservation equation \cite{abrilph,abrilphas}
\begin{eqnarray}
\frac{d}{dt}\left (\rho E_{\rm tot} a^3\right )=0.
\label{ge3a}
\end{eqnarray}
It expresses the conservation of the
charge of the SF.\footnote{For a spatially homogeneous SF, the density
of charge (or rest-mass density $\rho_m$) is proportional to $\rho E_{\rm tot}$
(see
Sec. \ref{sec_dmde} and Appendix \ref{sec_mtt}).} It can be
integrated into 
\begin{eqnarray}
\rho\frac{E_{\rm tot}}{mc^2}=\frac{Qm}{a^3},
\label{ge4}
\end{eqnarray}
where $Q$ is the charge of the SF.
These equations are equivalent to Eqs. (\ref{ext6}) and
(\ref{ext7}).\footnote{To make the
link between the SF variables and the hydrodynamical variables, we use
$|\varphi|=(\hbar/m)\sqrt{\rho}$,
$\theta=S_{\rm tot}/\hbar$ and $\omega=-\dot\theta=-\dot S_{\rm
tot}/\hbar=E_{\rm tot}/\hbar$.} 
Next, substituting Eq. (\ref{ge2}) into the KG equation
(\ref{hsf1}), taking the real part, and making the TF approximation
$\hbar\rightarrow 0$, we obtain the Hamilton-Jacobi (or Bernoulli) equation
\cite{abrilph,abrilphas} 
\begin{eqnarray}
E_{\rm tot}^2=m^2c^4+2m^2c^2V'(\rho).
\label{ge3b}
\end{eqnarray}
This equation is equivalent to Eq. (\ref{ext5}). It can be
rewritten as
\begin{eqnarray}
E_{\rm tot}=mc^2\sqrt{1+\frac{2}{c^2} V'(\rho)}. 
\label{ge5}
\end{eqnarray}
Note that Eq. (\ref{ge3b}) requires that
\begin{eqnarray}
V'_{\rm tot}(\rho)=1+\frac{2}{c^2} V'(\rho)>0.
\label{ge5b}
\end{eqnarray}
Combining Eqs. (\ref{ge4})  and (\ref{ge5}), we obtain
\begin{eqnarray}
\rho \sqrt{1+\frac{2}{c^2}V'(\rho)}=\frac{Qm}{a^3},
\label{ge6}
\end{eqnarray}
which corresponds to Eq. (\ref{suna2}).  Finally, writing Eqs. (\ref{hsf2}) and 
(\ref{hsf3}) in terms of hydrodynamic variables, making the TF approximation
$\hbar\rightarrow 0$, and using the Bernoulli equation (\ref{ge3b}), we get
\cite{abrilph,abrilphas} 
\begin{eqnarray}
\epsilon=\rho c^2+V(\rho)+\rho V'(\rho), 
\label{ge7}
\end{eqnarray}
\begin{eqnarray}
P=\rho
V'(\rho)-V(\rho),
\label{ge8}
\end{eqnarray}
in agreement with Eqs. (\ref{ext12}) and (\ref{ext13}).  Eq.
(\ref{ge8})
determine the equation of state $P(\rho)$ for a given potential $V(\rho)$.
Inversely, for a given equation of state, the potential is given by\footnote{We
note that the potential is defined from the
pressure up to a term of the form $A\rho$, where $A$ is a constant. If we add a
term $A\rho$ in the potential $V$, we do not change the pressure $P$ but we
introduce a term $2A\rho$ in the energy density. On the other hand, if we add a
constant term $C$ in the potential $V$ (cosmological constant), this adds a term
$-C$ in the pressure and a term $+C$ in the energy density. }
\begin{eqnarray}
V(\rho)=\rho\int \frac{P(\rho)}{\rho^2}\, d\rho.
\label{ge9b}
\end{eqnarray}

The various
correspondances between the results of this section and the results of the
previous sections
show that the fast oscillation regime ($\omega\gg H$) is equivalent to the TF
or semiclassical approximation ($\hbar\rightarrow 0$). We note that we cannot
directly take $\hbar=0$ in the KG equation (this is why we have to average
over the oscillations) while we can take 
$\hbar=0$ in the hydrodynamic equations (see Refs.
\cite{abrilph,abrilphas,action} for
details). This is an interest of the hydrodynamic representation of the SF. It
can be  shown  (see, e.g., \cite{action} and Appendix \ref{sec_ra}) that
 Eqs. (\ref{ge7}) and (\ref{ge8}) remain valid for a spatially inhomogeneous
SF in the TF limit. They determine the equation of
state
$P=P(\epsilon)$
in parametric form.  The equation of state
parameter can be written
as
\begin{eqnarray}
w=\frac{P}{\epsilon}=\frac{\rho
V'(\rho)-V(\rho)}{\rho c^2+V(\rho)+\rho V'(\rho)},
\label{ge8b}
\end{eqnarray}
which is equivalent to Eq.
(\ref{ext14}). We note that the condition from Eq. (\ref{ge5b}) implies $w>-1$ so that a universe described by a complex SF in the fast oscillation regime has never a phantom behavior. On the other hand, the universe is decelerating ($w>-1/3$) when $4\rho
V'(\rho)-2V(\rho)>-\rho c^2$ (or equivalently $2\rho V_{\rm tot}'(\rho)>V_{\rm
tot}(\rho)$) and accelerating in the opposite case. Finally, the pseudo squared
speed of sound is
\begin{eqnarray}
c_s^2=P'(\rho)=\rho V''(\rho),
\label{ge9}
\end{eqnarray}
while the
true  squared speed of sound is
\begin{eqnarray}
c_s^2=P'(\epsilon)=\frac{\rho V''(\rho)}{c^2+2V'(\rho)+\rho V''(\rho)}.
\label{ge9c}
\end{eqnarray}

\subsection{Cosmological evolution of a spatially homogeneous complex SF in the
fast
oscillation regime}
\label{sec_rm}

The differential equation determining the evolution of the scale
factor $a(t)$ of the Universe
induced by a spatially  homogeneous complex SF in the regime where its oscillations
are faster than the Hubble expansion is given by [see Eqs. (\ref{fe1}),
(\ref{ge6}) and (\ref{ge7})]
\begin{eqnarray}
\frac{3}{8\pi G}\left
(\frac{\dot a}{a}\right )^2=\rho+\frac{1}{c^2}\left\lbrack V(\rho)+\rho
V'(\rho)\right\rbrack
\label{b14}
\end{eqnarray}
with 
\begin{eqnarray}
\rho \sqrt{1+\frac{2}{c^2}V'(\rho)}=\frac{Qm}{a^3},
\label{b13}
\end{eqnarray}
Actually, it is not convenient to solve the differential equation (\ref{b14}) for the
scale factor $a$ because we have to inverse Eq. (\ref{b13}) in order to
express
$\rho$ as a function of $a$ in the r.h.s. of Eq. (\ref{b14}). It is
more convenient to view $a$ as a function of $\rho$, given by Eq.
(\ref{b13}), and
transform Eq. (\ref{b14}) into a differential equation for $\rho$. Taking the
logarithmic derivative of Eq. (\ref{b13}), we get
\begin{eqnarray}
\frac{\dot a}{a}=-\frac{1}{3}\frac{\dot \rho}{\rho}\left \lbrack
1+\frac{\rho V''(\rho)}{c^2+2V'(\rho)}\right\rbrack.
\label{b15}
\end{eqnarray}
Substituting this expression into Eq. (\ref{b14}), we obtain the differential
equation
\begin{eqnarray}
\frac{c^2}{24\pi G}\left
(\frac{\dot\rho}{\rho}\right )^2=\frac{\rho
c^2+V(\rho)+\rho V'(\rho)}{\left\lbrack 1+\frac{\rho
V''(\rho)}{c^2+2V'(\rho)}\right\rbrack^2}.
\label{b16}
\end{eqnarray}
For a given SF potential $V(\rho)$, this equation can be solved easily as it
is just a first order differential equation for $\rho$.
Therefore, $t(\rho)$ can be expressed in the form of an integral.
The temporal evolution
of the scale factor $a(t)$ is then obtained in parametric form $a=a(\rho)$ and
$t=t(\rho)$ from  Eqs. (\ref{b13}) and
(\ref{b16}).

\section{Effective dark matter and dark energy}
\label{sec_dmde}

In our model, there is no DM and no DE considered as two different and
independent species. There
is just a single SF which can be represented, through the de Broglie
transformation, as a single dark fluid. It therefore  provides a
unified dark matter and dark energy (UDM) model. In this section, we
interpret this
dark fluid as a barotropic gas at $T=0$ and determine its
rest-mass density $\rho_m$, internal energy density $u$ and equation of state
$P(\rho_m)$ in terms of the SF potential. This equivalence is valid
only in the TF approximation.
Following \cite{epjp,lettre,stiff}, we argue that the rest-mass density
$\rho_m$ 
plays the
role of DM and the internal energy density $u$ plays the role of DE. This
provides a
simple
physical interpretation of these two mysterious components.

\subsection{First principle of thermodynamics}

The first principle of
thermodynamics for a relativistic gas can be
written as
\begin{equation}
\label{mtd1}
d\left (\frac{\epsilon}{\rho_m}\right )=-Pd\left
(\frac{1}{\rho_m}\right )+Td\left (\frac{s}{\rho_m}\right ),
\end{equation}
where
\begin{eqnarray}
\label{mtd2}
\epsilon=\rho_m c^2+u(\rho_m)
\end{eqnarray}
is the energy density including the rest-mass energy density $\rho_m c^2$ (where
$\rho_m=n m$ is the rest-mass density) and the internal energy density
$u(\rho_m)$, $s$ is the entropy density, $P$ is the pressure, and $T$ is the
temperature. We assume that
$Td(s/\rho_m)=0$. This corresponds
to cold ($T=0$) or isentropic ($s/\rho_m={\rm cst}$) gases. In that
case, Eq.
(\ref{mtd1}) reduces to
\begin{equation}
\label{mtd3}
d\left (\frac{\epsilon}{\rho_m}\right )=-Pd\left
(\frac{1}{\rho_m}\right )=\frac{P}{\rho_m^2}\, d\rho_m.
\end{equation}
This equation can be rewritten as
\begin{equation}
\label{mtd4}
\frac{d\epsilon}{d\rho_m}=\frac{P+\epsilon}{\rho_m},
\end{equation}
where the term on the right hand side is the  enthalpy $h$  (or the chemical
potential $\mu=mh$). We have
\begin{equation}
\label{mtd4a}
h=\frac{P+\epsilon}{\rho_m},
\qquad h=\frac{d\epsilon}{d\rho_m},\qquad dh=\frac{dP}{\rho_m}. 
\end{equation}
Equation (\ref{mtd3}) can be
integrated
into
\begin{eqnarray}
\label{mtd5}
\epsilon=\rho_m c^2+\rho_m \int \frac{P(\rho_m)}{\rho_m^2}\, d\rho_m,
\end{eqnarray}
establishing that
\begin{eqnarray}
\label{mtd6}
u(\rho_m)=\rho_m \int \frac{P(\rho_m)}{\rho_m^2}\, d\rho_m.
\end{eqnarray}
This equation determines the internal energy density as a function of the
equation of
state $P(\rho_m)$. Inversely, the equation of state is determined by the
internal energy density $u(\rho_m)$ through the relation
\begin{equation}
\label{mtd7}
P(\rho_m)=-\frac{d(\epsilon/\rho_m)}{d(1/\rho_m)}=\rho_m^2\left\lbrack
\frac{u(\rho_m)}{\rho_m}\right\rbrack'=\rho_m u'(\rho_m)-u(\rho_m).
\end{equation}
We note that
\begin{eqnarray}
\label{mtd8}
P'(\rho_m)=\rho_m u''(\rho_m).
\end{eqnarray}

\subsection{Application to cosmology}

The previous results are general. We now consider a spatially homogeneous gas in
an expanding universe. Combining the energy conservation equation (\ref{hsf4})
with
the first
principle of thermodynamics [Eq. (\ref{mtd4})], we obtain
\begin{eqnarray}
\frac{d\rho_m}{dt}+3H\rho_m=0.
\label{mtd10}
\end{eqnarray}
This equation expresses the conservation of the particle number (or
rest-mass). It can be integrated into $\rho_m\propto a^{-3}$.
Inserting this relation into Eq. (\ref{mtd2}), we see that $\rho_m$ plays the
role of DM while $u$ plays the role of DE. This decomposition
provides therefore a simple interpretation of DM and DE in terms of a
single DF \cite{epjp,lettre,stiff}. Owing to this interpretation, we can write
\begin{eqnarray}
\rho_m c^2=\frac{\Omega_{\rm m,0}\epsilon_0}{a^3}
\label{mtd11}
\end{eqnarray}
and
\begin{eqnarray}
\label{mtd2b}
\epsilon=\frac{\Omega_{\rm m,0}\epsilon_0}{a^3}+u\left (\frac{\Omega_{\rm
m,0}\epsilon_0}{c^2a^3} \right ),
\end{eqnarray}
where $\epsilon_0$ is the present energy density of the universe and
$\Omega_{\rm m,0}$ is the present proportion of DM.
For given $P(\rho_m)$ or
$u(\rho_m)$ we can get  $\epsilon(a)$ from Eq.  (\ref{mtd2b}). We can then solve
the Friedmann equation (\ref{fe1}) to obtain the temporal evolution of the scale
factor $a(t)$ \cite{epjp,lettre,stiff}.

{\it Remark:} Eqs. (\ref{mtd2}) and (\ref{mtd7}) determine
the equation of state $P=P(\epsilon)$. As a result, we can obtain Eq.
(\ref{mtd11}) directly from Eqs.
(\ref{mtd2}), (\ref{mtd7}) and the energy conservation equation
(\ref{hsf4}).
Indeed, combining these equations we obtain Eq. (\ref{mtd10}) which  integrates
to give Eq. (\ref{mtd11}).

\subsection{Determination of the rest-mass density and internal energy density}

We now relate the rest-mass density $\rho_m$
(proportional to the charge density) and the
internal energy density $u$ of
the barotropic gas to the pseudo rest-mass density $\rho$ and potential $V$ of
the corresponding SF in the TF approximation. According to Eq. (\ref{mtd4}), we
have
\begin{equation}
\label{gest3}
\ln\rho_m=\int \frac{\epsilon'(\rho)}{\epsilon+P}\, d\rho.
\end{equation}
Using Eqs. (\ref{ge7}) and (\ref{ge8}), which are valid in the TF
approximation, we obtain
\begin{equation}
\label{gest4}
\ln\rho_m=\int \frac{c^2+\rho V''(\rho)+2V'(\rho)}{\rho c^2+2\rho
V'(\rho)}\, d\rho.
\end{equation}
This equation determines the function $\rho_m(\rho)$. It can be
explicitly integrated into
\begin{equation}
\label{gest4a}
\rho_m=\rho\sqrt{1+\frac{2}{c^2}V'(\rho)},
\end{equation}
where the constant of integration has been determined in 
order to obtain $\rho_m=\rho$ in the nonrelativistic limit. This relation
can also be established in the manner described in Appendix \ref{sec_ra} [see
Eq.
(\ref{charge10c})]. We emphasize that Eq. (\ref{gest4a}) is always valid in the
TF
approximation, even
for an inhomogeneous SF. Eliminating $\rho$
between $P(\rho)$ [see Eq. (\ref{ge8})] and $\rho_m(\rho)$ [see Eq.
(\ref{gest4a})]
we obtain the equation of state of the gas under the form $P(\rho_m)$. On the
other hand,
according to Eq. (\ref{mtd2}), its internal energy density can be obtained from
the
relation
\begin{equation}
\label{gest5}
u=\epsilon-\rho_m c^2.
\end{equation}
From Eqs.
(\ref{ge7}) and (\ref{gest4a}), we get
\begin{equation}
\label{gest6}
u=\rho c^2+\rho V'(\rho)+V(\rho)-\rho c^2\sqrt{1+\frac{2}{c^2}V'(\rho)}.
\end{equation}
where $\rho=\rho(\rho_m)$ can be obtained from Eq. (\ref{gest4a}). Therefore,
the rest-mass density (DM) is determined by Eq. (\ref{gest4a}) and the
internal energy density (DE) is determined by Eq. (\ref{gest6}). Eliminating
$\rho$ between Eqs. (\ref{gest4a}) and (\ref{gest6}) we obtain 
$u(\rho_m)$. We can also obtain $u(\rho_m)$ from $P(\rho_m)$, or the converse,
by using Eqs. (\ref{mtd6}) and (\ref{mtd7}). As mentioned above, $\rho_m$ mimics
DM and $u$ mimics DE. However, in our model, we have just one fluid (or just one
complex SF).

For a spatially homogeneous SF in a cosmological context
and in the TF approximation, combining  the Hamilton-Jacobi (or Bernoulli)
equation (\ref{ge5}) with Eq. (\ref{gest4a}), we find that the
relation between the rest-mass density $\rho_m$ and the pseudo rest-mass density
$\rho$ is 
\begin{eqnarray}
\rho_m=\rho\frac{E_{\rm
tot}}{mc^2}.
\label{rmd1}
\end{eqnarray}
Using Eqs. (\ref{gest4a}) and (\ref{rmd1}),  Eqs. (\ref{ge4}) and
(\ref{ge6}) can be rewritten as
\begin{eqnarray}
\rho_m=\frac{Qm}{a^3}.
\label{rmd2}
\end{eqnarray}
The rest-mass density (or the charge density) decreases as
$a^{-3}$. This  expresses the conservation
of the charge of the SF or, equivalently, the conservation of the boson
minus antiboson number. Comparing Eqs.
(\ref{mtd11}) and (\ref{rmd2}), we establish that
\begin{equation}
\label{gest6q}
Qmc^2=\Omega_{\rm m,0}\epsilon_0.
\end{equation}
We note that the constant $Qmc^2$ (which is proportional to the charge of the
SF) is equal to the
present energy density of DM $\epsilon_{{\rm m},0}=\Omega_{{\rm
m},0}\epsilon_0$.

{\it Remark:} In the homogeneous case, we can directly deduce the relation
\begin{eqnarray}
\rho_m=\rho\frac{E_{\rm
tot}}{mc^2}=\rho\sqrt{1+\frac{2}{c^2}V'(\rho)}
\end{eqnarray}
from Eqs. (\ref{ge4}) and (\ref{ge6}), but this derivation is less general than
the calculations detailed above [in particular Eq. (\ref{gest4a})] which remain
valid for inhomogeneous systems.

\subsection{Two-fluid model}
\label{sec_twofluids}

As mentioned above, in our model, we have a single dark fluid
with an equation of state $P=P(\rho_m)$. Still, the
energy density (\ref{mtd2}) is the sum of two terms, a rest-mass density term
$\rho_m$ which mimics DM and an internal energy density term $u(\rho_m)$ which
mimics
DE. It is interesting to consider a two-fluid model which leads to the
same results as our single dark fluid model, at least for what concerns the
evolution of the homogeneous background. In this two-fluid model, one fluid
corresponds to pressureless DM with an equation of state $P_{\rm m}=0$ and a
density $\rho_m c^2=\Omega_{\rm m,0}\epsilon_0/a^3$ determined by the energy
conservation equation for DM, and the other fluid corresponds to DE with an
equation of state $P_{\rm de}(\epsilon_{\rm de})$ and an energy density
$\epsilon_{\rm de}(a)$ determined by the energy
conservation equation for DE. We can obtain the equation of state of DE yielding
the same results as the one-fluid model by taking
\begin{eqnarray}
P_{\rm de}=P(\rho_m),\qquad \epsilon_{\rm de}=u(\rho_m)
\end{eqnarray}
or, equivalently, 
\begin{eqnarray}
P_{\rm de}=P(\rho),\qquad \epsilon_{\rm de}=u(\rho).
\end{eqnarray}
In other words, the  equation of state $P_{\rm de}(\epsilon_{\rm de})$ of DE
 in the two-fluid model corresponds to the
relation $P(u)$ in the single fluid (or SF) model. An explicit example of the
equivalence between the one and two-fluid models for the homogeneous background
is given in Secs. \ref{sec_kpg0} and \ref{sec_kpn0} in connection to the
(anti-)Chaplygin gas. We note that although
the one and two-fluid models are equivalent for the evolution  of the
homogeneous background, they may differ for what concerns the formation of the
large-scale structures of the Universe.

\section{Power-law potentials}
\label{sec_pwl}

The equations of Secs. \ref{sec_csf} and \ref{sec_dmde} are general. We now
consider 
specific potentials $V(|\varphi|^2)$ associated with polytropic and isothermal
equations of state.

\subsection{Polytropic equation of state}
\label{sec_pwlnon}

We first consider a power-law SF potential that we write under the form  (see
also Appendix I of \cite{abrilphas}) 
\begin{eqnarray}
V(|\varphi|^2)=\frac{K}{\gamma-1}\left (\frac{m}{\hbar}\right
)^{2\gamma}|\varphi|^{2\gamma}\qquad (\gamma\neq 1).
\label{plsf1}
\end{eqnarray}
For the sake of generality, we consider arbitrary values of $K$  (positive and
negative) 
and arbitrary values of $\gamma$. The total potential
including the rest-mass term is
\begin{eqnarray}
V_{\rm tot}(|\varphi|^2)=\frac{1}{2}\frac{m^2c^2}{\hbar^2}|\varphi|^2+\frac{K}{\gamma-1}\left (\frac{m}{\hbar}\right
)^{2\gamma}|\varphi|^{2\gamma}.
\label{plsf1b}
\end{eqnarray}
Introducing the pseudo rest-mass density defined by Eq. (\ref{ge1}), we have
\begin{eqnarray}
V(\rho)=\frac{K}{\gamma-1}\rho^{\gamma}
\label{plsf2}
\end{eqnarray}
and
\begin{eqnarray}
V_{\rm tot}(\rho)=\frac{1}{2}\rho c^2+\frac{K}{\gamma-1}\rho^{\gamma}.
\label{plsf2b}
\end{eqnarray}
Using Eqs. (\ref{ge8}),
(\ref{ge9}) and (\ref{plsf2}), we find that the
derivative of the potential, the pressure and the 
squared speed of sound are given by
\begin{eqnarray}
V'(\rho)=\frac{K\gamma}{\gamma-1}\rho^{\gamma-1},
\label{plsf2c}
\end{eqnarray}
\begin{eqnarray}
P=K\rho^{\gamma},
\label{plsf3}
\end{eqnarray}
\begin{eqnarray}
c_s^2=K\gamma\rho^{\gamma-1}.
\label{plsf3b}
\end{eqnarray}
We see that the equation of state (\ref{plsf3}) associated with the power-law
potential 
(\ref{plsf1}) is that of a polytrope with polytropic constant $K$ and polytropic
index $\gamma=1+1/n$. The pressure is positive when $K>0$ and negative when
$K<0$.
Negative pressures play an important role in cosmology. They are necessary to
account for the early
inflation and the present accelerating expansion of the universe (DE).
We note that the potential $V$ given by Eq. (\ref{plsf2}) is similar to the
Tsallis free
energy density $V=-Ks_\gamma$, where the polytropic constant $K$ plays the role
of
a generalized temperature and $s_\gamma=-\frac{1}{\gamma-1}(\rho^{\gamma}-\rho)$
is the Tsallis entropy density.

For the power-law potential from Eq. (\ref{plsf2}) the equations of the
problem [Eqs. (\ref{ge5}), (\ref{ge6}), (\ref{ge7}), (\ref{ge8b}) and
(\ref{b14})] become
\begin{eqnarray}
\rho
\sqrt{1+\frac{2}{c^2}\frac{K\gamma}{\gamma-1}\rho^{\gamma-1}}=\frac{Qm}{a^3},
\label{log3}
\end{eqnarray}
\begin{eqnarray}
\epsilon=\rho c^2+\frac{\gamma+1}{\gamma-1}K\rho^{\gamma},
\label{log4}
\end{eqnarray}
\begin{eqnarray}
\frac{E_{\rm
tot}}{mc^2}=\sqrt{1+\frac{2}{c^2}\frac{K\gamma}{\gamma-1}\rho^{
\gamma-1}},
\label{log5}
\end{eqnarray}
\begin{eqnarray}
\frac{3H^2}{8\pi G}=\rho+\frac{\gamma+1}{\gamma-1}\frac{K}{c^2}\rho^{\gamma},
\label{plsf5}
\end{eqnarray}
\begin{eqnarray}
w=\frac{\frac{K}{c^2}\rho^{\gamma-1}}{1+\frac{\gamma+1}{\gamma-1}\frac{K}{c^2}
\rho^{\gamma-1}}.
\label{plsf8}
\end{eqnarray}
From Eqs.
(\ref{plsf3}) and (\ref{log4}), we obtain the equation of state $P(\epsilon)$ of
the SF under the inverse form $\epsilon(P)$ as
\begin{eqnarray}
\epsilon=\left (\frac{P}{K}\right )^{1/\gamma}c^2+\frac{\gamma+1}{\gamma-1}P.
\label{plsf10}
\end{eqnarray}
On the other hand, the differential equation governing the
temporal
evolution of the pseudo rest-mass density [see Eq. (\ref{b16})] is
\begin{eqnarray}
\frac{c^2}{24\pi G}\left
(\frac{\dot\rho}{\rho}\right )^2=\frac{\rho
c^2+\frac{K(\gamma+1)}{\gamma-1}\rho^{\gamma}}{\left\lbrack
1+\frac{K\gamma\rho^{\gamma-1}}{c^2+\frac{2K\gamma}{\gamma-1}
\rho^{\gamma-1}}\right\rbrack^2 }.
\label{plsf11}
\end{eqnarray}

The rest-mass density (DM) and the internal energy density (DE) of the
SF are determined by
Eqs.
(\ref{gest4a}) and (\ref{gest6}) with
Eq. (\ref{plsf2}). We get
\begin{equation}
\rho_m=\rho\sqrt{1+\frac{2\gamma}{\gamma-1}\frac{K}{c^2}\rho^{\gamma-1}},
\label{theo2}
\end{equation}
\begin{equation}
u=\rho
c^2+\frac{\gamma+1}{\gamma-1}K\rho^{\gamma}-\rho
c^2\sqrt{1+\frac{2\gamma}{\gamma-1} \frac{K}{c^2}\rho^{\gamma-1}}.
\label{gpd9}
\end{equation}
We note that the rest-mass density (\ref{theo2}) can be read-off directly from
Eq. (\ref{log3}). Eqs. (\ref{plsf3}), (\ref{theo2}) and
(\ref{gpd9}) define $P(\rho_m)$ and
$u(\rho_m)$ 
in parametric form with parameter $\rho$. As we have recalled
in Sec. \ref{sec_dmde}, the rest mass density $\rho_m$ of the
SF mimics DM and the internal energy density $u$ of the SF mimics DE
\cite{epjp,lettre}.

The foregoing equations determine the cosmological evolution of a
spatially homogeneous complex SF described by the potential (\ref{plsf1}) in the
fast
oscillation regime for any values of $K$ and $\gamma$.
Some values of $\gamma$ are of particular interest.

(i) For $\gamma=-1$ ($n=-1/2$), we obtain
\begin{eqnarray}
V=-\frac{K}{2\rho},
\end{eqnarray}
\begin{eqnarray}
P=\frac{K}{\rho},
\label{chap1}
\end{eqnarray}
\begin{eqnarray}
\epsilon=\rho c^2,
\label{chap2}
\end{eqnarray}
leading to
\begin{eqnarray}
P=\frac{Kc^2}{\epsilon}.
\label{chap3}
\end{eqnarray}
This is the equation of state of the Chaplygin ($K<0$) or anti-Chaplygin ($K>0$)
gas \cite{kmp,btv,gkmp,cosmopoly2}. The total SF potential is
\begin{eqnarray}
V_{\rm tot}=\frac{1}{2}\rho c^2-\frac{K}{2\rho}
\label{cgpt1}
\end{eqnarray}
or, equivalently,
\begin{eqnarray}
V_{\rm tot}(|\varphi|^2)=\frac{1}{2}\frac{m^2c^2}{\hbar^2}|\varphi|^2-\frac{K}{2}\left (\frac{\hbar}{m}\right
)^{2}\frac{1}{|\varphi|^{2}}.
\label{chap4}
\end{eqnarray}
 It corresponds to an inverse square law self-interaction potential
$V(|\varphi|^2)\sim |\varphi|^{-2}$. The rest-mass
density and
the internal energy density are explicitly given by
\begin{eqnarray}
\rho_m c^2=\sqrt{(\rho c^2)^2+Kc^2},
\label{cgpu1dn}
\end{eqnarray}
\begin{eqnarray}
\rho c^2=\sqrt{(\rho_m c^2)^2-Kc^2},
\label{cgpu1cn}
\end{eqnarray}
\begin{eqnarray}
P=\frac{Kc^2}{\sqrt{(\rho_m c^2)^2-Kc^2}},
\label{cgpu1can}
\end{eqnarray}
\begin{eqnarray}
u=\rho c^2-\sqrt{(\rho c^2)^2+Kc^2},
\label{aswe}
\end{eqnarray}
\begin{eqnarray}
u=\sqrt{(\rho_m c^2)^2-Kc^2}-\rho_m c^2.
\label{cgpu1cbn}
\end{eqnarray}
In
the present context, the Chaplygin gas model is justified from a complex SF
theory. Furthermore, the pseudo rest-mass
density coincides with the energy density ($\epsilon=\rho c^2$). 

(ii) For $\gamma=2$ ($n=1$), we obtain
\begin{eqnarray}
V=K\rho^{2},
\end{eqnarray}
\begin{eqnarray}
P=K\rho^{2},
\label{bec1}
\end{eqnarray}
\begin{eqnarray}
\epsilon=\rho c^2+3K\rho^{2},
\label{bec2}
\end{eqnarray}
leading to
\begin{eqnarray}
\rho=\frac{-c^2\pm\sqrt{c^4+12K\epsilon}}{6K}
\label{bec3}
\end{eqnarray}
and
\begin{eqnarray}
P=\frac{1}{36K} \left (-c^2\pm\sqrt{c^4+12K\epsilon}\right )^2.
\label{bec4}
\end{eqnarray}
We must select the sign $+$ when $K>0$ while the two signs $\pm$ are allowed
when $K<0$. Alternatively, we can write the equation of state under the form
\begin{eqnarray}
\epsilon=\sqrt{\frac{P}{K}}c^2+3P.
\label{becpt3b}
\end{eqnarray}
The total SF potential is
\begin{eqnarray}
V_{\rm tot}=\frac{1}{2}\rho c^2+K\rho^{2}
\label{becpt1}
\end{eqnarray}
or, equivalently,
\begin{eqnarray}
V_{\rm tot}(|\varphi|^2)=\frac{1}{2}\frac{m^2c^2}{\hbar^2}|\varphi|^2+K\left (\frac{m}{\hbar}\right
)^{4}|\varphi|^{4}.
\label{bec5}
\end{eqnarray}
It corresponds to a quartic self-interaction potential
$V(|\varphi|^2)\sim|\varphi|^4$. This is the standard potential of a
relativistic BEC. It takes into account two-body interactions in a weakly
interacting microscopic theory of superfluidity. In that case,
$K={2\pi
a_s\hbar^2}/{m^3}$. The self-interaction is repulsive when
$K>0$ and attractive
when $K<0$. The
relativistic BEC model has been studied
in detail in the context of
boson stars \cite{colpi,partially,chavharko} and in
cosmology \cite{shapiro,abrilphas}.
For $\epsilon\rightarrow +\infty$ we have $P\sim
\epsilon/3$ (dark
radiation) and for $\epsilon\rightarrow +\infty$ we have $P\sim
K(\epsilon/c^2)^2$ (matter). The rest-mass density and the internal energy
density are
given by
\begin{eqnarray}
\rho_m=\rho\sqrt{1+\frac{4K}{c^2}\rho},
\label{becpu1cn}
\end{eqnarray}
\begin{eqnarray}
u=\rho c^2+3K\rho^2-\rho c^2\sqrt{1+\frac{4K}{c^2}\rho}.
\end{eqnarray}
With Eq. (\ref{bec1}), they define $P(\rho_m)$ and $u(\rho_m)$ in parametric
form. Actually, Eq.
(\ref{becpu1cn}) is a cubic equation for $\rho$ which can be solved by standard
means to get $\rho(\rho_m)$. We can then obtain  $P(\rho_m)$ and $u(\rho_m)$
explicitly.

(iii) For $\gamma=0$ ($n=-1$), we obtain
\begin{eqnarray}
V=-K,
\end{eqnarray}
\begin{eqnarray}
P=K,
\label{cdm1}
\end{eqnarray}
\begin{eqnarray}
\epsilon=\rho c^2-K.
\label{cdm2}
\end{eqnarray}
The pressure is
constant. This is the equation of state of the $\Lambda$CDM
($K<0$) 
or anti-$\Lambda$CDM  ($K>0$) model interpreted as an UDM
model \cite{sandvik,avelinoZ,cosmopoly2}. In that case $K=\mp\rho_{\Lambda}c^2$
where $\rho_{\Lambda}$ is
the cosmological
density. The pressure can be
rewritten as $P=\mp\rho_{\Lambda}c^2$ and the energy density as
$\epsilon=\rho c^2\pm\rho_{\Lambda}c^2$. In the present case, the pseudo
rest-mass
density $\rho$ plays the role
of DM and $\pm\rho_{\Lambda}c^2$ the role of DE. The total SF
potential is
\begin{eqnarray}
V_{\rm tot}=\frac{1}{2}\rho c^2-K
\label{troisL1}
\end{eqnarray}
or, equivalently,
\begin{eqnarray}
V_{\rm tot}(|\varphi|^2)=\frac{1}{2}\frac{m^2c^2}{\hbar^2}|\varphi|^2-K.
\label{cdm3}
\end{eqnarray}
It corresponds to a constant self-interaction potential
$V(|\varphi|^2)=-K=\pm\rho_\Lambda c^2$ equal to the cosmological energy
density.
The
rest-mass density and
the internal energy density are explicitly given by
\begin{eqnarray}
P=K, \qquad \rho_m=\rho, \qquad u=-K.
\label{troisL3}
\end{eqnarray}
In
the present context, the $\Lambda$CDM model is justified from a complex SF
theory. Furthermore, the pseudo rest-mass
density coincides with the rest-mass density
($\rho_m=\rho$).

(iv)  For $\gamma=3$ ($n=1/2$), we obtain
\begin{eqnarray}
V=\frac{1}{2} K\rho^3,
\end{eqnarray}
\begin{eqnarray}
P=K\rho^3,
\label{sup1}
\end{eqnarray}
\begin{eqnarray}
\epsilon=\rho c^2+2K\rho^3.
\label{sup2}
\end{eqnarray}
Equation (\ref{sup2}) is a cubic equation for
$\rho$ which can be solved by standard means to get
$\rho(\epsilon)$. Using Eq. (\ref{sup1}), we can then obtain the equation of
state $P(\epsilon)$ explicitly. Alternatively, we can write the equation of
state under the form 
\begin{eqnarray}
\epsilon=\left (\frac{P}{K}\right )^{1/3} c^2+2P.
\end{eqnarray}
The
total SF potential is
\begin{eqnarray}
V_{\rm tot}=\frac{1}{2}\rho c^2+\frac{1}{2} K\rho^3,
\label{sup3}
\end{eqnarray}
or, equivalently,
\begin{eqnarray}
V_{\rm
tot}(|\varphi|^2)=\frac{1}{2}\frac{m^2c^2}{\hbar^2}|\varphi|^2+\frac{1}{2}
K \left (\frac{m}{\hbar}\right )^6 |\varphi|^6.
\label{sup4}
\end{eqnarray}
It corresponds to a sextic self-interaction potential
$V(|\varphi|^2)\sim |\varphi|^{6}$ (see, e.g., \cite{phi6}). It takes into
account three-body
interactions in a weakly
interacting microscopic theory of superfluidity. It can also
describe an exotic DM superfluid with a different interpretation
\cite{ferreira}. The rest-mass density and
the internal energy density are explicitly given by
\begin{eqnarray}
\rho_m =\rho \sqrt{1+\frac{3K}{c^2}\rho^2},
\label{sup5}
\end{eqnarray}
\begin{eqnarray}
\rho=\left(
-\frac{c^2}{6K}\pm\frac{c^2}{6K}\sqrt{1+\frac{12K}{c^2}\rho_m^2}\right )^{1
/2},
\label{sup6}
\end{eqnarray}
\begin{eqnarray}
P=K\left (
-\frac{c^2}{6K}\pm\frac{c^2}{6K}\sqrt{1+\frac{12K}{c^2}\rho_m^2}
\right )^{3/2},
\label{sup7}
\end{eqnarray}
\begin{eqnarray}
u=\left (
-\frac{c^2}{6K}\pm\frac{c^2}{6K}\sqrt{1+\frac{12K}{c^2}\rho_m^2}
\right )^{1/2}\nonumber\\
\times\left
(\frac{2}{3}\pm\frac{1}{3}\sqrt{1+\frac{12K}{c^2}\rho_m^2}\right )c^2-\rho_m
c^2.
\label{sup8}
\end{eqnarray}

(v)  For $\gamma=1/2$ ($n=-2$),\footnote{An interpretation of this index is
given in Sec. \ref{sec_kn0g1}.} we obtain
\begin{eqnarray}
V=-2 K\sqrt{\rho},
\end{eqnarray}
\begin{eqnarray}
P=K\sqrt{\rho},
\label{gsup1}
\end{eqnarray}
\begin{eqnarray}
\epsilon=\rho c^2-3K\sqrt{\rho}.
\label{gsup2}
\end{eqnarray}
Equation (\ref{gsup2}) can be reversed to give
\begin{eqnarray}
\rho=\left (\frac{3K}{2c^2}\pm\frac{1}{2c^2}\sqrt{9K^2+4c^2\epsilon}\right )^2
\label{gsup5}
\end{eqnarray}
and
\begin{eqnarray}
P=\frac{3K^2}{2c^2}\pm\frac{K}{2c^2}\sqrt{9K^2+4c^2\epsilon}.
\label{gsup6}
\end{eqnarray}
Alternatively, we can write the equation of state under the form
\begin{eqnarray}
\epsilon=\left (\frac{P}{K}\right )^2 c^2-3P.
\end{eqnarray}
The total SF potential is
\begin{eqnarray}
V_{\rm tot}=\frac{1}{2}\rho c^2-2 K\sqrt{\rho},
\label{gsup3}
\end{eqnarray}
or, equivalently,
\begin{eqnarray}
V_{\rm
tot}(|\varphi|^2)=\frac{1}{2}\frac{m^2c^2}{\hbar^2}|\varphi|^2-2
K\frac{m}{\hbar}|\varphi|.
\label{gsup4}
\end{eqnarray}
It corresponds to a linear self-interaction potential $V(|\varphi|^2)\sim
|\varphi|$. The rest-mass density and
the internal energy density are given by
\begin{eqnarray}
\rho_m=\sqrt{\rho^2-\frac{2K}{c^2}\rho^{3/2}},
\label{gsup7}
\end{eqnarray}
\begin{eqnarray}
u=\rho c^2-3K\sqrt{\rho}-\rho_m c^2.
\label{gsup8}
\end{eqnarray}
With Eq. (\ref{gsup1}) they define $P(\rho_m)$ and $u(\rho_m)$ in parametric
form but we cannot have more explicit results.

{\it Remark:} We emphasize that the previous equations are valid for a possibly
inhomogeneous SF. We also 
note that the polytropic indices $\gamma=0$ and $\gamma=-1$
are the only ones for which the polytropic equation
of state $P=K\rho^{\gamma}$ yields a polytropic equation of state $P=K
(\epsilon/c^2)^{\gamma}$.

\subsection{Isothermal equation of state}
\label{sec_pwloui}

We now consider  a potential of the form 
\begin{eqnarray}
V(|\varphi|^2)=\frac{k_B T m}{\hbar^2}|\varphi|^2\left\lbrack \ln\left
(\frac{m^2|\varphi|^2}{\rho_*\hbar^2}\right )-1\right\rbrack.
\label{iplsf1}
\end{eqnarray}
For the sake of generality, we consider arbitrary values of $T$  (positive and negative). 
The total potential including the rest-mass term is
\begin{equation}
V_{\rm tot}(|\varphi|^2)=\frac{1}{2}\frac{m^2c^2}{\hbar^2}|\varphi|^2+\frac{k_B
T m}{\hbar^2}|\varphi|^2\left\lbrack \ln\left
(\frac{m^2|\varphi|^2}{\rho_*\hbar^2}\right )-1\right\rbrack.
\label{iso1}
\end{equation}
Introducing the pseudo rest-mass density defined by Eq. (\ref{ge1}), we have
\begin{eqnarray}
V(\rho)=\frac{k_B T}{m}\rho\left\lbrack \ln\left (\frac{\rho}{\rho_*}\right
)-1\right\rbrack
\label{iplsf2}
\end{eqnarray}
and
\begin{eqnarray}
V_{\rm tot}(\rho)=\frac{1}{2}\rho c^2+\frac{k_B T}{m}\rho\left\lbrack \ln\left (\frac{\rho}{\rho_*}\right )-1\right\rbrack.
\label{iso2b}
\end{eqnarray}
Using Eqs. (\ref{ge8}), (\ref{ge9}) and (\ref{iplsf2}), we find that the
derivative of the potential, the pressure and the squared speed of sound are
given by
\begin{eqnarray}
V'(\rho)=\frac{k_B T}{m} \ln\left (\frac{\rho}{\rho_*}\right ),
\label{iplsf2b}
\end{eqnarray}
\begin{eqnarray}
P(\rho)=\rho\frac{k_B T}{m},
\label{iplsf3}
\end{eqnarray}
\begin{eqnarray}
c_s^2=\frac{k_B T}{m}.
\label{iplsf3b}
\end{eqnarray}
We see that the equation 
of state (\ref{iplsf3}) associated with the potential (\ref{iplsf1}) is the
isothermal equation
of state with an effective temperature $T$. It corresponds to a polytrope of
index $\gamma=1$ ($n\rightarrow +\infty$). The pressure is positive when $T>0$
and negative when $T<0$.
Negative pressures play an important role in cosmology in
relation to the early inflation and the present accelerating expansion of the
universe (DE). A positive temperature ($T>0$) can account for
thermal effects in DM. We note that the potential $V$ given by Eq.
(\ref{iplsf2}) is similar to the
Boltzmann free
energy density $V=-Ts_B$, where $T$ is the temperature and
$s_B=-k_B(\rho/m)[\ln(\rho/\rho_*)-1]$ is the Boltzmann entropy density.

For the 
potential (\ref{iplsf1}) the equations of the
problem [Eqs. (\ref{ge5}), (\ref{ge6}), (\ref{ge7}), (\ref{ge8b}) and
(\ref{b14})] become
\begin{eqnarray}
\rho
\sqrt{1+\frac{2k_B T}{m c^2}\ln\left (\frac{\rho}{\rho_*}\right )}
=\frac{Qm}{a^3},
\label{iso1b}
\end{eqnarray}
\begin{eqnarray}
\epsilon=\rho c^2+\frac{2k_B T}{m}\rho\ln\left (\frac{\rho}{\rho_*}\right
)-\frac{k_B T}{m}\rho,
\label{iso2}
\end{eqnarray}
\begin{eqnarray}
\frac{E_{\rm tot}}{mc^2}=\sqrt{1+\frac{2k_B T}{m c^2}\ln\left
(\frac{\rho}{\rho_*}\right )},
\label{iso3}
\end{eqnarray}
\begin{eqnarray}
\frac{3H^2}{8\pi G}=\rho+\frac{2k_B T}{mc^2}\rho\ln\left
(\frac{\rho}{\rho_*}\right )-\frac{k_B T}{mc^2}\rho,
\label{iso4}
\end{eqnarray}
\begin{eqnarray}
w=\frac{1}{\frac{mc^2}{k_B T}+2\ln\left (\frac{\rho}{\rho_*}\right )-1}.
\label{iso5}
\end{eqnarray}
From Eqs.
(\ref{iplsf3}) and (\ref{iso2}), we obtain the equation of state
$P(\epsilon)$ of
the SF under the inverse
form $\epsilon(P)$ as
\begin{eqnarray}
\epsilon=\frac{mc^2}{k_B T}P+2P\ln\left (\frac{mP}{\rho_* k_B T}\right )-P.
\label{iso6}
\end{eqnarray}
Finally, the differential equation governing the temporal
evolution of the pseudo rest-mass density [see Eq. (\ref{b16})] is
\begin{eqnarray}
\frac{c^2}{24\pi G}\left
(\frac{\dot\rho}{\rho}\right )^2=\frac{\rho c^2+\frac{2k_B T}{m}\rho\ln\left (\frac{\rho}{\rho_*}\right )-\frac{k_B T}{m}\rho}{\left\lbrack 1+\frac{\frac{k_B T}{mc^2}}{1+\frac{2k_B T}{mc^2}\ln\left (\frac{\rho}{\rho_*}\right )}\right \rbrack^2}.
\label{iso7}
\end{eqnarray}

The rest-mass density (DM) and the internal energy density (DE) of the
SF are determined by
Eqs.
(\ref{gest4a}) and (\ref{gest6}) with
Eq. (\ref{iplsf2}). We get
\begin{equation}
\rho_m=\rho\sqrt{1+\frac{2k_B T}{m c^2}\ln\left
(\frac{\rho}{\rho_*}\right )},
\label{theo1}
\end{equation}
\begin{eqnarray}
u=\rho c^2+\frac{2 k_B T}{m}\rho \ln\left(\frac{\rho}{\rho_*}\right
)-\frac{k_B T}{m}\rho\nonumber\\
-\rho
c^2\sqrt{1+\frac{2k_B T}{mc^2}\ln\left (\frac{\rho}{\rho_*}\right
)}.
\label{gpd8}
\end{eqnarray}
Eqs. (\ref{iplsf3}), (\ref{theo1}) and (\ref{gpd8})  determine  
$P(\rho_m)$ and
$u(\rho_m)$ in parametric form with parameter $\rho$.  As we
have recalled
in Sec. \ref{sec_dmde}, the rest mass density $\rho_m$ of the
SF mimics DM and the internal energy density $u$ of the SF mimics DE
\cite{epjp,lettre}.

The foregoing equations determine the cosmological evolution of a
spatially homogeneous complex SF described by the potential (\ref{iplsf1}) in
the fast
oscillation regime for any value of $T$.

{\it Remark:} We emphasize that the previous equations are valid for a possibly
inhomogeneous SF. The results of
this section (isothermal systems) can be recover from
the results of Sec. \ref{sec_pwlnon} (polytropes) in the limit
$\gamma\rightarrow 1$. For example,
\begin{eqnarray}
V=\frac{K}{\gamma-1}\rho^{\gamma}=\frac{K\rho}{\gamma-1}e^{(\gamma-1)\ln\rho}\nonumber\\
\simeq \frac{K\rho}{\gamma-1}\left \lbrack 1+(\gamma-1)\ln\rho\right
\rbrack+...=K\rho\ln\rho+{\rm cst}.
\label{iso7b}
\end{eqnarray}
This corresponds to the passage from the Tsallis to the Boltzmann free energy
when $\gamma\rightarrow 1$.

\section{The case $K>0$}
\label{sec_kp}

In this section, we consider the case of a positive polytropic 
constant ($K>0$), corresponding to a positive pressure. Since $P>0$, the
universe is always decelerating. In the figures, we take
$c=Qm=4\pi G=1$ and $K=1$.\footnote{The case $K=0$ corresponds to pure DM with
$P=V=u=0$ and $\epsilon=\rho c^2=\rho_m c^2=\Omega_{\rm m,0}\epsilon_0/a^3$.}

\subsection{The case $\gamma>1$}
\label{sec_kpg1}

For $\gamma>1$ the equations determining the pseudo rest-mass density and the energy density as a function of the scale factor can be written as
\begin{eqnarray}
\rho
\sqrt{1+\frac{2}{c^2}\frac{K\gamma}{\gamma-1}\rho^{\gamma-1}}
=\frac{Qm}{a^3},
\label{e1}
\end{eqnarray}
\begin{eqnarray}
\epsilon=\rho c^2+\frac{\gamma+1}{\gamma-1}K\rho^{\gamma}.
\label{e2}
\end{eqnarray}

When $\rho\rightarrow +\infty$, Eqs. (\ref{e1}) and (\ref{e2}) reduce to
\begin{eqnarray}
\rho\sim \left(\frac{Q^2m^2c^2}{2}\frac{\gamma-1}{K\gamma}\right
)^{1/(1+\gamma)} \frac{1}{a^{6/(1+\gamma)}},
\label{e3}
\end{eqnarray}
\begin{eqnarray}
\epsilon\sim
\frac{\gamma+1}{\gamma-1}K\rho^{\gamma},
\label{e4}
\end{eqnarray}
\begin{eqnarray}
\epsilon\sim
\frac{\gamma+1}{\gamma-1}K\left(\frac{Q^2m^2c^2}{2}\frac{\gamma-1}{K\gamma}
\right
)^{\gamma/(1+\gamma)} \frac{1}{a^{6\gamma/(1+\gamma)}}.
\label{e5}
\end{eqnarray}
This corresponds to the ultrarelativistic regime valid for $a\rightarrow 0$.
Starting from $+\infty$ when $a\rightarrow 0$, the pseudo rest-mass density 
and the energy density decrease as $a$ increases. Using Eqs. (\ref{plsf3}) and
(\ref{e4}), we obtain the equation of state
\begin{eqnarray}
P\sim \frac{\gamma-1}{\gamma+1}\epsilon.
\label{e7}
\end{eqnarray}
The pressure is a linear function $P\sim\alpha\epsilon$ of the energy density
 with coefficient $\alpha=(\gamma-1)/(\gamma+1)$. For
$\gamma=2$, we recover the equation of state of the dark radiation
$P=\epsilon/3$ due to a complex SF with a repulsive $|\varphi|^4$
self-interaction \cite{shapiro,abrilphas}.

When $\rho\rightarrow 0$, Eqs. (\ref{e1}) and (\ref{e2}) reduce to
\begin{eqnarray}
\rho\sim\frac{Q m}{a^3},
\label{e8}
\end{eqnarray}
\begin{eqnarray}
\epsilon\sim \rho c^2.
\label{e9}
\end{eqnarray}
This corresponds to the nonrelativistic regime $P/\epsilon\ll 1$ (matterlike
era) valid for
$a\rightarrow +\infty$. The pseudo rest-mass
density and the energy density decrease to zero as $a$ increases to $+\infty$.

\begin{figure}[!h]
\begin{center}
\includegraphics[clip,scale=0.3]{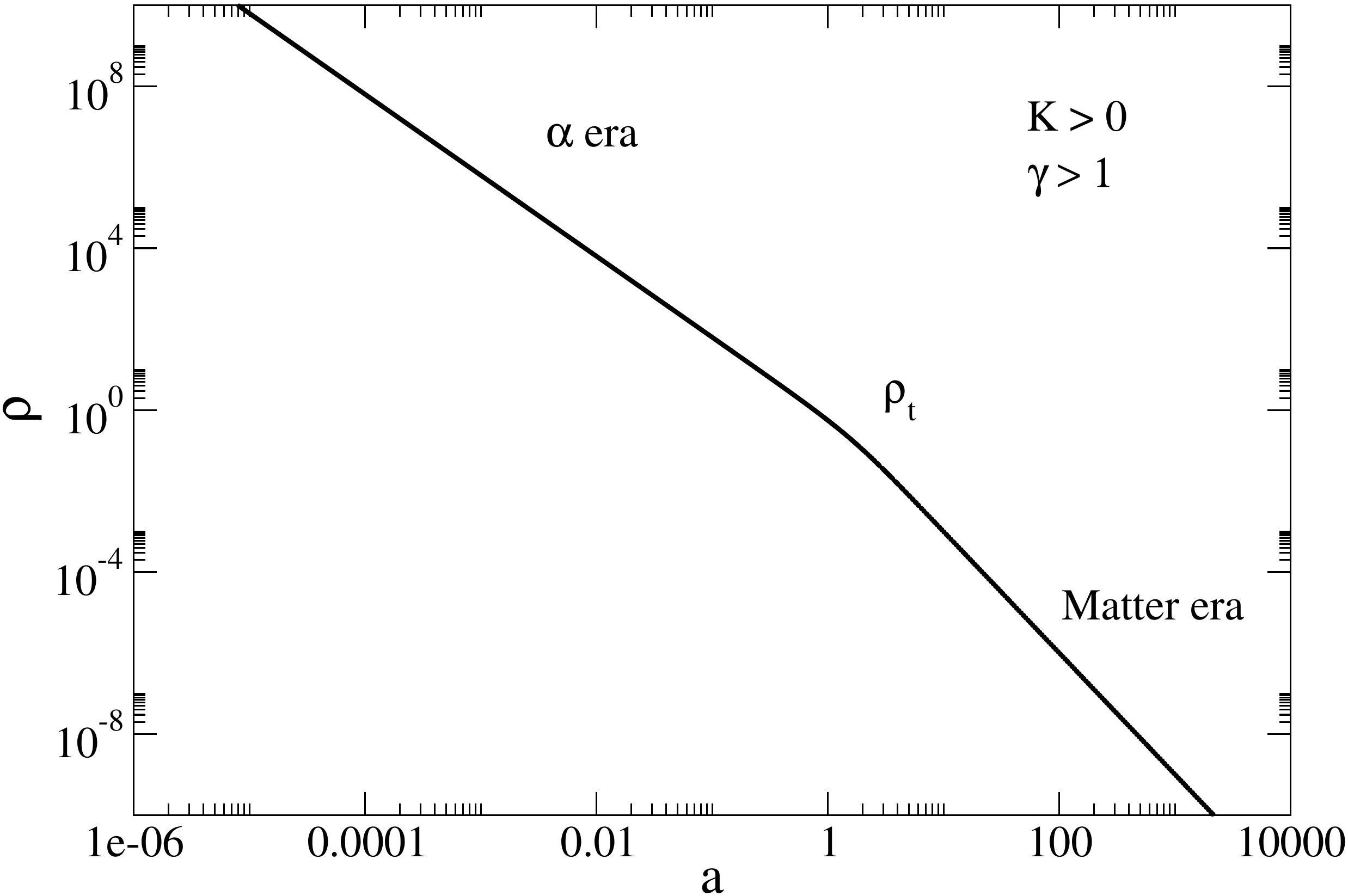}
\caption{Evolution of the pseudo rest-mass density as a function of the scale
factor for $\gamma>1$ (specifically $\gamma=2$). The pseudo rest-mass density
decreases more rapidly in the matterlike era than in the $\alpha$-era.}
\label{arKposg2}
\end{center}
\end{figure}

\begin{figure}[!h]
\begin{center}
\includegraphics[clip,scale=0.3]{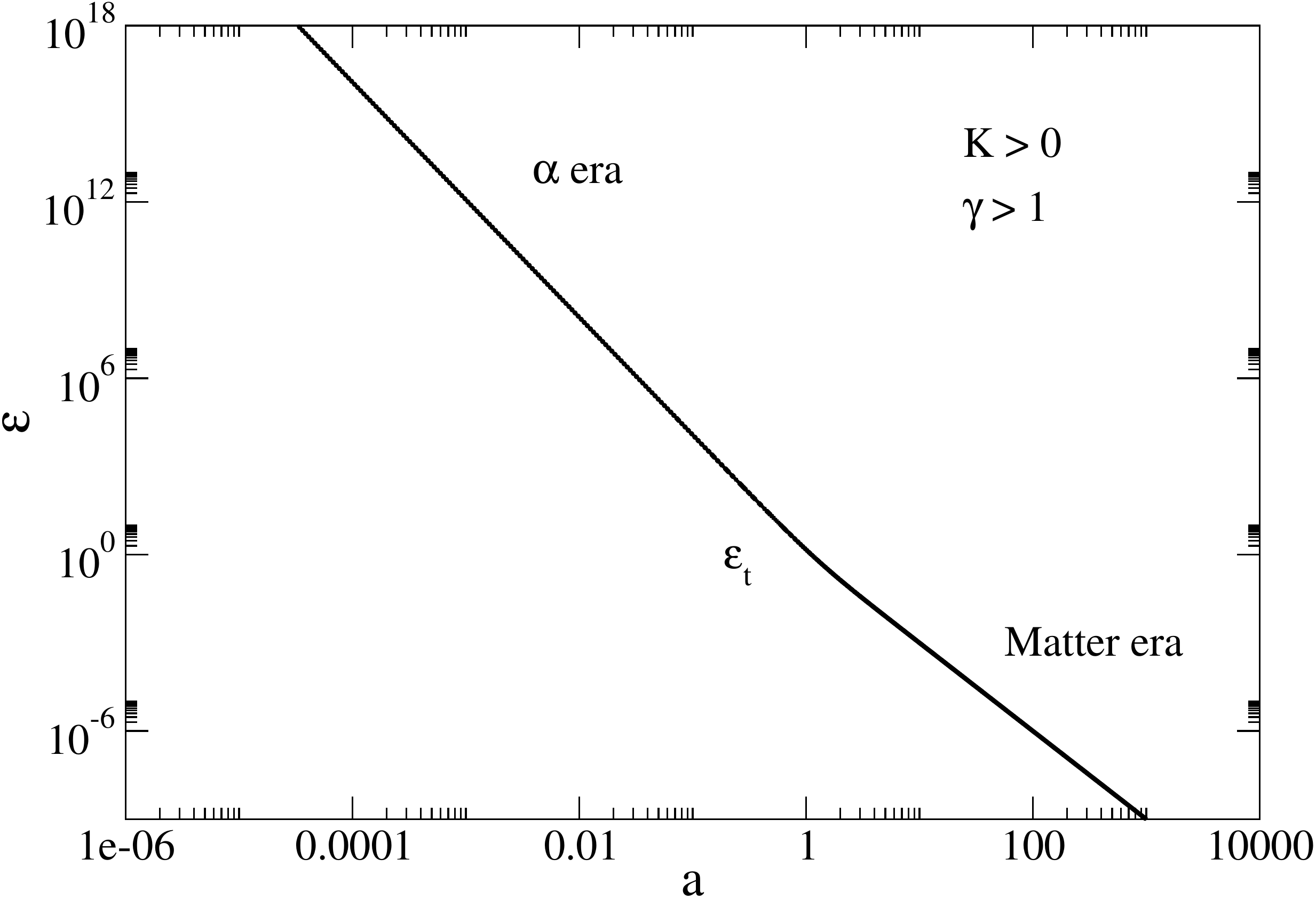}
\caption{Evolution of the energy density as a function of the scale factor
for $\gamma>1$ (specifically $\gamma=2$). The energy density
decreases more rapidly in the $\alpha$-era than in the matterlike era.}
\label{aepsKposg2}
\end{center}
\end{figure}

When $K>0$ and $\gamma>1$, the universe evolves from an $\alpha$-era 
($P\sim\alpha\epsilon$) in the early universe to a
matterlike era ($P\simeq 0$) in the late universe. The case $\gamma=2$ and
$K>0$, corresponding to a relativistic  BEC
with a repulsive $|\varphi|^4$ self-interaction, is treated in detail in Sec.
III of \cite{abrilphas}. In that case, the universe
evolves from a dark radiation era to a matterlike era.  The curves $\rho(a)$ and
$\epsilon(a)$ are plotted in Figs. \ref{arKposg2} and \ref{aepsKposg2}.

\begin{figure}[!h]
\begin{center}
\includegraphics[clip,scale=0.3]{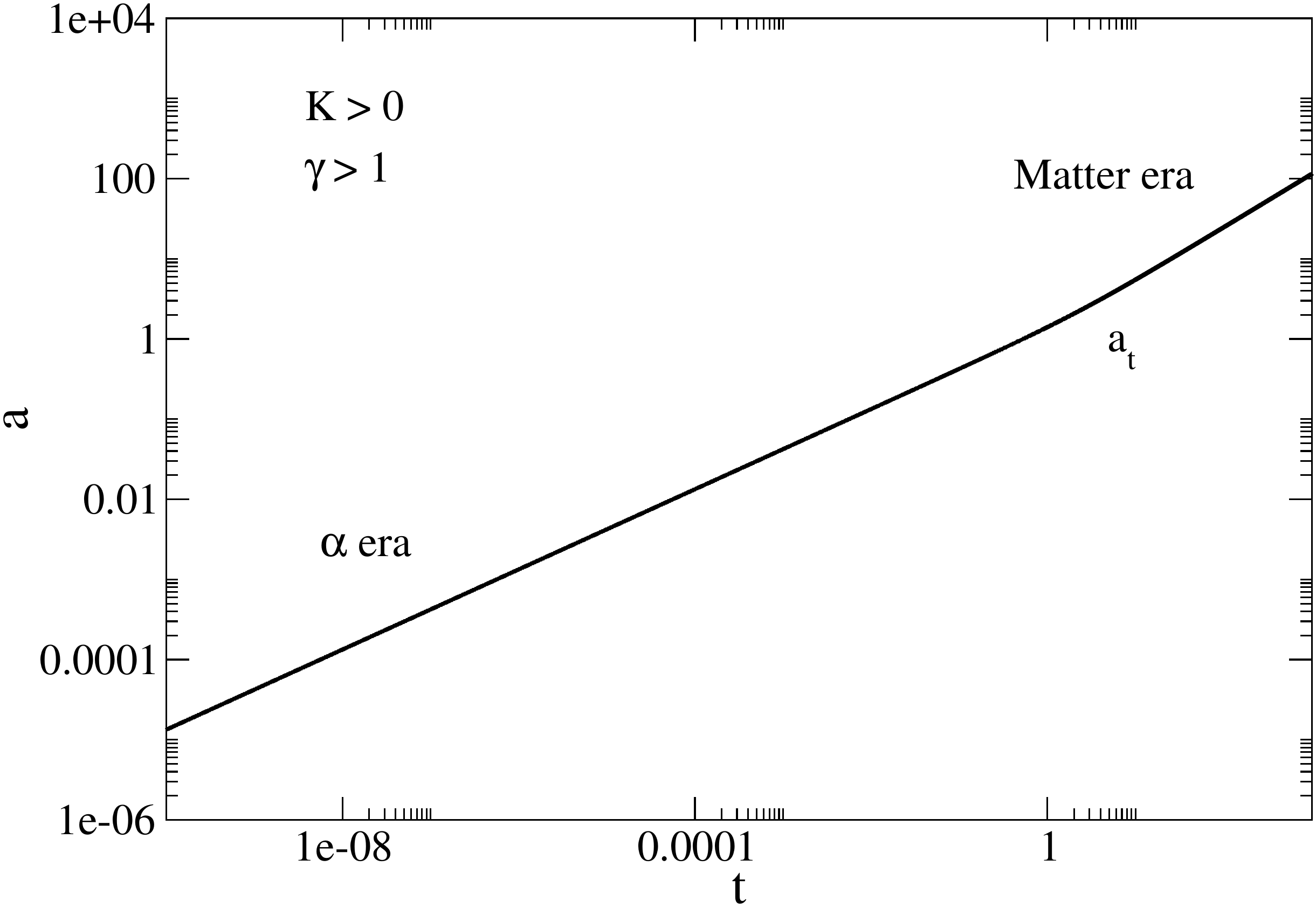}
\caption{Evolution of the scale factor as a function of time for
$\gamma>1$ (specifically $\gamma=2$). The scale factor increases more rapidly
in the matter era than in the $\alpha$-era.}
\label{taKposg2}
\end{center}
\end{figure}

The temporal evolution $a(t)$ of the scale factor is represented in Fig.
\ref{taKposg2}. It is obtained by integrating Eq. (\ref{plsf11})
numerically. 
Starting
from a singularity at $t=0$ where $a=0$ and $\epsilon\rightarrow +\infty$
(big bang) the scale factor first grows as $a\propto t^{(\gamma+1)/3\gamma}$
(corresponding to $a\propto
t^{2/[3(1+\alpha)]}$) in
the $\alpha$-era then as $a\propto t^{2/3}$ in the matterlike era
[Einstein-de Sitter (EdS) solution].

The transition between the two regimes  typically occurs at 
\begin{eqnarray}
a_t=(Qm)^{\frac{1+\gamma}{3(\gamma-1)}}\left (\frac{2}{Q^2m^2c^2}\frac{K\gamma}{\gamma-1}\right )^{\frac{1}{3(\gamma-1)}},
\label{e9b}
\end{eqnarray}
\begin{eqnarray}
\epsilon_t=\rho_t c^2=\left (\frac{c^2}{2}\frac{\gamma-1}{K\gamma}\right )^{\frac{1}{\gamma-1}}c^2.
\label{e9c}
\end{eqnarray}

\subsection{The case $\gamma=1$}
\label{sec_kpeq1}

The case $\gamma=1$ must be treated specifically. It corresponds to the
SF potential from Eq. (\ref{iplsf1}) associated with the isothermal
equation of state (\ref{iplsf3}).
In
the present case $T>0$. The equations determining the pseudo rest-mass density
and the energy density as a function of the scale factor are
\begin{eqnarray}
\rho
\sqrt{1+\frac{2k_B T}{m c^2}\ln\left (\frac{\rho}{\rho_*}\right )}
=\frac{Qm}{a^3},
\label{ie1}
\end{eqnarray}
\begin{eqnarray}
\epsilon=\rho c^2+\frac{2k_B T}{m}\rho\ln\left (\frac{\rho}{\rho_*}\right )-\frac{k_B T}{m}\rho.
\label{ie2}
\end{eqnarray}

When $\rho\rightarrow +\infty$, Eqs. (\ref{ie1}) and (\ref{ie2}) reduce to
\begin{eqnarray}
\rho
\sqrt{\frac{2k_B T}{m c^2}\ln\rho}
\sim\frac{Qm}{a^3},
\label{ie3}
\end{eqnarray}
\begin{eqnarray}
\epsilon\sim \frac{2k_B T}{m}\rho\ln\rho.
\label{ie4}
\end{eqnarray}
This regime is valid for $a\rightarrow 0$. Starting from $+\infty$ when
$a\rightarrow 0$, 
the pseudo rest-mass density and the energy density decrease as $a$ increases.
This corresponds to a matterlike era ($\epsilon\simeq a^{-3}$) modified by
logarithmic corrections. Using Eqs. (\ref{iplsf3}) and (\ref{ie4}), we
obtain the
equation of state
\begin{eqnarray}
P\simeq \frac{\epsilon}{2\ln\epsilon}.
\label{ie5}
\end{eqnarray}
The pressure is an approximately linear function of the energy density
$P\simeq \alpha\epsilon$ with a logarithmic correction yielding a small
effective coefficient  $\alpha\sim 1/(2\ln\epsilon)\ll 1$. For $a\rightarrow 0$,
we have
$P/\epsilon\ll 1$.

Considering now small values of $\rho$, we see that Eq. (\ref{ie1}) imposes
the condition $\rho\ge \rho_{\rm Min}$ with
\begin{eqnarray}
\rho_{\rm Min}=\rho_* e^{-\frac{mc^2}{2k_B T}}.
\label{ie6}
\end{eqnarray}
The pseudo rest-mass density decreases as $a$ increases and tends to $\rho_{\rm
Min}$ when $a\rightarrow +\infty$. However, we must be careful that the energy
density vanishes before $\rho$ reaches its absolute minimum value. Indeed,
$\epsilon_{\rm Min}=-\rho_{\rm Min}{k_B T/m}<0$ when $\rho=\rho_{\rm Min}$.
According to Eq. (\ref{ie2}), the energy density vanishes ($\epsilon=0$) at
\begin{eqnarray}
\rho_{\rm min}=\rho_* e^{\frac{1}{2}-\frac{mc^2}{2k_B T}}.
\label{ie7}
\end{eqnarray}
This happens when the scale factor reaches the value
\begin{equation}
a_{\rm max}=\left (\frac{Qm}{\rho_*}\sqrt{\frac{mc^2}{k_B
T}}e^{-\frac{1}{2}+\frac{mc^2}{2k_B T}}\right )^{1/3}.
\label{ie8}
\end{equation}
The solution of Eqs. (\ref{ie1}) and (\ref{ie2}) is defined only for $a\le
a_{\rm max}$. The pseudo rest-mass density and the energy density decrease as
$a$ increases. When $a=a_{\rm max}$, the energy density vanishes while the
pseudo rest-mass density reaches its minimum accessible value $\rho_{\rm min}$. 
This evolution is illustrated in the following section which 
presents similar features.

\subsection{The case $0<\gamma<1$}
\label{sec_kp0g1}

For $0<\gamma<1$ the equations determining the pseudo rest-mass density and the energy density as a function of the scale factor can be written as
\begin{eqnarray}
\rho
\sqrt{1-\frac{2}{c^2}\frac{K\gamma}{1-\gamma}\frac{1}{\rho^{1-\gamma}}}=\frac{
Qm } { a^3 } ,
\label{e10}
\end{eqnarray}
\begin{eqnarray}
\epsilon=\rho c^2-\frac{\gamma+1}{1-\gamma}K\rho^{\gamma}.
\label{e11}
\end{eqnarray}

When $\rho\rightarrow +\infty$, Eqs. (\ref{e10}) and (\ref{e11}) reduce to
\begin{eqnarray}
\rho\sim\frac{Qm}{a^3},
\label{e12}
\end{eqnarray}
\begin{eqnarray}
\epsilon\sim \rho c^2.
\label{e13}
\end{eqnarray}
This corresponds to the nonrelativistic regime $P/\epsilon\ll 1$ (matterlike
era) valid for
$a\rightarrow 0$. Starting from $+\infty$ when
$a\rightarrow 0$, the pseudo rest-mass density and the energy density decrease
as $a$ increases.

\begin{figure}[!h]
\begin{center}
\includegraphics[clip,scale=0.3]{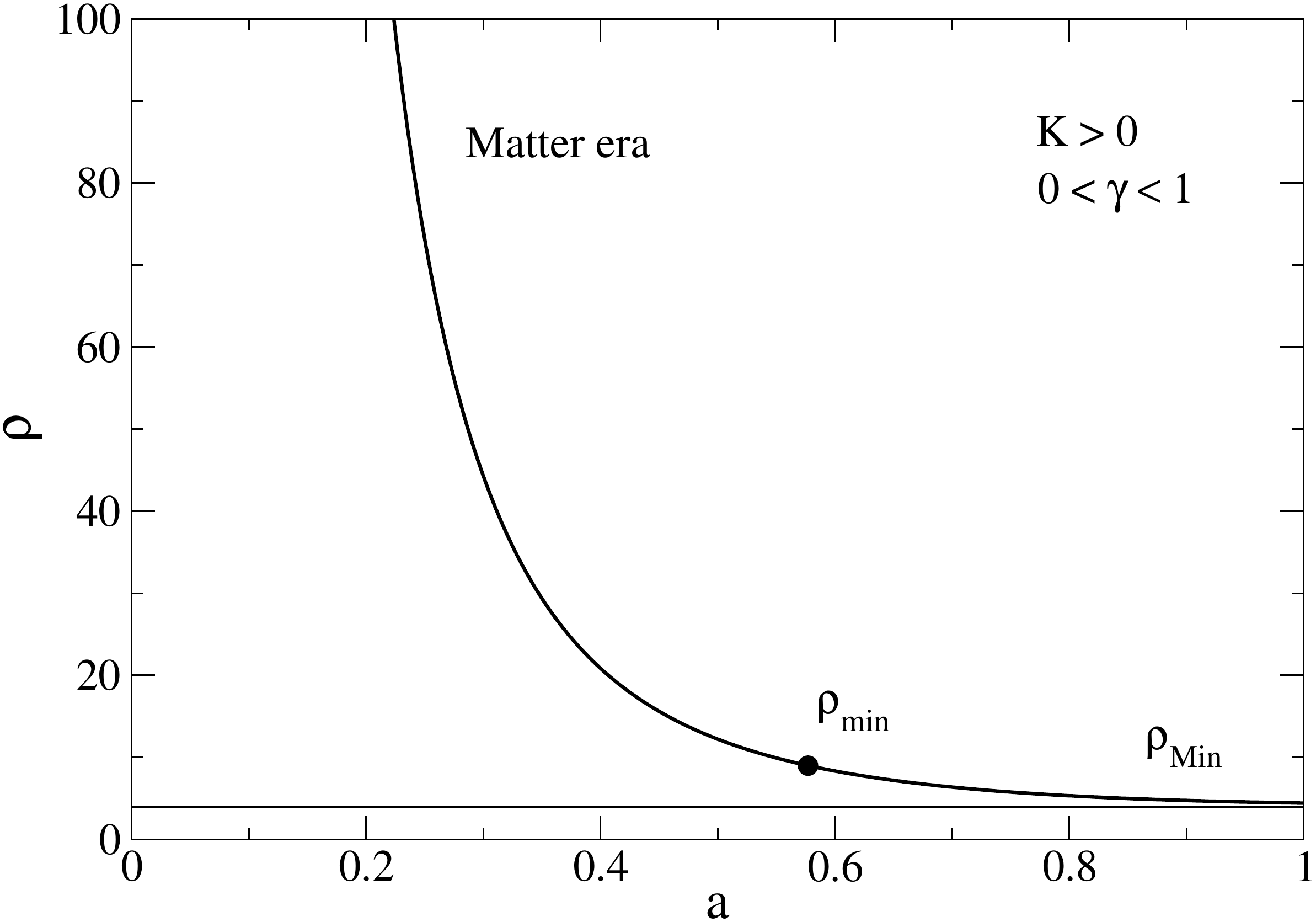}
\caption{Evolution of the pseudo rest-mass density as a function of the scale
factor for $0<\gamma<1$ (specifically $\gamma=0.5$). The region $\rho<\rho_{\rm
min}$ is inaccessible. }
\label{arKpos0g1}
\end{center}
\end{figure}

\begin{figure}[!h]
\begin{center}
\includegraphics[clip,scale=0.3]{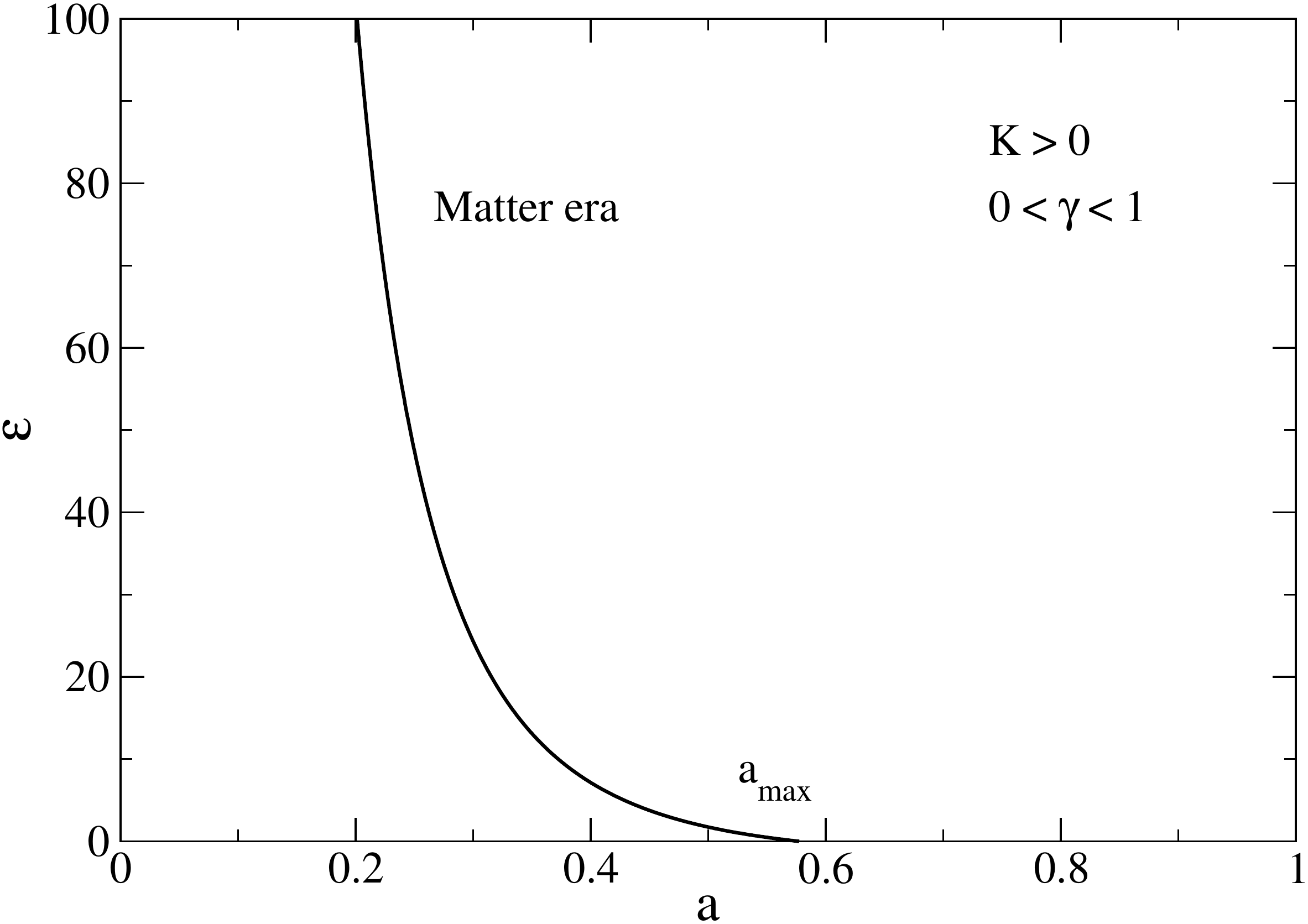}
\caption{Evolution of the energy density as a function of the scale factor
for $0<\gamma<1$ (specifically $\gamma=0.5$). The energy density vanishes at
$a_{\rm max}$. }
\label{aepsKpos0g1}
\end{center}
\end{figure}

\begin{figure}[!h]
\begin{center}
\includegraphics[clip,scale=0.3]{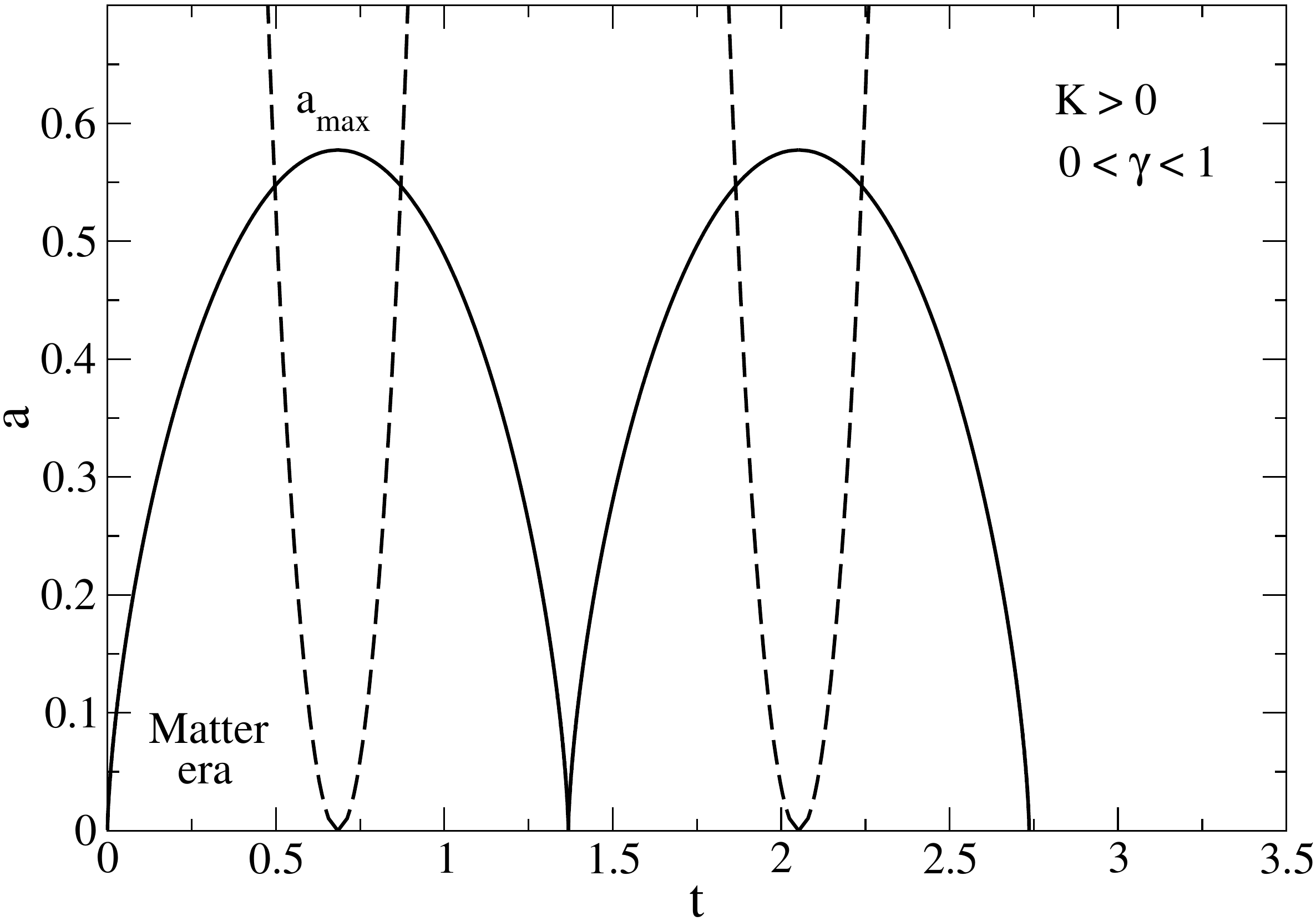}
\caption{
Evolution of the scale factor as a function of time for
$0<\gamma<1$ (specifically $\gamma=0.5$). The dashed line corresponds to the
energy density. }
\label{taKpos0g1}
\end{center}
\end{figure}

Considering now small values of $\rho$ (ultrarelativistic regime), we see that
Eq. (\ref{e10}) imposes the condition $\rho>\rho_{\rm Min}$ with
\begin{eqnarray}
\rho_{\rm Min}=\left (\frac{2}{c^2}\frac{K\gamma}{1-\gamma}\right
)^{1/(1-\gamma)}.
\label{e14}
\end{eqnarray}
The pseudo rest-mass density decreases as $a$ increases and tends to $\rho_{\rm
Min}$ when $a\rightarrow +\infty$. However, as in the previous section, the
energy density vanishes before $\rho$ reaches its absolute minimum value. 
Indeed, $\epsilon_{\rm
Min}=-\frac{1-\gamma}{2\gamma}\rho_{\rm Min}c^2<0$ when $\rho=\rho_{\rm Min}$.
According to Eq. (\ref{e11}), the energy density vanishes ($\epsilon=0$) at
\begin{eqnarray}
\rho_{\rm min}=\left (\frac{\gamma+1}{1-\gamma}\frac{K}{c^2}\right
)^{1/(1-\gamma)}.
\label{e15}
\end{eqnarray}
This happens when the scale factor reaches the value
\begin{equation}
a_{\rm max}=(Qm)^{1/3}\left (\frac{1-\gamma}{\gamma+1}\right
)^{(\gamma+1)/[6(1-\gamma)]}\left (\frac{c^2}{K}\right )^{1/[3(1-\gamma)]}.
\label{e16}
\end{equation}
The solution of Eqs. (\ref{e10}) and (\ref{e11}) is defined only for $a\le
a_{\rm max}$. The pseudo rest-mass density and the energy density decrease as
$a$ increases. When $a=a_{\rm max}$, the energy density vanishes while the
pseudo rest-mass density reaches its minimum accessible value $\rho_{\rm min}$.
The curves $\rho(a)$ and $\epsilon(a)$ are plotted in
Figs. \ref{arKpos0g1} and \ref{aepsKpos0g1}.

The temporal evolution  $a(t)$ of the scale factor is represented 
in Fig. \ref{taKpos0g1}. It is obtained by integrating Eq. (\ref{plsf11})
numerically. Starting
from a singularity at $t=0$ where
$a=0$ and
$\epsilon\rightarrow +\infty$ (big bang), the scale factor first grows as
$a\propto t^{2/3}$ in the matterlike era (EdS solution) until it reaches a
maximum value
$a_{\rm
max}$ at which the energy density vanishes
($\epsilon=0$). After that moment, the universe collapses and forms a
singularity at $t_{\rm bc}$ where $a=0$ and $\epsilon\rightarrow +\infty$ (big
crunch). This
process repeats itself periodically in time. This solution describes a cyclic
universe presenting phases of expansion and contraction separated
by critical points  where the energy density is either infinite (when $a=0$)
or zero (when $a=a_{\rm max}$). At that point the universe ``disappears''.

\subsection{The case $\gamma=0$ (anti-$\Lambda$CDM model)}
\label{sec_kp0g}

For $\gamma=0$, the equations determining the pseudo rest-mass 
density and the energy density as a function of the scale factor reduce to
\begin{eqnarray}
\rho=\frac{Qm}{a^3},
\label{e17}
\end{eqnarray}
\begin{eqnarray}
\epsilon=\rho c^2-K.
\label{e18}
\end{eqnarray}
They can be combined to give
\begin{eqnarray}
\epsilon=\frac{Qmc^2}{a^3}-K.
\label{e19}
\end{eqnarray}
This equation is equivalent to the one obtained in the anti-$\Lambda$CDM
model which assumes that the universe is filled with pressureless DM ($P=0$) and
that the cosmological constant is negative ($\Lambda<0$) or
that it is represented by a fluid with an equation of state $P_{\rm
de}=-\epsilon_{\rm de}$ yielding a constant energy density $\epsilon_{\rm
de}=-\rho_\Lambda c^2<0$. This agreement is
expected since, when $\gamma=0$, the pressure is constant ($P=K$) and we know
that a constant positive pressure $P=\rho_\Lambda c^2$
returns the
anti-$\Lambda$CDM model \cite{cosmopoly2}. In the present context, the
anti-$\Lambda$CDM model is obtained from a complex SF theory with a constant
negative potential $V(|\varphi|^2)=-\rho_\Lambda c^2$.
For this particular model, we
see that the pseudo rest-mass density decreases as $\rho\propto a^{-3}$ and
behaves as DM (see the Remark below).

The anti-$\Lambda$CDM model has been studied in
Sec. 6
of \cite{cosmopoly2}. To make the connection with this study, we set
$K=\rho_{\Lambda}c^2$ and $a_2=(Qm/\rho_{\Lambda})^{1/3}$. Equation
(\ref{e19}) can then be rewritten as
\begin{eqnarray}
\epsilon=\rho_{\Lambda}c^2\left\lbrack \left (\frac{a_2}{a}\right )^{3}-1\right\rbrack.
\label{antil1b}
\end{eqnarray}
Starting from $+\infty$ when $a\rightarrow 0$,
the energy density decreases as $a$ increases and vanishes at 
\begin{eqnarray}
a_{\rm max}= \left (\frac{Qmc^2}{K}\right )^{1/3}=a_2.
\label{antil2}
\end{eqnarray}
This is the maximum scale factor. At that point, the
pseudo rest-mass density reaches its minimum accessible value
\begin{eqnarray}
\rho_{\rm min}=\frac{K}{c^2}=\rho_{\Lambda}.
\label{antil3}
\end{eqnarray}
In
the nonrelativistic regime $a\ll a_2$, corresponding to the matterlike era
where $P/\epsilon\ll 1$, we
have
\begin{eqnarray}
\epsilon\sim \frac{Qmc^2}{a^3}\sim \rho_{\Lambda}c^2 \left
(\frac{a_2}{a}\right )^{3}.
\label{e33}
\end{eqnarray}
The temporal evolution of the scale factor is obtained by solving the Friedmann
equation (\ref{fe1}) with Eq. (\ref{antil1b}). In that case, the solution can be
obtained
analytically yielding \cite{cosmopoly2} 
\begin{eqnarray}
\frac{a}{a_2}=\sin^{2/3}\left (\sqrt{6\pi}\frac{t}{t_\Lambda}\right ),
\label{antil4}
\end{eqnarray}
\begin{eqnarray}
\epsilon=\frac{\rho_{\Lambda}c^2}{\tan^{2}\left (\sqrt{6\pi}\frac{t}{t_\Lambda}\right )},
\label{antil5}
\end{eqnarray}
\begin{eqnarray}
\rho=\frac{\rho_\Lambda c^2}{\sin^{2}\left (\sqrt{6\pi}\frac{t}{t_\Lambda}\right
)},
\label{antil5b}
\end{eqnarray}
where $t_{\Lambda}=1/\sqrt{G\rho_{\Lambda}}$ is the cosmological time. The
evolution of 
the universe is similar to the one  described in Sec. \ref{sec_kp0g1} with
the
particularity that $\rho_{\rm Min}=0$.
The anti-$\Lambda$CDM model is studied in more detail in Sec. 6
of \cite{cosmopoly2}.

{\it Remark:} In Eq. (\ref{e19}) the first term is the rest-mass energy density
$\rho_m c^2$ [see
Eqs. (\ref{mtd11}) and (\ref{rmd2})] and the second term is the internal energy
density $u$
[see Eq. (\ref{mtd2b})]. As discussed in Sec. \ref{sec_dmde}, the
rest-mass density $\rho_m$ can be interpreted as DM and the internal energy
density $u$ can be interpreted as DE \cite{epjp,lettre}. In the present case,
the
pseudo rest-mass density coincides with the rest-mass density ($\rho=\rho_m$)
and the internal energy density is constant ($u=-K=-\rho_\Lambda c^2$).
In the two-fluid model associated with the anti-$\Lambda$CDM model
(see Sec.
\ref{sec_twofluids}), DM has an  equation of state $P_{\rm
m}(\epsilon_{\rm m})=0$ and DE has an equation of state $P_{\rm
de}(\epsilon_{\rm
de})=-\epsilon_{\rm de}$.

\subsection{The case $\gamma<0$}
\label{sec_kpinf0}

For $\gamma<0$ the equations determining the pseudo rest-mass density and the energy density as a function of the scale factor can be written as
\begin{eqnarray}
\rho
\sqrt{1+\frac{2}{c^2}\frac{K|\gamma|}{1-\gamma}\frac{1}{\rho^{1-\gamma}}}=\frac{
Qm } { a^3 } ,
\label{e20}
\end{eqnarray}
\begin{eqnarray}
\epsilon=\rho c^2-\frac{\gamma+1}{1-\gamma}K\frac{1}{\rho^{|\gamma|}}.
\label{e21}
\end{eqnarray}

When $\rho\rightarrow +\infty$, Eqs. (\ref{e20}) and (\ref{e21}) reduce to
\begin{eqnarray}
\rho\sim\frac{Qm}{a^3},
\label{e22}
\end{eqnarray}
\begin{eqnarray}
\epsilon\sim \rho c^2.
\label{e23}
\end{eqnarray}
This corresponds to the nonrelativistic regime $P/\epsilon\ll 1$
(matterlike era) valid for $a\rightarrow 0$. Starting from $+\infty$ when
$a\rightarrow 0$, the pseudo rest-mass density and the energy density decrease
as $a$ increases.

When $\rho\rightarrow 0$ (ultrarelativistic regime), Eqs. (\ref{e20}) and
(\ref{e21}) reduce to
\begin{eqnarray}
\rho\sim \left(\frac{Q^2m^2c^2}{2}\frac{1-\gamma}{K|\gamma|}\right
)^{1/(1+\gamma)} \frac{1}{a^{6/(1+\gamma)}}.
\label{e24}
\end{eqnarray}
We have to distinguish three subcases:

(i) When $\gamma>-1$, the asymptotic behavior from Eq. (\ref{e24}) is valid for
large values of $a$. The pseudo rest-mass density decreases as  $a$ increases
and tends to zero as $\rho\propto a^{-6/(1+\gamma)}$ when $a\rightarrow
+\infty$. However, the energy
vanishes before. According to Eq. (\ref{e21}), it
vanishes at
\begin{eqnarray}
\rho_{\rm min}=\left (\frac{\gamma+1}{1-\gamma}\frac{K}{c^2}\right
)^{1/(1-\gamma)}.
\label{e25}
\end{eqnarray}
This happens when the scale factor reaches the value
\begin{equation}
a_{\rm max}=(Qm)^{1/3}\left (\frac{1-\gamma}{\gamma+1}\right
)^{(\gamma+1)/[6(1-\gamma)]}\left (\frac{c^2}{K}\right )^{1/[3(1-\gamma)]}.
\label{e26}
\end{equation}
In that case, the evolution of the universe is similar to the one described in
Sec. \ref{sec_kp0g1} with the particularity that $\rho_{\rm Min}=0$.

(ii) The case $\gamma=-1$ (anti-Chaplygin gas) is specifically treated in the
next subsection.

\begin{figure}[!h]
\begin{center}
\includegraphics[clip,scale=0.3]{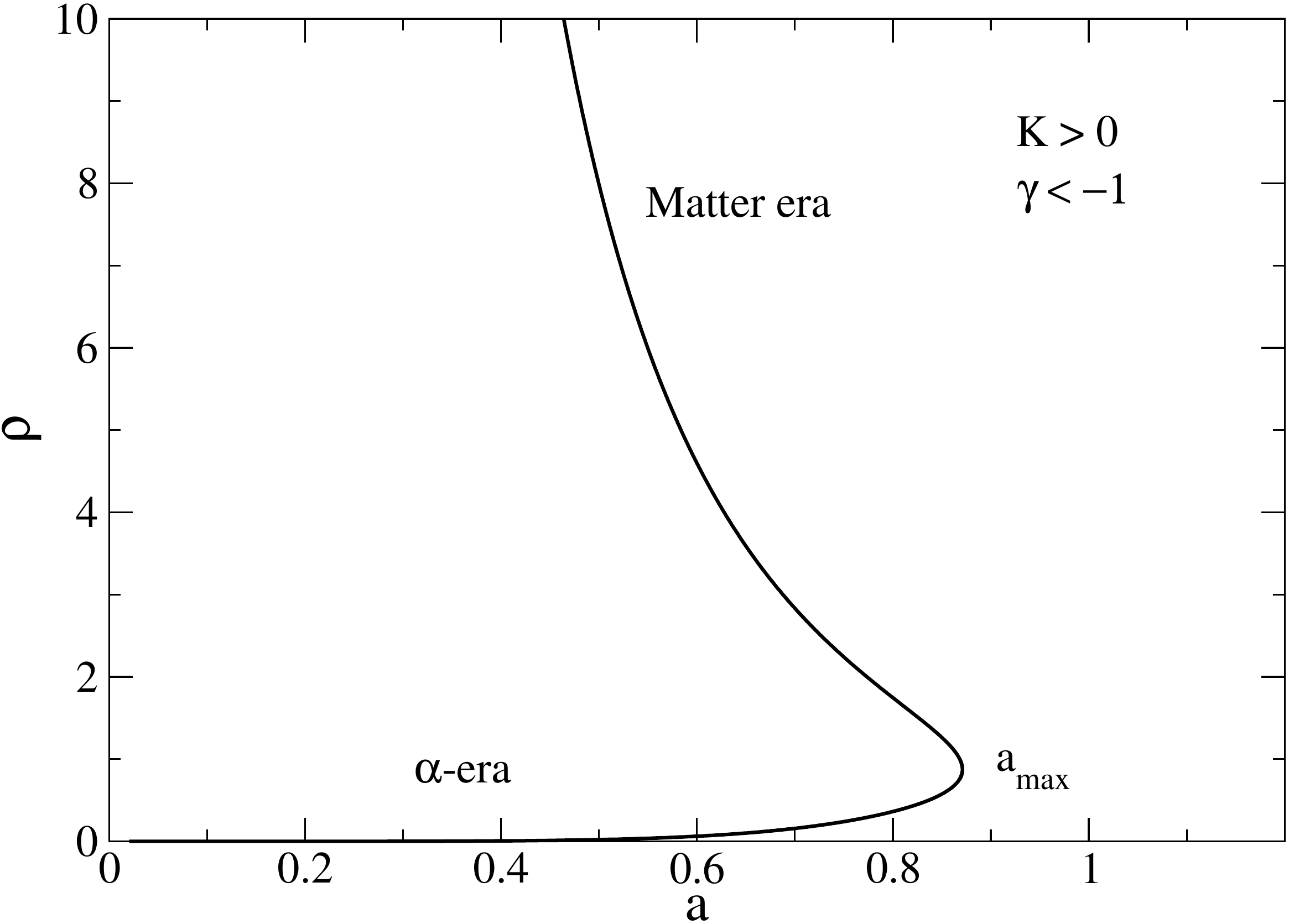}
\caption{Evolution of the pseudo-rest mass density as a
function of the scale factor for $\gamma<-1$ (specifically $\gamma=-2$).}
\label{arKposgminus2}
\end{center}
\end{figure}

\begin{figure}[!h]
\begin{center}
\includegraphics[clip,scale=0.3]{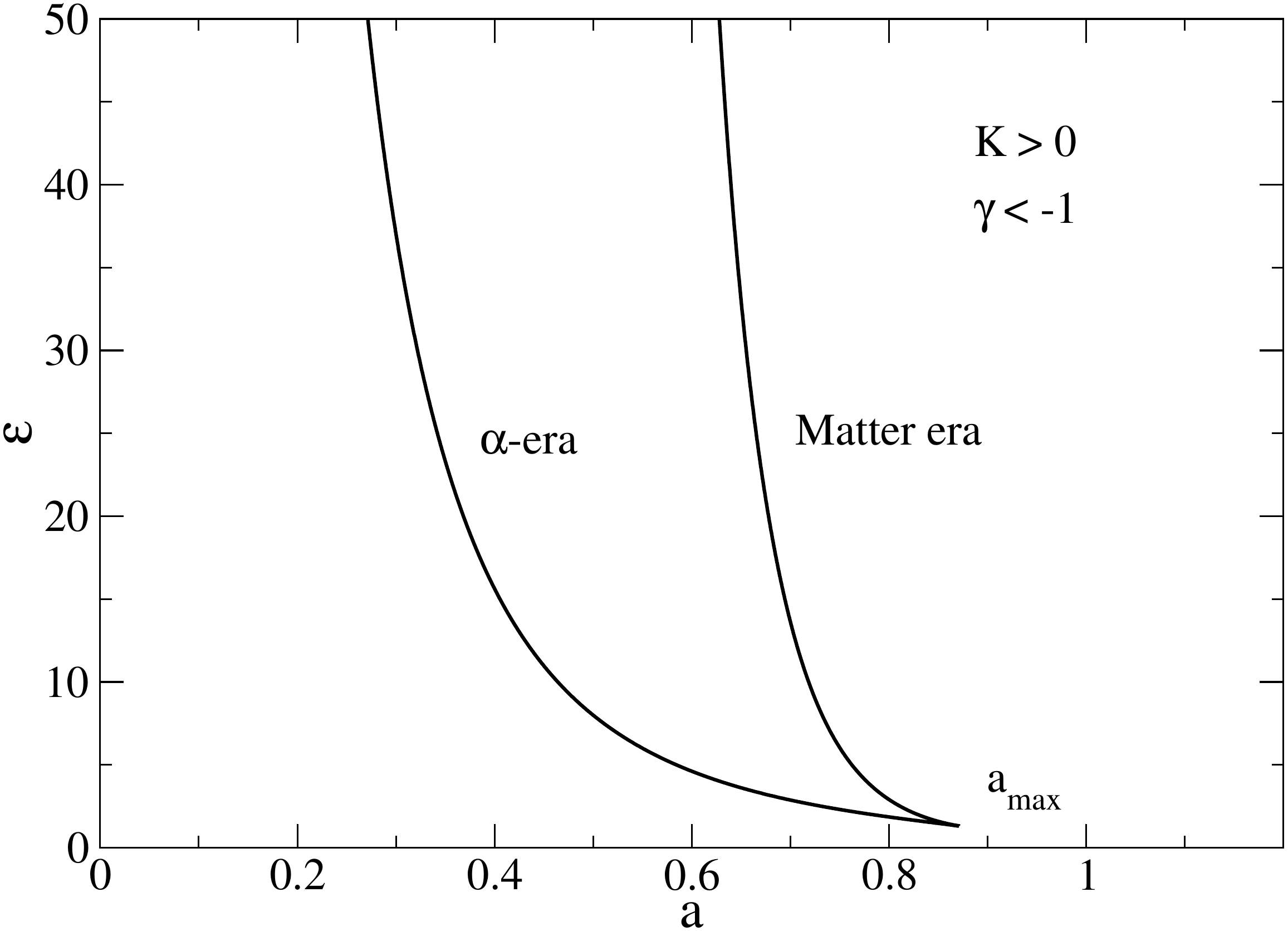}
\caption{Evolution of the energy density as a
function of the scale factor for $\gamma<-1$ (specifically
$\gamma=-2$). }
\label{aepsKposgminus2}
\end{center}
\end{figure}

(iii) When $\gamma<-1$, the asymptotic behavior from Eq. (\ref{e24}) is valid
for small values of $a$. The pseudo rest-mass
density decreases as $a$ decreases and tends
to zero as $\rho\propto a^{6/|1+\gamma|}$ when $a\rightarrow 0$. In parallel,
the energy density increases as $a$ decreases and tends to $+\infty$ as
\begin{eqnarray}
\epsilon\sim \frac{|\gamma+1|}{1-\gamma}K\frac{1}{\rho^{|\gamma|}}\propto
\frac{1}{a^{6|\gamma|/|1+\gamma|}}
\label{e27}
\end{eqnarray}
when $a\rightarrow 0$. Using Eqs. (\ref{plsf3}) and (\ref{e27}), we obtain the
equation of
state
\begin{eqnarray}
P=\frac{1-\gamma}{|\gamma+1|}\epsilon.
\label{e27b}
\end{eqnarray}
The pressure is a linear function $P\sim\alpha\epsilon$ of the energy density
with coefficient $\alpha=(1-\gamma)/|\gamma+1|$. This determines  an
$\alpha$-era.  Strikingly, we have two
branches of solutions for sufficiently small values of the scale factor: the
nonrelativistic
branch from Eqs. (\ref{e22}) and (\ref{e23}) and the
ultrarelativistic branch described above. The two
branches
merge at a  maximum scale factor
 \begin{eqnarray}
a_{\rm max}=(Qm)^{1/3}\left (\frac{|\gamma+1|}{1-\gamma}\right
)^{|\gamma+1|/[6(1-\gamma)]}\nonumber\\
\times \frac{1}{|\gamma|^{1/[3(1-\gamma)]}}\left
(\frac{c^2}{K}\right )^{1/[3(1-\gamma)]},
\label{e29}
\end{eqnarray}
corresponding to
\begin{eqnarray}
\rho_c=\left (c^2\frac{1-\gamma}{K|\gamma|}\frac{1}{|1+\gamma|}\right
)^{1/(\gamma-1)}
\label{e28}
\end{eqnarray}
and 
\begin{eqnarray}
\epsilon_c=\frac{1-\gamma}{|\gamma|}\rho_c c^2.
\label{e28b}
\end{eqnarray} 
The curves $\rho(a)$ and $\epsilon(a)$ are
plotted in Figs. \ref{arKposgminus2} and \ref{aepsKposgminus2}.

\begin{figure}[!h]
\begin{center}
\includegraphics[clip,scale=0.3]{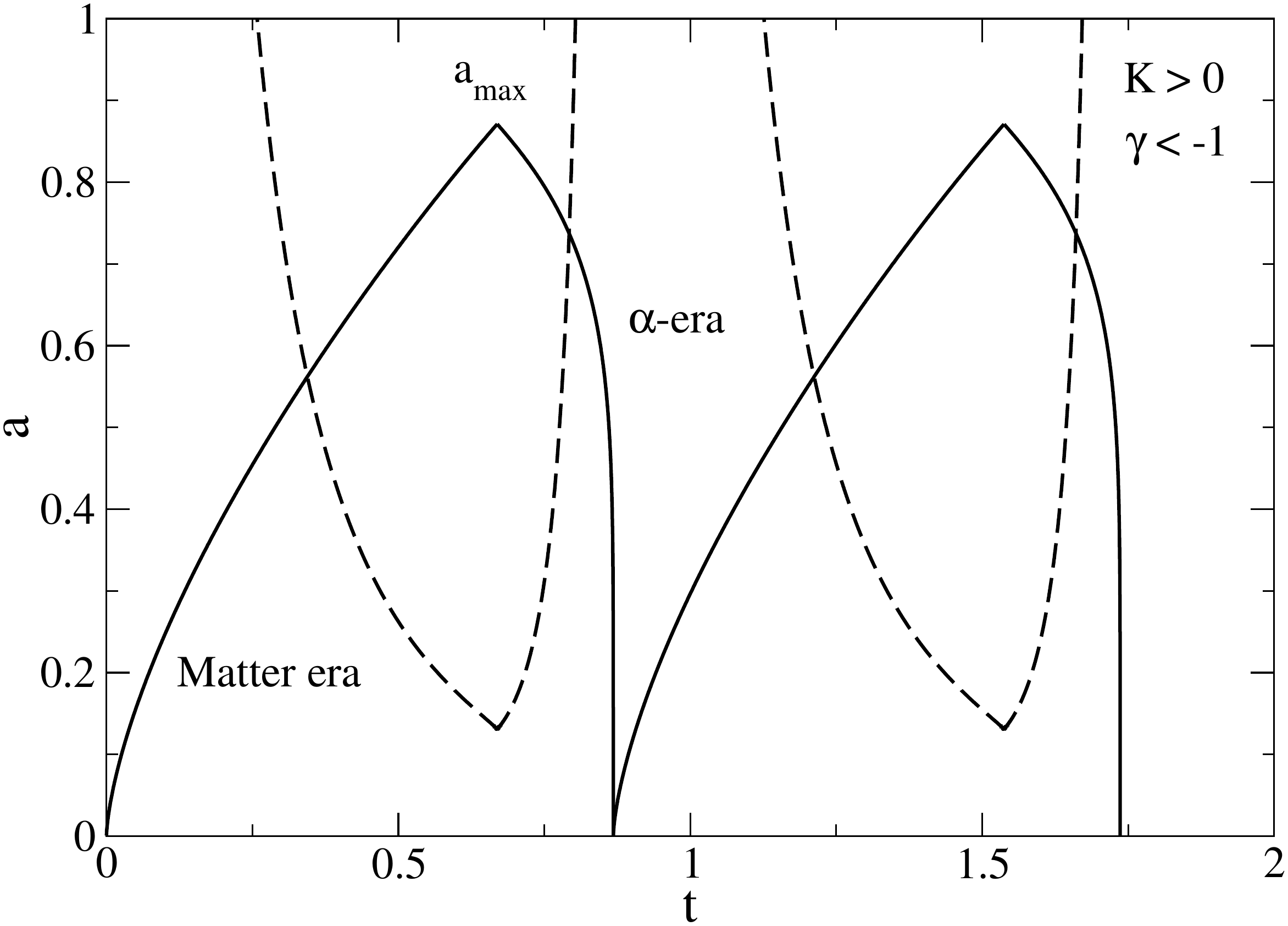}
\caption{Evolution of the scale
factor
as a
function of time for
$\gamma<-1$ (specifically
$\gamma=-2$). The dashed line corresponds to the energy density divided by $10$.
We note that $\dot a\neq 0$ at $a=a_{\rm max}$ (contrary to the case of Fig.
\ref{taKpos0g1}) because $\epsilon_c\neq 0$ at that point (recall that $\dot
a/a\propto \sqrt{\epsilon}$ according to the Friedmann equation (\ref{fe1})).}
\label{taKpos0gminus2}
\end{center}
\end{figure}

The temporal evolution  $a(t)$ of the scale factor is
represented 
in Fig. \ref{taKpos0gminus2}. It is obtained by integrating Eq. (\ref{plsf11})
numerically, assuming that the SF follows the curve of Fig. \ref{arKposgminus2}
from $\rho=+\infty$ to $\rho=0$. Starting from a singularity at $t=0$ where
$a=0$ and
$\epsilon\rightarrow +\infty$ (big bang), the energy density decreases as
$a^{-3}$ and the
scale factor grows as $a\propto t^{2/3}$ in the matter era (EdS solution) until
it reaches a maximum value $a_{\rm
max}$ at which the energy density equals $\epsilon_c$. This phase of
expansion  corresponds to the nonrelativistic  branch
of Fig. \ref{arKposgminus2}. After that
moment, the universe collapses ($\alpha$-era) and forms a
singularity at $t_{\rm bc}$ where $a=0$ and $\epsilon\rightarrow +\infty$ (big
crunch). This
phase of contraction corresponds to the 
ultrarelativistic branch  of Fig. \ref{arKposgminus2}. The energy density
increases as $a^{-6|\gamma|/|1+\gamma|}$ while the scale factor
decreases as $a\propto (t_{\rm bc}-t)^{|\gamma+1|/3|\gamma|}$
(corresponding to $a\propto
(t_{\rm bc}-t)^{2/[3(1+\alpha)]}$)  in the $\alpha$-era.
This process repeats itself periodically in time. This solution describes an
asymmetric  cyclic
universe presenting phases of expansion (explosion) and
contraction  (implosion) separated
by critical points  where the energy density is either infinite (when $a=0$)
or equal to $\epsilon_c$ (when $a=a_{\rm max}$).

{\it Remark:} Other possible evolutions can be contemplated.
The SF could follow the curve of Fig. \ref{arKposgminus2} in the reverse sense, 
from $\rho=0$ to $\rho=+\infty$. This would
lead to an asymmetric  cyclic universe where  the phase of
expansion  corresponds to the  ultrarelativistic 
branch ($\alpha$-era) and the phase of contraction   corresponds
to the nonrelativistic branch (matterlike era). We
could also consider the case of a symmetric  cyclic
universe (like in Sec. \ref{sec_kp0g1}) where the SF follows
the same
branch (nonrelativistic or ultrarelativistic)
during the phases of expansion and contraction. In that case, there would be two
possible
evolutions.

\subsection{The case $\gamma=-1$ (anti-Chaplygin gas) }
\label{sec_kpg0}

For  $\gamma=-1$, the equations determining the 
pseudo rest-mass density and the energy density as a function of the scale
factor reduce to
\begin{eqnarray}
\rho=\sqrt{\frac{Q^2m^2}{a^6}-\frac{K}{c^2}},
\label{e30}
\end{eqnarray}
\begin{eqnarray}
\epsilon=\rho c^2.
\label{e31}
\end{eqnarray}
They can be combined to give
\begin{eqnarray}
\epsilon= \sqrt{\frac{Q^2m^2c^4}{a^6}-K c^2}.
\label{e32}
\end{eqnarray}
This equation is equivalent to the one obtained in  the anti-Chaplygin gas
model. This agreement is expected because, when $\gamma=-1$, we have
$\epsilon=\rho c^2$. Therefore, the equation of state
$P=K/\rho$ from Eq. (\ref{plsf3})
can be written as $P=Kc^2/\epsilon$ which is the equation of state of the
anti-Chaplygin gas when $K>0$ \cite{gkmp,cosmopoly2}. In
the present context, the
anti-Chaplygin gas model is obtained from a complex SF theory with a potential
$V(|\varphi|^2)=-\frac{1}{2}K(\frac{\hbar}{m})^2\frac{1}{|\varphi|^2}$. For this
particular model, the pseudo rest-mass
density coincides with the energy density ($\rho=\epsilon/c^2$).

The anti-Chaplygin gas has been studied in Sec.
4.3 of \cite{cosmopoly2}. To
make the connection with this study, we set $\rho_*=\sqrt{K/c^2}$ and
$a_*=(Q^2m^2c^2/K)^{1/6}$. Eq. (\ref{e19}) can then be rewritten as
\begin{eqnarray}
\epsilon=\rho_{*}c^2\sqrt{\left (\frac{a_*}{a}\right )^{6}-1}.
\label{antil1}
\end{eqnarray}
Starting from $+\infty$ when $a\rightarrow 0$,
the energy density decreases as $a$ increases and vanishes at 
\begin{eqnarray}
a_{\rm max}= \left (\frac{Q^2m^2c^2}{K}\right )^{1/6}=a_*.
\label{antil2b}
\end{eqnarray}
This is the maximum scale factor. In
the nonrelativistic regime $a\ll a_*$, corresponding to the matterlike era
where $P/\epsilon\ll 1$, we
have
\begin{eqnarray}
\epsilon=\rho c^2\sim \frac{Q m c^2}{a^3}=\rho_{*}c^2\left
(\frac{a_*}{a}\right )^{3}.
\label{e33b}
\end{eqnarray}
The temporal evolution of the scale factor
is obtained by solving the Friedmann equation (\ref{fe1}) with Eq.
(\ref{antil1}). This yields \cite{cosmopoly2}
\begin{eqnarray}
\sqrt{96\pi G\rho_*}\, t=\int_{\left (\frac{a_*}{a}\right)^6}^{+\infty}
\frac{dx}{x(x-1)^{1/4}}.
\label{antil4b}
\end{eqnarray}
The integral can be calculated explicitly
and is
given by Eq. (30) of \cite{cosmopoly2}. The
evolution of the universe is similar to the one described in
Sec. \ref{sec_kp0g1} with the particularity that $\rho$ vanishes at $a_{\rm
max}$ (i.e. $\rho_{\rm min}=0$). The 
anti-Chaplygin gas model is
studied in more detail in Sec. 4.3 of \cite{cosmopoly2}.

{\it Remark:} We can rewrite Eq. (\ref{e32}) as
\begin{eqnarray}
\epsilon=\frac{Qmc^2}{a^3}+\left\lbrack \sqrt{\frac{Q^2m^2c^4}{a^6}-K
c^2}-\frac{Qmc^2}{a^3}\right\rbrack,
\label{ne32}
\end{eqnarray} 
where the first term is the rest-mass energy density $\rho_m c^2$ [see
Eqs. (\ref{mtd11}) and (\ref{rmd2})] and the second term is the internal energy
density $u$
[see Eqs. (\ref{mtd2b}) and (\ref{gest6})]. As discussed in Sec. \ref{sec_dmde},
the
rest-mass density $\rho_m$ can be interpreted as DM and the internal energy
density $u$ can be interpreted as DE \cite{epjp,lettre}.
In
the two-fluid model associated with the
anti-Chaplygin gas (see Sec.
\ref{sec_twofluids}), DM has an  equation of state $P_{\rm
m}(\epsilon_{\rm m})=0$ and DE has an equation of state 
\begin{eqnarray}
P_{\rm de}(\epsilon_{\rm de})=\frac{2Kc^2\epsilon_{\rm de}}{\epsilon_{\rm
de}^2-Kc^2},
\label{twg}
\end{eqnarray}
which is obtained by eliminating $\rho$ between Eqs. (\ref{chap1}) and
(\ref{aswe}), and by identifying $P(u)$ with $P_{\rm de}(\epsilon_{\rm de})$.
Solving the energy conservation equation (\ref{hsf4}) with the equation of
state (\ref{twg}), we recover the expression of DE from Eq. (\ref{ne32}).

\section{The case $K<0$}
\label{sec_kn}

In this section, 
we consider the case of a negative polytropic constant ($K<0$), corresponding to
a negative pressure. This allows the universe to experience a phase of
accelerating expansion. In the figures, we take $c=Qm=4\pi G=1$ and $K=-1$.

\subsection{The case $\gamma>1$}
\label{sec_kng1}

For $\gamma>1$ the equations determining the pseudo rest-mass density 
and the energy density as a function of the scale factor can be written as
\begin{eqnarray}
\rho
\sqrt{1-\frac{2}{c^2}\frac{|K|\gamma}{\gamma-1}\rho^{\gamma-1}}
=\frac{Qm}{a^3},
\label{f1}
\end{eqnarray}
\begin{eqnarray}
\epsilon=\rho c^2-\frac{\gamma+1}{\gamma-1}|K|\rho^{\gamma}.
\label{f2}
\end{eqnarray}

When $\rho\rightarrow 0$, Eqs. (\ref{f1}) and (\ref{f2}) reduce to
\begin{eqnarray}
\rho\sim\frac{Q m}{a^3},
\label{f3}
\end{eqnarray}
\begin{eqnarray}
\epsilon\sim \rho c^2.
\label{f4}
\end{eqnarray}
This corresponds to the nonrelativistic regime $P/\epsilon\ll 1$ (matterlike
era) 
valid for $a\rightarrow +\infty$. The pseudo rest-mass density and the energy
density decrease to zero as $a$ increases.

Considering now large values of $\rho$ (ultrarelativistic regime), we see that
Eq. (\ref{f1}) imposes the condition $\rho\le \rho_{\rm max}$
with
\begin{eqnarray}
\rho_{\rm max}=\left (\frac{c^2}{2}\frac{\gamma-1}{|K|\gamma}\right
)^{1/(\gamma-1)}.
\label{f5}
\end{eqnarray}
The pseudo rest-mass density increases as $a$ increases and tends 
to $\rho_{\rm max}$ when  $a\rightarrow +\infty$. In parallel, the energy
density decreases  as $a$ increases and tends to a
constant
\begin{eqnarray}
\epsilon_{\rm min}= \frac{\gamma-1}{2\gamma}\rho_{\rm max}c^2
\label{f6}
\end{eqnarray}
when  $a\rightarrow +\infty$. This leads to a DE era corresponding to a de
Sitter evolution.

\begin{figure}[!h]
\begin{center}
\includegraphics[clip,scale=0.3]{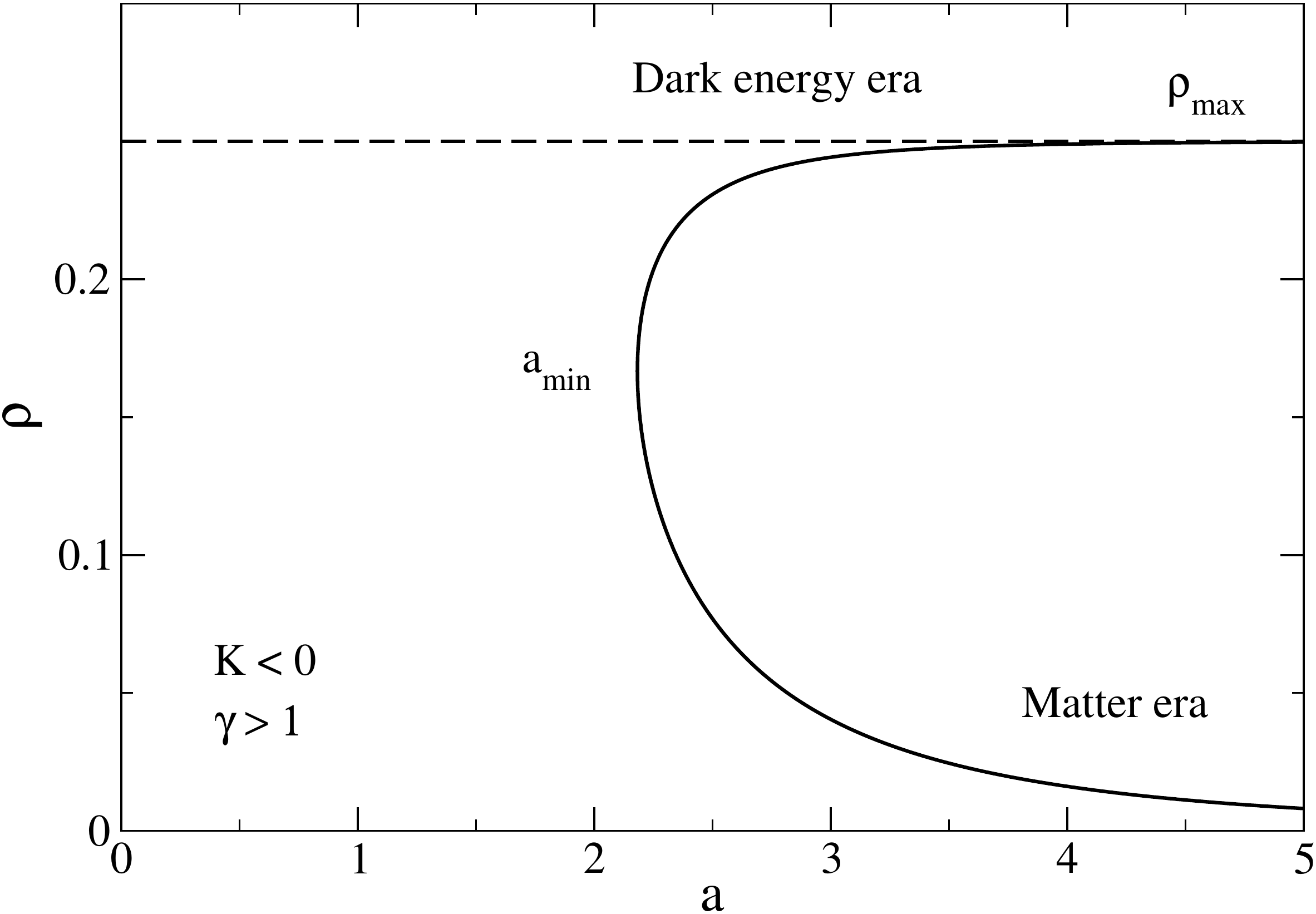}
\caption{Evolution of the pseudo rest-mass density as a function of the scale
factor for $\gamma>1$ (specifically $\gamma=2$).}
\label{arKnegg2}
\end{center}
\end{figure}

\begin{figure}[!h]
\begin{center}
\includegraphics[clip,scale=0.3]{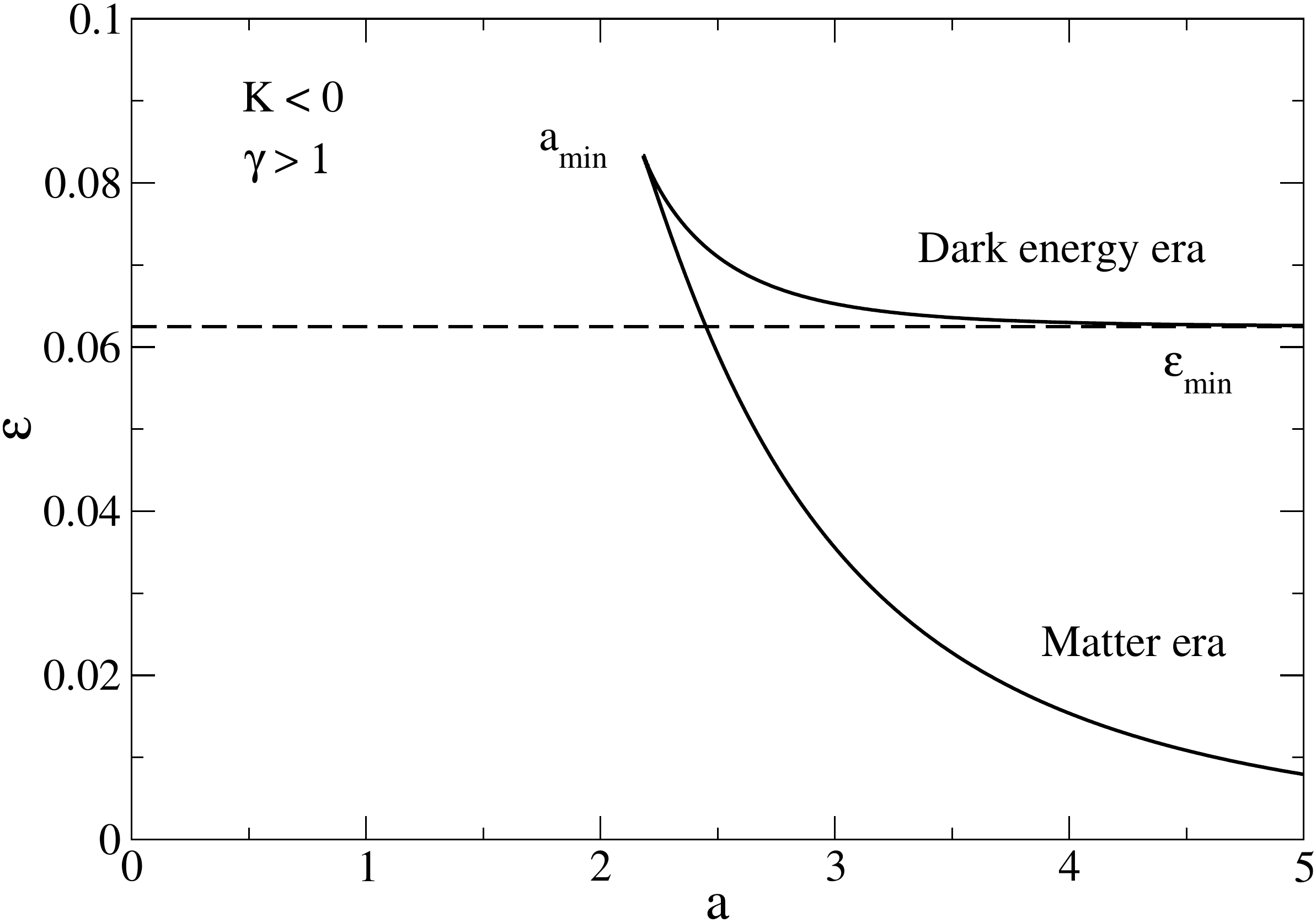}
\caption{Evolution of the energy density as a function of the scale
factor for $\gamma>1$ (specifically $\gamma=2$). 
}
\label{aepsKnegg2}
\end{center}
\end{figure}

Therefore, we have two branches of solutions for sufficiently large values of
the 
scale factor: the nonrelativistic branch from Eqs. (\ref{f3}) and (\ref{f4}) and
the ultrarelativistic branch described above. The two branches merge at a 
minimum scale factor
\begin{eqnarray}
a_{\rm min}=(Qm)^{1/3}\left (\frac{\gamma+1}{\gamma-1}\right
)^{(\gamma+1)/[6(\gamma-1)]}\nonumber\\
\times \gamma^{1/[3(\gamma-1)]}\left
(\frac{|K|}{c^2}\right )^{1/[3(\gamma-1)]}
\label{f7}
\end{eqnarray}
corresponding to 
\begin{eqnarray}
\rho_i=\left (c^2\frac{\gamma-1}{|K|\gamma}\frac{1}{1+\gamma}\right
)^{1/(\gamma-1)}
\label{f8}
\end{eqnarray}
and
\begin{eqnarray}
\epsilon_i=\frac{\gamma-1}{\gamma}\rho_i c^2.
\label{f8eps}
\end{eqnarray}
The case $\gamma=2$ and $K<0$, corresponding to a relativistic BEC with an
attractive $|\varphi|^4$ self-interaction,  is treated in detail in Sec. IV of
\cite{abrilphas}. 
The curves  $\rho(a)$ and $\epsilon(a)$  are plotted in Figs. \ref{arKnegg2}
and \ref{aepsKnegg2}.

\begin{figure}[!h]
\begin{center}
\includegraphics[clip,scale=0.3]{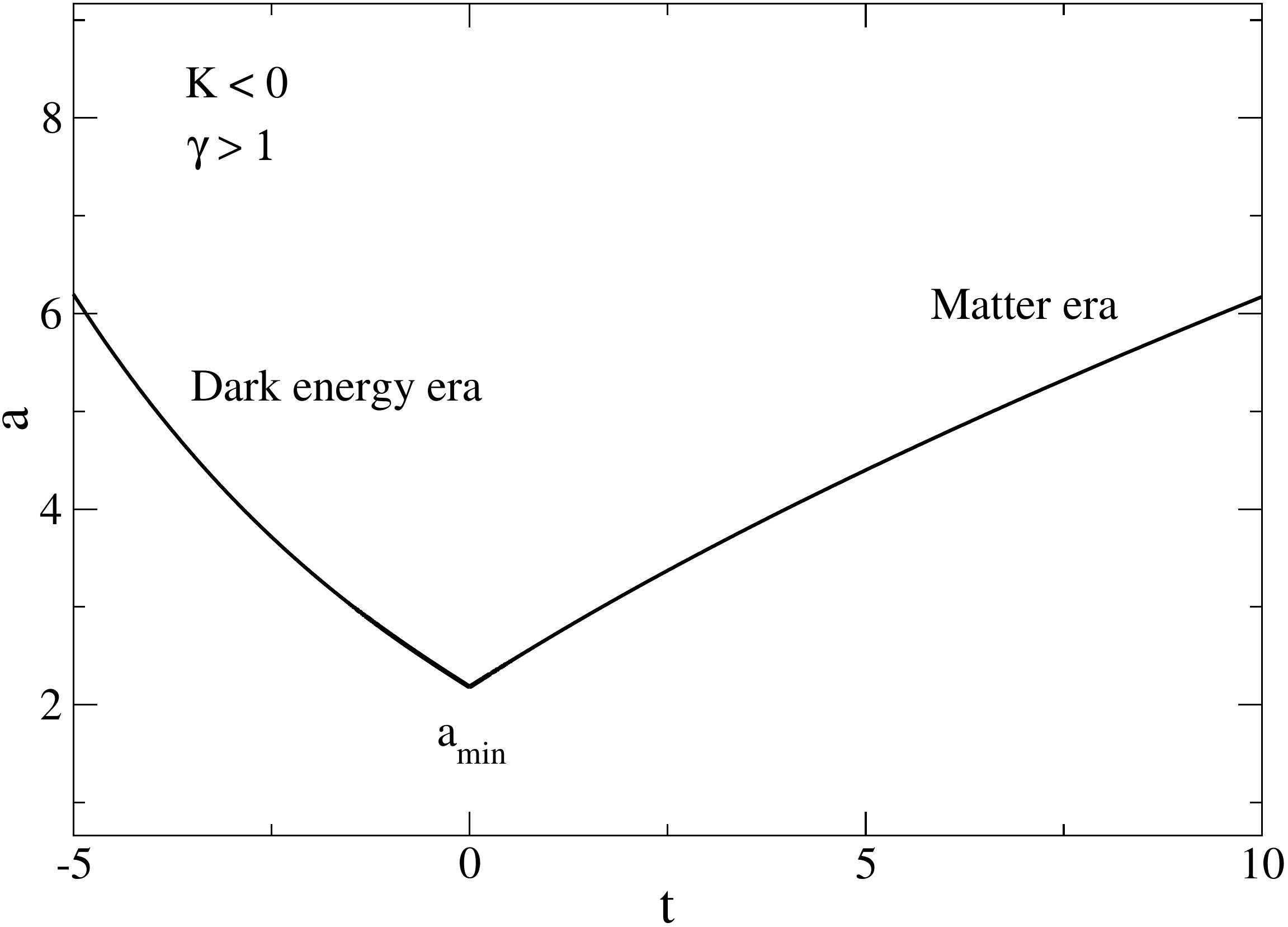}
\caption{Evolution of the scale factor as a function of time for  $\gamma>1$
(specifically $\gamma=2$). }
\label{taKnegg2}
\end{center}
\end{figure}

The temporal evolution  $a(t)$ of the scale factor is
represented 
in Fig. \ref{taKnegg2}. It is obtained by integrating Eq. (\ref{plsf11})
numerically, assuming that the SF follows the curve of Fig. \ref{arKnegg2}
from $\rho_{\rm max}$ to $\rho=0$. Starting from a DE era at $t=-\infty$
where
$a=+\infty$ and
$\epsilon=\epsilon_{\rm min}$, the energy density slowly increases while the 
scale factor decreases exponentially rapidly as $a\propto e^{-(8\pi
G\epsilon_{\rm min}/3c^2)^{1/2}t}$ (de Sitter) until
it reaches a minimum  value $a_{\rm
min}$ at which the energy density equals $\epsilon_i$. This phase of
contraction  corresponds to the ultrarelativistic 
branch
of Fig. \ref{arKnegg2}. After that
moment, the universe expands as it enters into the matterlike era.  This phase
of expansion corresponds to the 
nonrelativistic branch  of Fig. \ref{arKnegg2}. The energy density
decreases as $\epsilon\propto a^{-3}$ while the scale factor
increases as $a\propto t^{2/3}$  (EdS). This solution
describes an asymmetric  bouncing
universe presenting a phase of contraction and a phase of expansion. There is
no big bang singularity in this model.

{\it Remark:} Other possible evolutions can be contemplated.
The SF could follow the curve of Fig. \ref{arKnegg2} in the reverse sense, 
from $\rho=0$ to $\rho_{\rm max}$. This would
lead to an asymmetric bouncing universe where  the phase of
contraction  corresponds to the  nonrelativistic 
branch (matterlike era) and the phase of expansion   corresponds
to the  ultrarelativistic branch (DE era).
We
could also consider the case of a purely expanding
universe where the SF follows
the same
branch (nonrelativistic or
ultrarelativistic). In that case, the SF suddently appears
at a finite scale factor $a_i$ and behaves either as DM (normal branch) or
as DE (peculiar branch). This corresponds to the two possible evolutions (DM
and DE)
described in \cite{abrilphas}.

\subsection{The case $\gamma=1$}
\label{sec_kneq1}

The case $\gamma=1$ must be treated specifically. It corresponds to the SF
potential from Eq. (\ref{iplsf1}) associated with the isothermal equation of
state (\ref{iplsf3}).
In the present case $T<0$. The equations determining the pseudo rest-mass
density and the energy density as a function of the scale factor are
\begin{eqnarray}
\rho
\sqrt{1-\frac{2k_B |T|}{m c^2}\ln\left (\frac{\rho}{\rho_*}\right )}
=\frac{Qm}{a^3},
\label{nie1}
\end{eqnarray}
\begin{eqnarray}
\epsilon=\rho c^2-\frac{2k_B |T|}{m}\rho\ln\left (\frac{\rho}{\rho_*}\right )+\frac{k_B |T|}{m}\rho.
\label{nie2}
\end{eqnarray}

When $\rho\rightarrow 0$, Eqs. (\ref{nie1}) and (\ref{nie2}) reduce to
\begin{eqnarray}
\rho
\sqrt{\frac{2k_B |T|}{m c^2}|\ln\rho|}
\sim\frac{Qm}{a^3},
\label{nie3}
\end{eqnarray}
\begin{eqnarray}
\epsilon\sim \frac{2k_B |T|}{m}\rho|\ln\rho|.
\label{nie4}
\end{eqnarray}
This regime is valid for $a\rightarrow +\infty$. The pseudo rest-mass density
and the energy density decrease to zero as $a$ increases. This corresponds to a
matterlike era ($\epsilon\simeq a^{-3}$) modified by logarithmic corrections. 
Using Eqs. (\ref{iplsf3}) and (\ref{nie4}), we obtain the equation of
state
\begin{eqnarray}
P\simeq -\frac{\epsilon}{2|\ln\epsilon|}.
\label{nie5}
\end{eqnarray}
The pressure is an approximately linear function of the energy density
$P\simeq \alpha\epsilon$ with a logarithmic correction yiedling a small
effective coefficient  $\alpha\sim -1/(2|\ln\epsilon|)\ll 1$. For $a\rightarrow
+\infty$,
we have
$|P|/\epsilon\ll 1$.

Considering now large values of $\rho$, we see that Eq. (\ref{nie1}) imposes
the condition $\rho\le \rho_{\rm max}$ with
\begin{eqnarray}
\rho_{\rm max}=\rho_* e^{\frac{mc^2}{2k_B |T|}}.
\label{nie6}
\end{eqnarray}
The pseudo rest-mass density increases as $a$ increases and tends to $\rho_{\rm
max}$ when $a\rightarrow +\infty$. In parallel, the energy density decreases
 as $a$ increases and tends to a constant
\begin{eqnarray}
\epsilon_{\rm min}=\rho_{\rm max}\frac{k_B |T|}{m}
\label{nie6b}
\end{eqnarray}
when $a\rightarrow +\infty$.

Therefore, we have two branches of solutions for sufficiently large values of
the scale
factor: the nonrelativistic branch from Eqs.
(\ref{nie3})
and
(\ref{nie4}) and the ultrarelativistic branch described
above.
The two branches merge at a minimum scale factor
\begin{equation}
a_{\rm min}=\left (\frac{Qm}{\rho_*}\sqrt{\frac{mc^2}{k_B |T|}}e^{\frac{1}{2}-\frac{mc^2}{2k_B |T|}}\right )^{1/3}.
\label{nie8}
\end{equation}
corresponding to 
\begin{eqnarray}
\rho_{i}=\rho_* e^{-\frac{1}{2}+\frac{mc^2}{2k_B |T|}}
\label{nie7}
\end{eqnarray}
and
\begin{eqnarray}
\epsilon_{i}=\frac{2k_B |T|}{m}\rho_i.
\label{nie7eps}
\end{eqnarray}
This situation is similar to the case
described previously.

\subsection{The case $0<\gamma<1$}
\label{sec_kn0g1}

For $0<\gamma<1$ the equations determining 
the pseudo rest-mass density and the energy density as a function of the scale
factor can be written as
\begin{eqnarray}
\rho
\sqrt{1+\frac{2}{c^2}\frac{|K|\gamma}{1-\gamma}\frac{1}{\rho^{1-\gamma}}}=\frac{
Qm } { a^3 },
\label{f9}
\end{eqnarray}
\begin{eqnarray}
\epsilon=\rho c^2+\frac{\gamma+1}{1-\gamma}|K|\rho^{\gamma}.
\label{f10}
\end{eqnarray}

When $\rho\rightarrow +\infty$, Eqs. (\ref{f9}) and (\ref{f10}) reduce to
\begin{eqnarray}
\rho\sim\frac{Qm}{a^3},
\label{f11}
\end{eqnarray}
\begin{eqnarray}
\epsilon\sim \rho c^2.
\label{f12}
\end{eqnarray}
This corresponds to the nonrelativistic regime $P/\epsilon\ll 1$ (matterlike
era) valid for $a\rightarrow 0$.  Starting from $+\infty$ when $a\rightarrow 0$,
the pseudo rest-mass density and the energy density decrease as $a$ increases.

\begin{figure}[!h]
\begin{center}
\includegraphics[clip,scale=0.3]{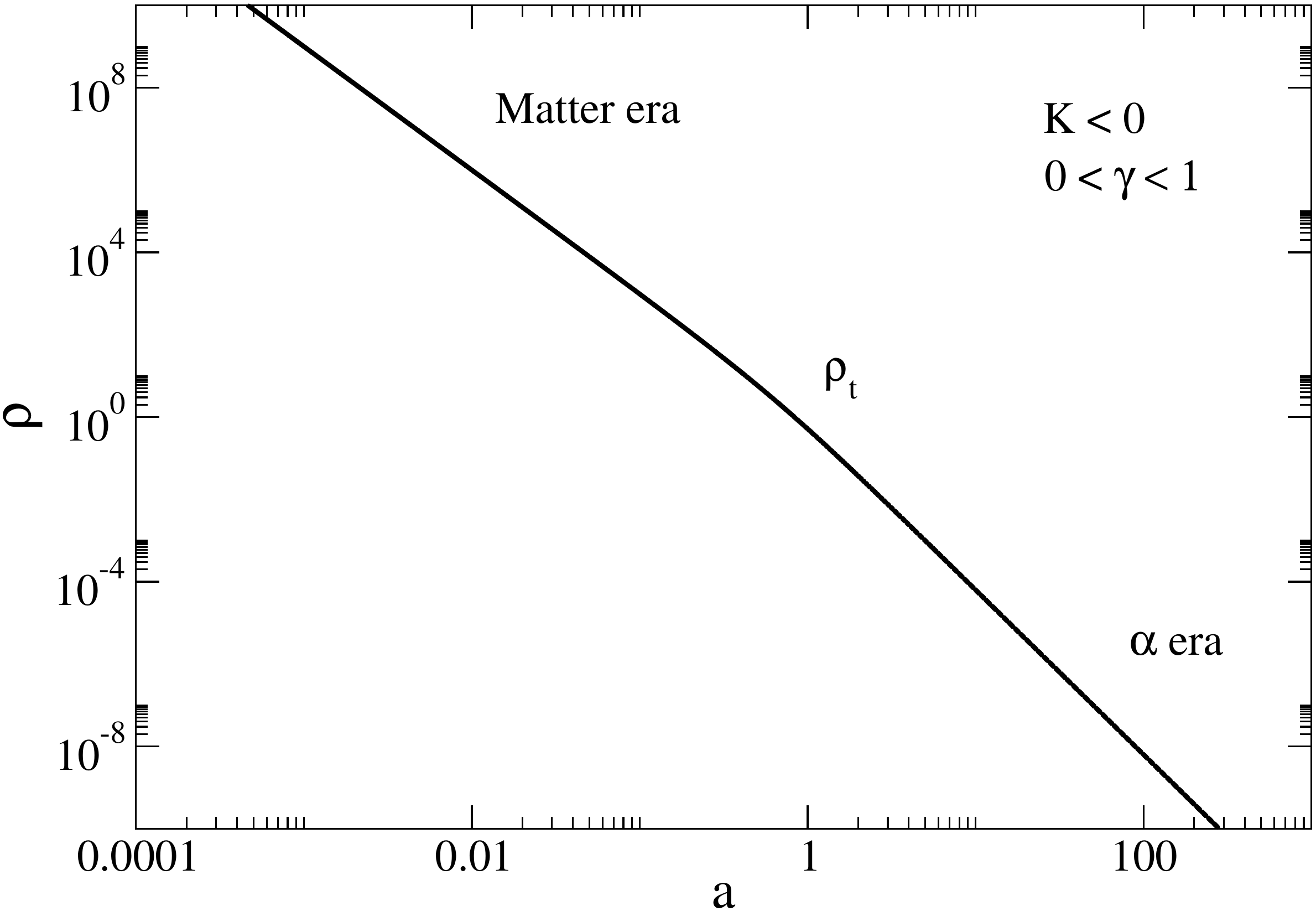}
\caption{Evolution of the pseudo rest-mass density as a function of the scale
factor for $0<\gamma<1$ (specifically $\gamma=0.5$). The
pseudo rest-mass density 
decreases more rapidly in the $\alpha$-era than in the matterlike era.}
\label{arKneg0g1}
\end{center}
\end{figure}

\begin{figure}[!h]
\begin{center}
\includegraphics[clip,scale=0.3]{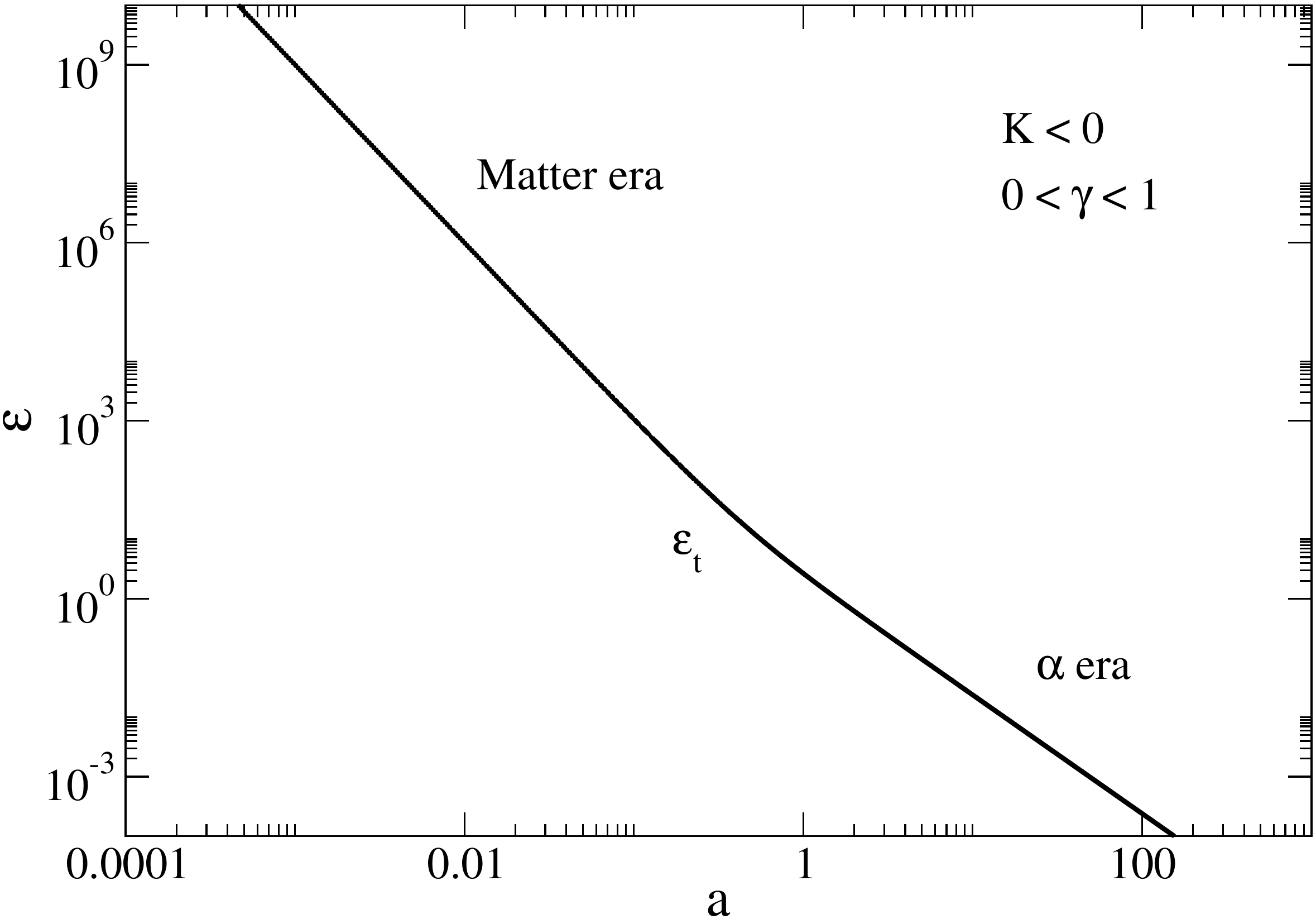}
\caption{Evolution of the energy density as a function of the scale factor for
$0<\gamma<1$ (specifically $\gamma=0.5$). The energy density 
decreases more rapidly in the matterlike era than in the $\alpha$-era.}
\label{aepsKneg0g1}
\end{center}
\end{figure}

\begin{figure}[!h]
\begin{center}
\includegraphics[clip,scale=0.3]{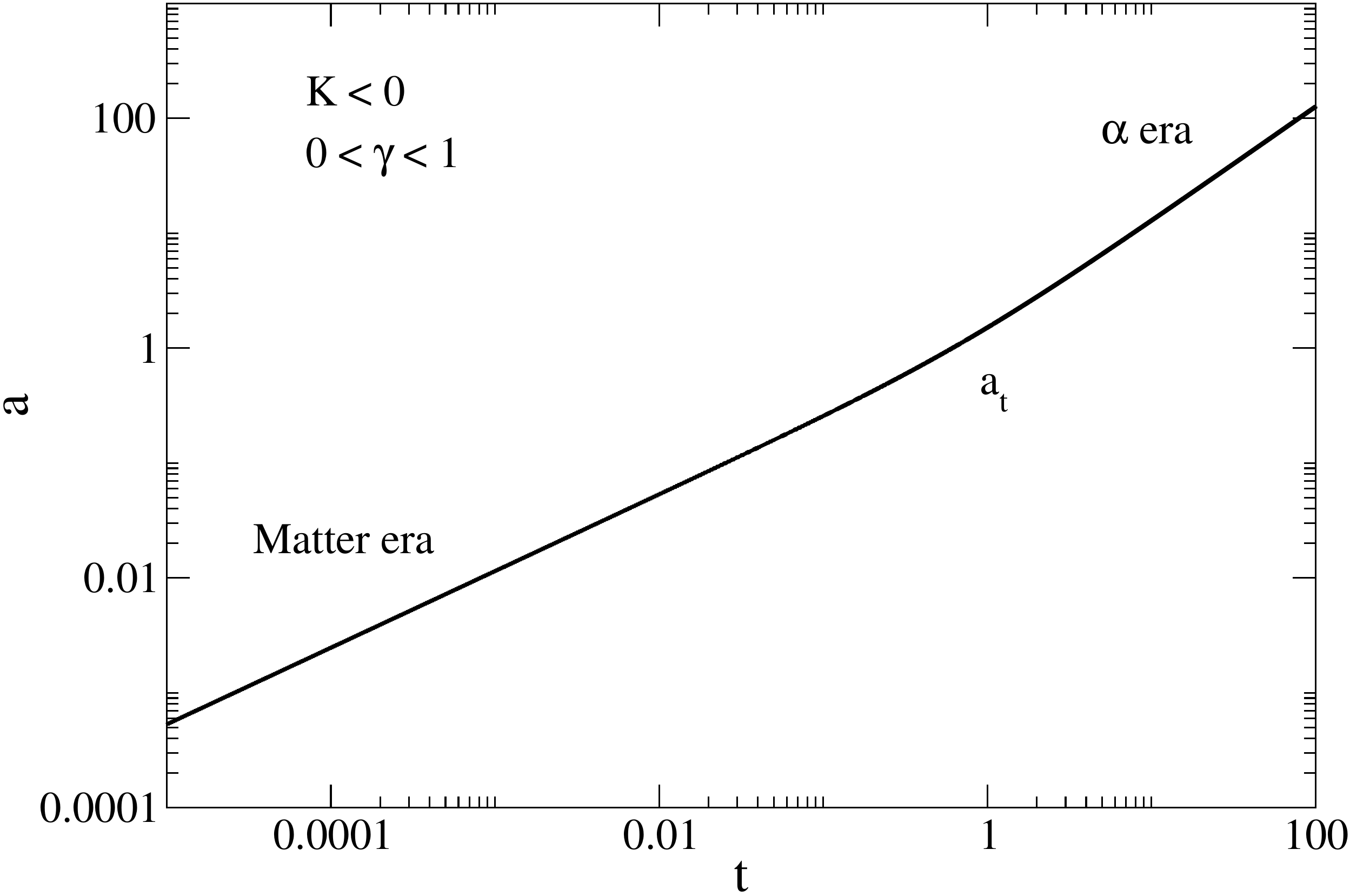}
\caption{Evolution of the scale factor as a function of time for $0<\gamma<1$
(specifically $\gamma=0.5$). The scale factor increases more
rapidly in the $\alpha$-era than in the matterlike era.}
\label{taKneg0g1}
\end{center}
\end{figure}

When $\rho\rightarrow 0$, Eqs. (\ref{f9}) and (\ref{f10})  reduce to
\begin{eqnarray}
\rho\sim \left(\frac{Q^2m^2c^2}{2}\frac{1-\gamma}{|K|\gamma}\right
)^{1/(1+\gamma)} \frac{1}{a^{6/(1+\gamma)}},
\label{f13}
\end{eqnarray}
\begin{eqnarray}
\epsilon\sim
\frac{\gamma+1}{1-\gamma}|K|\rho^{\gamma},
\label{f14b}
\end{eqnarray}
\begin{eqnarray}
\epsilon\sim
\frac{\gamma+1}{1-\gamma}|K|\left(\frac{Q^2m^2c^2}{2}\frac{1-\gamma}{|K|\gamma}
\right
)^{\gamma/(1+\gamma)} \frac{1}{a^{6\gamma/(1+\gamma)}}.
\label{f14}
\end{eqnarray}
This corresponds to the ultrarelativistic regime valid for  $a\rightarrow
+\infty$. The pseudo rest-mass density and the energy density decrease with $a$
and tend to zero when $a\rightarrow +\infty$. Using Eqs. (\ref{plsf3}) and
(\ref{f14b}) we obtain
the equation of state
\begin{eqnarray}
P=-\frac{1-\gamma}{\gamma+1}\epsilon.
\label{f15}
\end{eqnarray}
The pressure is a linear function $P\sim\alpha\epsilon$ of the energy density
with 
coefficient $\alpha=-(1-\gamma)/(\gamma+1)$.

When $K<0$ and $0<\gamma<1$, the universe evolves from a
 matterlike era
($P\simeq 0$)  in the early universe to an
$\alpha$-era ($P\sim\alpha\epsilon$)
in the late universe.  The curves $\rho(a)$ and $\epsilon(a)$ 
are plotted in Figs. \ref{arKneg0g1} and \ref{aepsKneg0g1}.

The temporal evolution  $a(t)$ of the scale factor is represented in Fig.
\ref{taKneg0g1}. It is obtained by integrating Eq. (\ref{plsf11})
numerically. Starting
from a singularity at $t=0$ where $a=0$ and $\epsilon\rightarrow +\infty$
(big bang) the scale factor first grows as $a\propto t^{2/3}$ in the matterlike
era
(EdS solution) then as $a\propto t^{(\gamma+1)/3\gamma}$
(corresponding to $a\propto
t^{2/[3(1+\alpha)]}$) in
the $\alpha$-era.

The transition between the two regimes typically  occurs at 
\begin{eqnarray}
a_t=\frac{1}{(Qm)^{\frac{1+\gamma}{3(1-\gamma)}}}\left (\frac{Q^2m^2c^2}{2}\frac{1-\gamma}{|K|\gamma}\right )^{\frac{1}{3(1-\gamma)}},
\label{ne9b}
\end{eqnarray}
\begin{eqnarray}
\epsilon_t=\rho_t c^2=\left (\frac{2}{c^2}\frac{|K|\gamma}{1-\gamma}\right )^{\frac{1}{1-\gamma}}c^2.
\label{ne9c}
\end{eqnarray}

The expansion of the universe is decelerating ($\alpha>-1/3$) in the
$\alpha$-era when $1/2<\gamma<1$  and accelerating ($\alpha<-1/3$)
when $0<\gamma<1/2$. This provides a physical interpretation
of the particular index $\gamma=1/2$ considered in Sec. \ref{sec_pwlnon}.

\subsection{The case $\gamma=0$ ($\Lambda$CDM model)}
\label{sec_kn0g}

For $\gamma=0$, the equations determining
the
pseudo 
rest-mass density and the energy density as a function of the scale factor
reduce to
\begin{eqnarray}
\rho=\frac{Qm}{a^3},
\label{ie17}
\end{eqnarray}
\begin{eqnarray}
\epsilon=\rho c^2+|K|.
\label{ie18}
\end{eqnarray}
They can be combined to give
\begin{eqnarray}
\epsilon=\frac{Qmc^2}{a^3}+|K|.
\label{ie19}
\end{eqnarray}
This equation is equivalent to the one obtained in the  $\Lambda$CDM model which
assumes that the universe is filled with pressureless DM ($P=0$) and that the
cosmological constant is positive  ($\Lambda>0$) or
that it is represented by a fluid with an equation of state $P_{\rm
de}=-\epsilon_{\rm de}$ yielding a constant energy density $\epsilon_{\rm
de}=\rho_\Lambda c^2>0$. This agreement is expected
since, when $\gamma=0$, the pressure is constant ($P=K$) and we know that a
constant negative pressure $P=-\rho_\Lambda c^2$  returns the $\Lambda$CDM model
\cite{sandvik,avelinoZ,cosmopoly2}. In the present context, the
$\Lambda$CDM model is obtained
from a complex SF theory with a constant positive potential
$V(|\varphi|^2)=\rho_\Lambda c^2$. For this particular model, we
see that the pseudo rest-mass density decreases as $\rho\propto a^{-3}$ and
behaves as DM (see the Remark below). 

The $\Lambda$CDM model has been studied in Sec.
5 of \cite{cosmopoly2}. To make the connection with this study,  we set
$K=-\rho_{\Lambda}c^2$ and $a_2=(Qm/\rho_{\Lambda})^{1/3}$. Eq.
(\ref{ie19}) can then be rewritten as
\begin{eqnarray}
\epsilon=\rho_{\Lambda}c^2\left\lbrack \left (\frac{a_2}{a}\right )^{3}+1\right\rbrack.
\label{iantil1}
\end{eqnarray}
Starting from $+\infty$ when $a\rightarrow 0$, the energy
density decreases as $a$ increases and tends to a constant value
$K=\rho_{\Lambda}c^2$ when $a\rightarrow +\infty$. The temporal evolution of the
scale factor is obtained by solving the Friedmann equation (\ref{fe1}) with
Eq. (\ref{iantil1}). In
that case, the solution can be obtained analytically yielding \cite{cosmopoly2}
\begin{eqnarray}
\frac{a}{a_2}=\sinh^{2/3}\left (\sqrt{6\pi}\frac{t}{t_\Lambda}\right ),
\label{iantil4}
\end{eqnarray}
\begin{eqnarray}
\epsilon=\frac{\rho_{\Lambda}c^2}{\tanh^{2}\left (\sqrt{6\pi}\frac{t}{t_\Lambda}\right )},
\label{iantil5}
\end{eqnarray}
\begin{eqnarray}
\rho=\frac{\rho_{\Lambda}}{\sinh^{2}\left
(\sqrt{6\pi}\frac{t}{t_\Lambda}\right )},
\label{iantil5rho}
\end{eqnarray}
where $t_{\Lambda}=1/\sqrt{G\rho_{\Lambda}}$ is the cosmological time. The
evolution of the universe is similar to the one described in Sec.
\ref{sec_kng0} below with the particularity that $\rho\rightarrow 0$
for
$a\rightarrow +\infty$. The $\Lambda$CDM model is
studied in more detail in Sec. 5 of \cite{cosmopoly2}.

{\it Remark:} In Eq. (\ref{ie19}) the first
term is the rest-mass energy density
$\rho_m c^2$ [see
Eqs. (\ref{mtd11}) and (\ref{rmd2})] and the second term is the internal energy
density $u$ [see Eq. (\ref{mtd2b})]. As discussed in Sec. \ref{sec_dmde}, the
rest-mass density $\rho_m$ can be interpreted as DM and the internal energy
density $u$ can be interpreted as DE \cite{epjp,lettre}. In the present case,
the
pseudo rest-mass density coincides with the rest-mass density ($\rho=\rho_m$)
and the internal energy density is constant ($u=|K|=\rho_\Lambda c^2$).
In the two-fluid model associated with the $\Lambda$CDM model
(see Sec.
\ref{sec_twofluids}), DM has an  equation of state $P_{\rm
m}(\epsilon_{\rm m})=0$ and DE has an equation of state $P_{\rm
de}(\epsilon_{\rm
de})=-\epsilon_{\rm de}$.

\subsection{The case $\gamma<0$}
\label{sec_kng0}

For $\gamma<0$ the equations determining the pseudo rest-mass density 
and the energy density as a function of the scale factor can be written as
\begin{eqnarray}
\rho
\sqrt{1-\frac{2}{c^2}\frac{|K||\gamma|}{1-\gamma}\frac{1}{\rho^{1-\gamma}}}
=\frac {
Qm } { a^3 } ,
\label{f19}
\end{eqnarray}
\begin{eqnarray}
\epsilon=\rho c^2+\frac{\gamma+1}{1-\gamma}|K|\frac{1}{\rho^{|\gamma|}}.
\label{f20}
\end{eqnarray}

When $\rho\rightarrow +\infty$, Eqs. (\ref{f19}) and (\ref{f20}) reduce to
\begin{eqnarray}
\rho\sim\frac{Qm}{a^3},
\label{f21}
\end{eqnarray}
\begin{eqnarray}
\epsilon\sim \rho c^2.
\label{f22}
\end{eqnarray}
This corresponds to the nonrelativistic regime $|P|/\epsilon\ll 1$
(matterlike era) valid for $a\rightarrow 0$.  Starting from $+\infty$ when
$a\rightarrow 0$, the pseudo rest-mass density and the energy density decrease
as $a$ increases.

\begin{figure}[!h]
\begin{center}
\includegraphics[clip,scale=0.3]{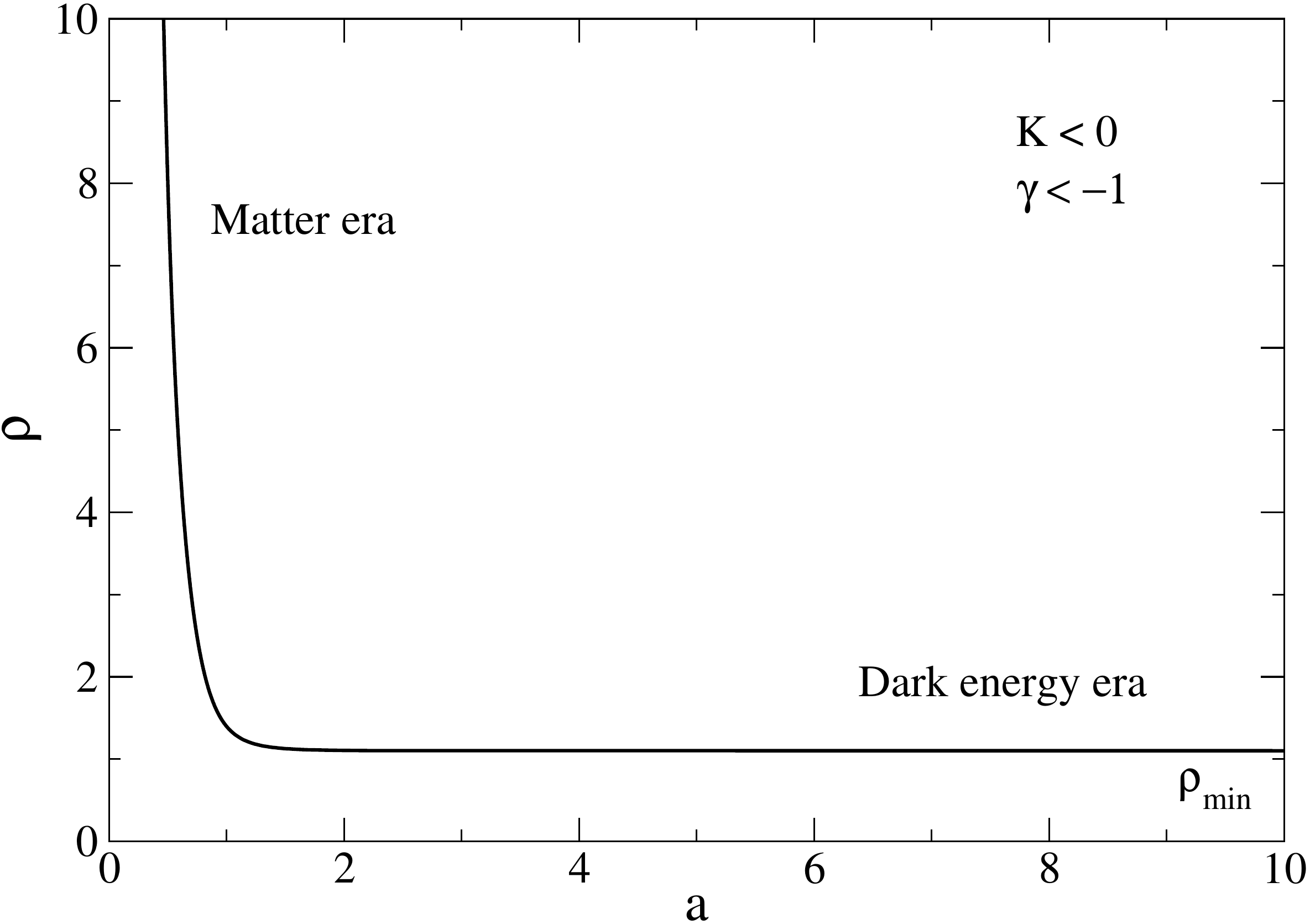}
\caption{Evolution of the pseudo rest-mass density as a function of the 
scale factor for $\gamma=-2$.}
\label{arKneggminus2}
\end{center}
\end{figure}

\begin{figure}[!h]
\begin{center}
\includegraphics[clip,scale=0.3]{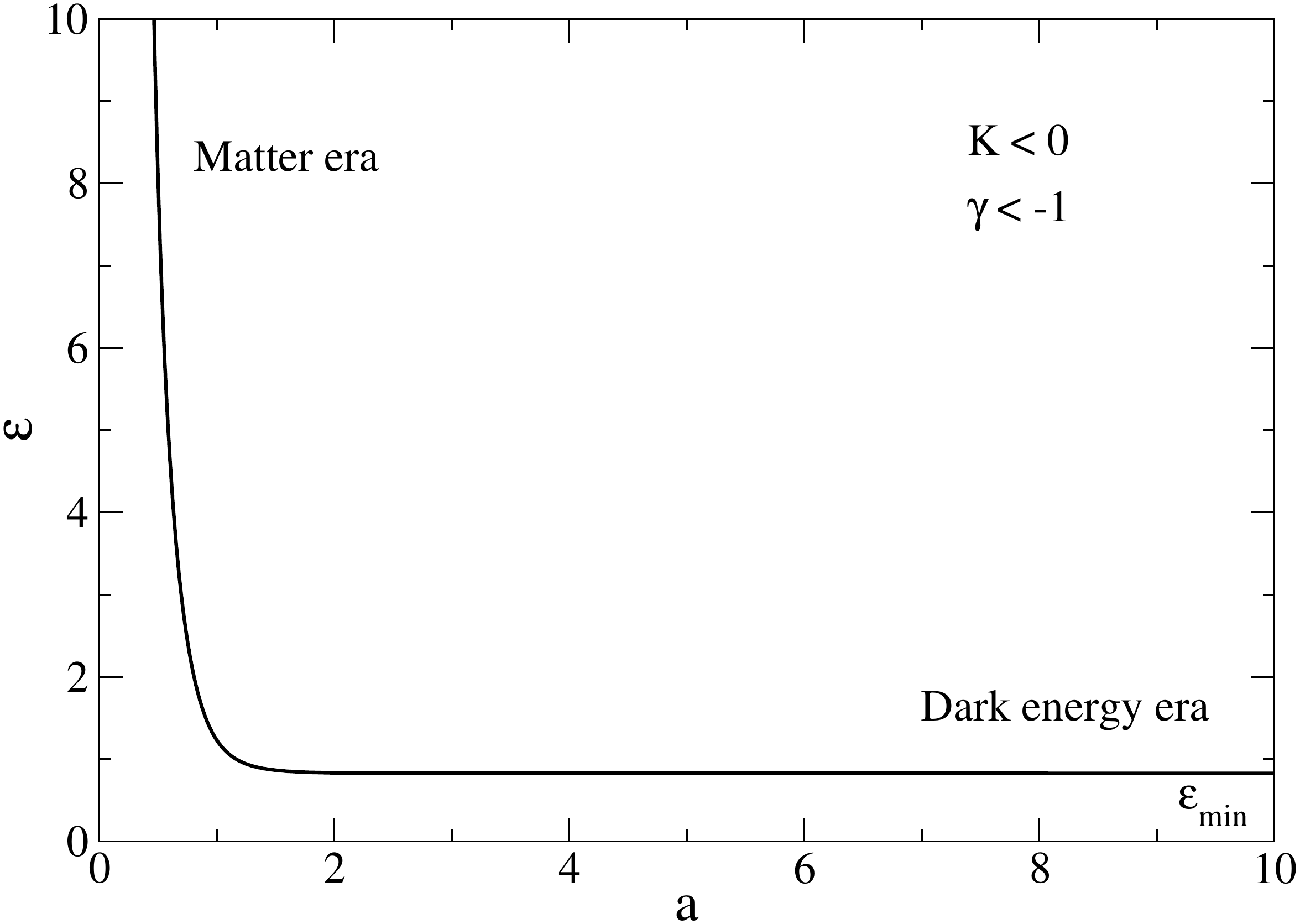}
\caption{Evolution of the energy density as a function of the scale factor 
for $\gamma=-2$. }
\label{aepsKneggminus2}
\end{center}
\end{figure}

Considering now  small values of $\rho$ (ultrarelativistic
regime) we see that Eq. (\ref{f19}) imposes the condition
$\rho\ge \rho_{\rm min}$ with
\begin{eqnarray}
\rho_{\rm min}=\left (\frac{2}{c^2}\frac{|K||\gamma|}{1-\gamma}\right
)^{1/(1-\gamma)}.
\label{f23}
\end{eqnarray}
The pseudo rest-mass density decreases as $a$ increases
and tends to $\rho_{\rm min}$ when $a\rightarrow +\infty$. Similarly, the energy
density decreases as $a$ increases and tends to a constant
\begin{eqnarray}
\epsilon_{\rm min}= \frac{1-\gamma}{2|\gamma|}\rho_{\rm min}c^2
\label{f24}
\end{eqnarray}
when $a\rightarrow +\infty$. We see that the
evolution of the universe in this regime is similar to the one induced by a
cosmological constant or by a constant energy density.  This
corresponds to
the DE era.  The curves
$\rho(a)$ and  $\epsilon(a)$  are plotted in Figs.
\ref{arKneggminus2} and \ref{aepsKneggminus2}.

\begin{figure}[!h]
\begin{center}
\includegraphics[clip,scale=0.3]{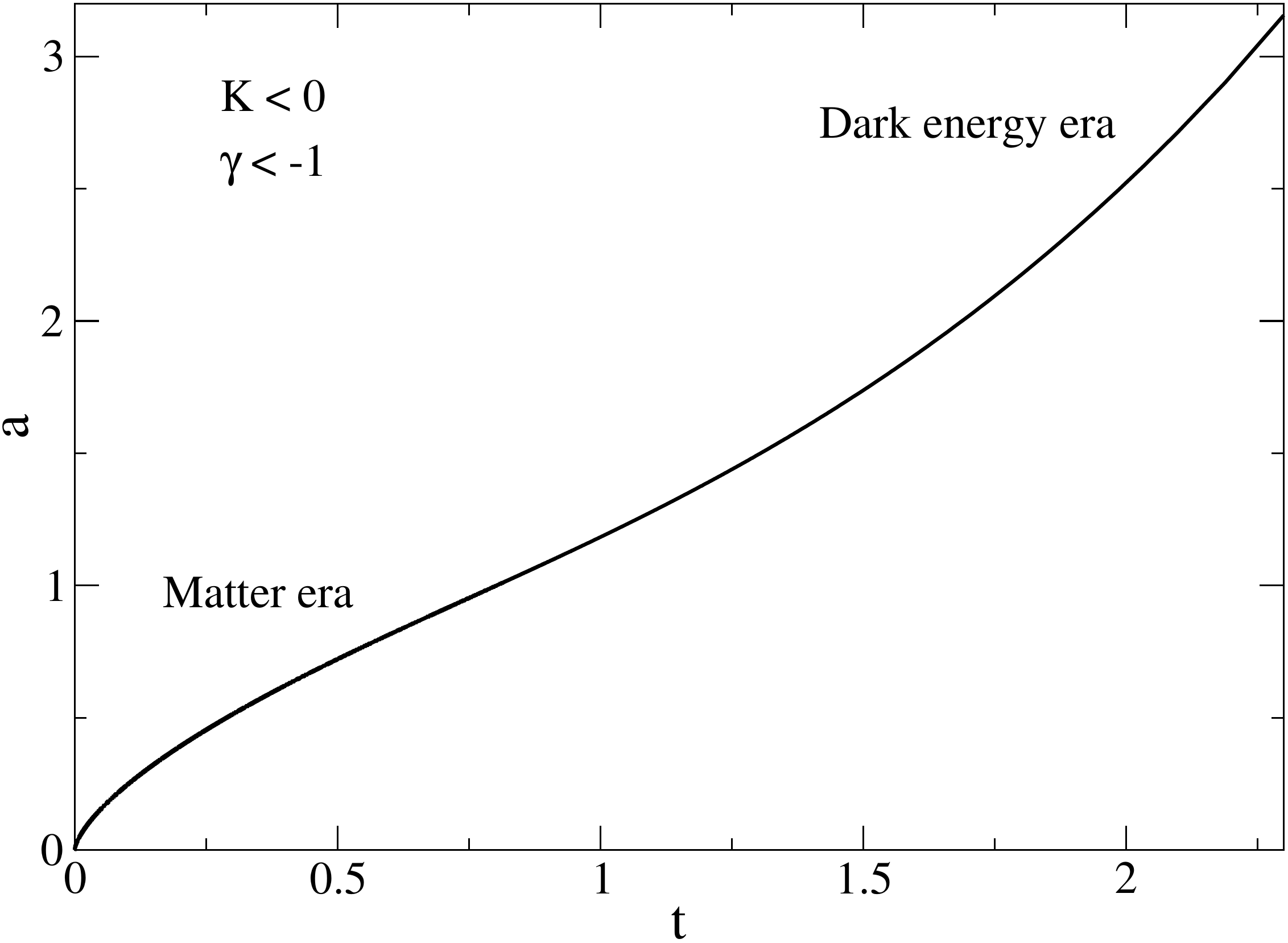}
\caption{Evolution of
the scale factor as a function of 
time for $\gamma=-2$. }
\label{taKneggminus2}
\end{center}
\end{figure}

The temporal evolution $a(t)$ of the scale factor is represented in Fig.
\ref{taKneggminus2}. It is obtained by integrating Eq. (\ref{plsf11})
numerically. Starting
from a singularity at $t=0$ where $a=0$ and $\epsilon\rightarrow +\infty$
(big bang) the scale factor first grows as $a\propto t^{2/3}$ in the matterlike
era (EdS) then as  $a\propto e^{(8\pi
G\epsilon_{\rm min}/3c^2)^{1/2}t}$ in the DE era (de Sitter).

The transition between the two regimes  typically occurs when $Qmc^2/a^3\sim
\epsilon_{\rm min}$  yielding
\begin{eqnarray}
a_t=\left (\frac{2|\gamma|Qm}{1-\gamma}\right )^{1/3}\left (\frac{c^2}{2}\frac{1-\gamma}{|K||\gamma|}\right )^{\frac{1}{3(1-\gamma)}}.
\label{atg}
\end{eqnarray}

\subsection{The case $\gamma=-1$ (Chaplygin gas)}
\label{sec_kpn0}

For  $\gamma=-1$, the equations determining the pseudo rest-mass
density and the energy density as a function of the scale factor reduce to
\begin{eqnarray}
\rho=\sqrt{\frac{Q^2m^2}{a^6}+\frac{|K|}{c^2}},
\label{e30b}
\end{eqnarray}
\begin{eqnarray}
\epsilon=\rho c^2.
\label{e31b}
\end{eqnarray}
They can be combined to give
\begin{eqnarray}
\epsilon= \sqrt{\frac{Q^2m^2c^4}{a^6}+|K| c^2}.
\label{e32c}
\end{eqnarray}
This equation is equivalent to the
one obtained in the Chaplygin gas model. This agreement is expected because,
when
$\gamma=-1$, we have $\epsilon=\rho c^2$. Therefore, the
equation of state $P=K/\rho$
from Eq. (\ref{plsf3}) can be written as $P=Kc^2/\epsilon$ which is the
equation of state of the Chaplygin gas when $K<0$
\cite{kmp,btv,gkmp,cosmopoly2}. In the present
context, the Chaplygin gas model is obtained from a complex SF theory with a
potential $V(|\varphi|^2)=-\frac{1}{2}K(\frac{\hbar}{m})^2\frac{1}{
|\varphi|^2}$. For this particular model, the pseudo rest-mass
density coincides with the energy density ($\rho=\epsilon/c^2$).

The Chaplygin gas has been studied in Sec. 4.4 of \cite{cosmopoly2}. To make
the connection with this study, we set $\rho_*=\sqrt{|K|/c^2}$ and
$a_*=(Q^2m^2c^2/|K|)^{1/6}$. Eq. (\ref{e32c}) can then  be rewritten as
\begin{eqnarray}
\epsilon=\rho_{*}c^2\sqrt{\left (\frac{a_*}{a}\right )^{6}+1}.
\label{antil1c}
\end{eqnarray}
In the nonrelativistic regime where
$\rho,\epsilon\rightarrow +\infty$, we have
\begin{eqnarray}
\epsilon=\rho c^2 \sim\frac{Q m c^2}{a^3}=\rho_{*}c^2\left (\frac{a_*}{a}\right
)^{3}.
\label{e33c}
\end{eqnarray}
This corresponds to the matterlike era ($a\ll a_*$) where $|P|/\epsilon\ll 1$.
When $a\rightarrow
+\infty$, the energy density tends to a constant
\begin{eqnarray}
\epsilon_{\rm min}=\sqrt{|K|c^2}=\rho_* c^2.
\label{f29}
\end{eqnarray}
This corresponds to the  DE era ($a\gg a_*$). The Chaplygin equation of state
can mimic 
the effect of a cosmological constant at late times. The temporal evolution of
the scale factor is obtained by solving the Friedmann equation (\ref{fe1})
with Eq. (\ref{antil1c}).
This yields \cite{cosmopoly2}
\begin{eqnarray}
\sqrt{96\pi G\rho_*}\, t=\int_{\left (\frac{a_*}{a}\right)^6}^{+\infty} \frac{dx}{x(x+1)^{1/4}}.
\label{antil4c}
\end{eqnarray}
The integral can be calculated explicitly and is given by Eq. (38) of
\cite{cosmopoly2}. The
evolution of the universe has been described in the previous section. The
Chaplygin gas model is
studied in more detail in Sec. 4.4 of \cite{cosmopoly2}.

{\it Remark:} We can rewrite Eq. (\ref{e32c}) as
\begin{eqnarray}
\epsilon=\frac{Qmc^2}{a^3}+\left\lbrack \sqrt{\frac{Q^2m^2c^4}{a^6}+|K|
c^2}-\frac{Qmc^2}{a^3}\right\rbrack,
\label{ne32c}
\end{eqnarray} 
where the first term is the rest-mass energy density $\rho_m c^2$ [see
Eqs. (\ref{mtd11}) and (\ref{rmd2})] and the second term is the internal energy
density $u$
[see Eq. (\ref{mtd2b})]. As discussed in Sec. \ref{sec_dmde}, the
rest-mass density $\rho_m$ can be interpreted as DM and the internal energy
density $u$ can be interpreted as DE \cite{epjp,lettre}. In the two-fluid
model associated with the
Chaplygin gas (see Sec.
\ref{sec_twofluids}),  DM has an  equation of state $P_{\rm
m}(\epsilon_{\rm m})=0$ and DE has an equation of state 
\begin{eqnarray}
P_{\rm de}(\epsilon_{\rm de})=\frac{-2|K|c^2\epsilon_{\rm de}}{\epsilon_{\rm
de}^2+|K|c^2},
\label{twgg}
\end{eqnarray}
obtained by eliminating $\rho$ between Eqs. (\ref{chap1}) and
(\ref{aswe}), and by identifying $P(u)$ with $P_{\rm de}(\epsilon_{\rm de})$.
Solving the energy conservation equation (\ref{hsf4}) with the equation of
state (\ref{twgg}), we recover the expression of DE in Eq. (\ref{ne32c}).

\section{Conclusion}

In this paper, we have considered a complex SF with a self-interaction
potential $V(|{\varphi}|^2)$ described by the KGE equations (\ref{csf4})
and (\ref{ak21}). In the nonrelativistic regime, these equations reduce to the
GPP equations (\ref{nr1}) and (\ref{nr2}). We have determined the equation of
state $P(\rho)$ associated with the self-interaction potential
$V(|{\varphi}|^2)$ in the TF approximation. Some examples of SF potentials, and
the corresponding equations of state, are given in Appendix \ref{sec_ex}. In
this paper, we have specifically considered the cosmological evolution of a
spatially homogeneous complex SF with a polytropic or an isothermal equation of
state in the fast oscillation regime (equivalent to the TF approximation).

In the case $K>0$ corresponding to a positive pressure, we
have found the following results. The
models with $\gamma>1$ describe the transition between an $\alpha$-era and a
pressureless DM
era. For a $|\varphi|^4$ potential ($\gamma=2$),
corresponding to the standard BEC with a repulsive self-interaction,  there is a
transition between a dark radiation era (due to the SF) and a matter
era. This model is consistent
with
the observations \cite{shapiro,abrilphas}. The models with $\gamma\le 1$
describe a cyclic
universe presenting periods of expansion (big bang) and contraction (big
crunch). The evolution is symmetric for $-1\le\gamma\le 1$ (with a vanishing
energy density at the maximum scale factor) and asymmetric for
$\gamma<-1$ (with a nonvanishing energy density at the maximum scale
factor).
Symmetric cyclic universes include the anti-$\Lambda$CDM model ($\gamma=0$) and
the anti-Chaplygin gas model ($\gamma=-1$). These models are not consistent
with
the observations. 

In the case $K<0$ corresponding to a negative pressure, we
have found the following results. The models with
$\gamma\ge 1$ describe a bouncing universe dominated by DE when $t<0$ and
by pressureless DM when $t>0$, or
the converse. At $t=0$ the universe  achieves its minimum
radius $a_{\rm min}$ and its maximum energy density $\epsilon_{i}$. There is no
singularity. This model
can also describe a peculiar evolution with two branches \cite{abrilphas}.
The SF emerges suddently at some finite scale factor  $a_{\rm min}$  and energy
density $\epsilon_{i}$ and follows either the nonrelativistic branch (DM) or the
ultrarelativistic branch (DE). Therefore, we get either an
asymmetric 
bouncing universe or a universe with two branches of solutions. These models
are
not consistent with the observations. The
models with $\gamma<1$ describe the transition between a pressureless DM era and
a DE era. They provide UDM models.
For $0<\gamma<1$, the energy density increases indefinitely in the DE era
($\alpha$-era) like in
quintessence models. For $\gamma\le 0$, the energy density tends to a constant
in
the DE era like in the presence of a cosmological constant. This gives rise to
a de Sitter era at late times. These UDM models include the $\Lambda$CDM model
($\gamma=0$) and the Chaplygin gas model ($\gamma=-1$). The Chaplygin gas model
does
not give a good agreement with the observations. Only polytropic models with
$\gamma$ sufficiently close to $0$, i.e., sufficiently close to the $\Lambda$CDM
model are consistent with
the observations.

A limitiation of our study is that we have not studied in detail (case by case)
the validity of the fast oscillation regime (this is done in \cite{abrilphas}
for the
quartic potential). This is partly due to reasons of conciseness and partly due
to the fact
that most models are not consistent with the observations so it may not be
necessary
to perform a more detailed study than the one given here. However, in a
companion paper \cite{graal}, we consider the logotropic model
which is consistent with the observations. In
that case we study the validity of the fast oscillation regime in detail.

In Ref. \cite{action} we have shown that the equation of state of a 
relativistic barotropic fluid can be specified in different manners depending
on whether the pressure is expressed in terms of the energy density
$\epsilon$ (model I), the rest-mass density $\rho_m$ (model II), or the pseudo
rest-mass density $\rho$ (model III). In the present paper, we have considered
the cosmological evolution of fluids described by a polytropic equation of
state of type III. The cosmological evolution of fluids described by a
polytropic equation of state of type I has been studied in
\cite{cosmopoly1,cosmopoly2,cosmopoly3} and the cosmological evolution of fluids
described by a
polytropic equation of state of type II has been studied in \cite{stiff}.

\appendix

\section{Inhomogeneous relativistic complex SF in a curved spacetime}
\label{sec_ra}

\subsection{Klein-Gordon-Einstein equations}
\label{sec_csfb}

The evolution of a possibly spatially inhomogeneous relativistic  complex SF
$\varphi(x^\mu)=\varphi(x,y,z,t)$,  which may represent the wavefunction of a
relativistic BEC, is governed by the KG equation
\begin{equation}
\label{csf4}
\square\varphi+2\frac{dV_{\rm tot}}{d|\varphi|^2}\varphi=0,
\end{equation}
where
$\square=D_{\mu}\partial^{\mu}=\frac{1}{\sqrt{-g}}\partial_{\mu}(\sqrt{-g}\, g^{
\mu\nu} \partial_{\nu})$ is the d'Alembertian
operator in a curved spacetime.  The potential $V_{\rm
tot}(|\varphi|^2)$ can be
decomposed into a rest-mass energy term and a 
self-interaction energy term as
\begin{equation}
\label{csf3}
V_{\rm
tot}(|\varphi|^2)=\frac{m^2c^2}{2\hbar^2}|\varphi|^2+V(|\varphi|^2).
\end{equation}
The KG equation is coupled to the Einstein
field equations 
\begin{equation}
R_{\mu\nu}-\frac{1}{2}g_{\mu\nu}R=\frac{8\pi G}{c^4}T_{\mu\nu},
\label{ak21}
\end{equation}
where
\begin{eqnarray}
\label{em2}
T_{\mu\nu}&=&\frac{1}{2}(\partial_{\mu}\varphi^*\partial_{\nu}\varphi+\partial_{
\nu}\varphi^*\partial_{\mu}\varphi)
\nonumber\\
&-&g_{\mu\nu}\left\lbrack
\frac{1}{2}g^{\rho\sigma}\partial_{\rho}\varphi^*\partial_{\sigma}
\varphi-V_{\rm tot}(|\varphi|^2)\right\rbrack
\end{eqnarray}
is the energy-momentum tensor of the SF. This leads to the
Klein-Gordon-Einstein (KGE) equations (the term in brackets
corresponds to the Lagrangian density of the SF). The energy-momentum tensor
satisfies
the equation ${D}_{\mu}T^{\mu\nu}=0$ expressing the local conservation of energy
and momentum. On the other hand, the quadricurrent 
\begin{eqnarray}
\label{charge1}
J_{\mu}=-\frac{m}{2i\hbar}
(\varphi^*\partial_\mu\varphi-\varphi\partial_\mu\varphi^*)
\end{eqnarray}
satisfies the equation
${D}_{\mu}J^{\mu}=0$ expressing the local conservation of
charge (see, e.g., \cite{action} for details). The charge 
\begin{eqnarray}
Q=\frac{e}{mc}\int
J^0\sqrt{-g}\,
d^3x
\end{eqnarray}
of the SF is proportional to the number $N$
of bosons ($Q=Ne$)
provided that antibosons are counted 
negatively \cite{landaulifshitz}.  Therefore, the
equation ${D}_{\mu}J^{\mu}=0$ also
expresses the local conservation of the boson number. For a real SF the
quadricurrent vanishes implying that the particle number is not
conserved.

\subsection{The de Broglie transformation}
\label{sec_db}

We can write the KG equation (\ref{csf4}) under the form of 
hydrodynamic equations by making the de Broglie
\cite{broglie1927a,broglie1927b,broglie1927c}  transformation. To that
purpose, we write the SF as
\begin{equation}
\varphi=\frac{\hbar}{m}\sqrt{\rho}e^{i S_{\rm tot}/\hbar},
\label{db1}
\end{equation}
where $\rho$ is the pseudo rest-mass density\footnote{We
stress that
$\rho$ is {\it not} the rest-mass density $\rho_m=nm$ (see below). It is only in
the nonrelativistic regime $c\rightarrow +\infty$ that $\rho$ coincides with the
rest-mass density $\rho_m$.} and $S_{\rm tot}$ is the action. They are given by
\begin{eqnarray}
\rho=\frac{m^2}{\hbar^2}|\varphi|^2\quad {\rm and}\quad S_{\rm
tot}=\frac{\hbar}{2i}\ln \left (\frac{\varphi}{\varphi^*}\right ).
\label{db2}
\end{eqnarray}
We also have
\begin{equation}
\label{db10}
V_{\rm
tot}(\rho)=\frac{1}{2}\rho c^2+V(\rho).
\end{equation}
For convenience, we define $\theta=S_{\rm tot}/m$. In that case, Eq.
(\ref{db1}) becomes
\begin{equation}
\varphi=\frac{\hbar}{m}\sqrt{\rho}e^{i m \theta/\hbar}.
\label{db4}
\end{equation}
The angle (phase) and the pulsation of the SF
are given by\footnote{The angle was noted $\theta$
instead of $\Theta$ in Sec. \ref{sec_csf} and in \cite{abrilphas}.}
\begin{equation}
\Theta=\frac{S_{\rm tot}}{\hbar}=\frac{m\theta}{\hbar},\qquad
\omega=-\dot\Theta=-\frac{{\dot S}_{\rm
tot}}{\hbar}=-\frac{m\dot\theta}{\hbar}=\frac{E_{\rm tot}}{\hbar},
\label{db4ap}
\end{equation}
where $E_{\rm tot}=-{\dot S}_{\rm tot}$ is the energy.

Substituting the de Broglie transformation from Eq. (\ref{db4}) into the
KG equation (\ref{csf4}), and separating the real and the imaginary parts we get
\begin{eqnarray}
D_{\mu}\left ( \rho \partial^{\mu}\theta\right )=0,
\label{db8}
\end{eqnarray}
\begin{eqnarray}
\frac{1}{2}\partial_{\mu}\theta\partial^{\mu}\theta-\frac{\hbar^2}{2m^2}\frac{
\square\sqrt { \rho } } { \sqrt { \rho } } -  V'_{\rm
tot}(\rho)=0.
\label{db9}
\end{eqnarray}
Equation (\ref{db8}) can be
interpreted as a continuity equation and Eq.
(\ref{db9}) can be interpreted as a  quantum relativistic Hamilton-Jacobi (or
Bernoulli)
equation with a relativistic covariant quantum potential
\begin{eqnarray}
Q_{\rm dB}=\frac{\hbar^2}{2m}\frac{\square\sqrt{\rho}}{\sqrt{\rho}}.
\label{db11}
\end{eqnarray}
Introducing the pseudo quadrivelocity\footnote{The pseudo quadrivelocity $v_\mu$
does not 
satisfy $v_\mu v^\mu=c^2$ so it is not guaranteed to be always timelike.
Nevertheless, $v_\mu$ can be introduced as a convenient notation.}
\begin{eqnarray}
v_\mu=-\frac{\partial_{\mu}S_{\rm tot}}{m}=-\partial_\mu\theta,
\label{db12}
\end{eqnarray}
we can rewrite Eqs. (\ref{db8}) and  (\ref{db9}) as
\begin{eqnarray}
D_{\mu}\left ( \rho v^{\mu}\right )=0,
\label{db13}
\end{eqnarray}
\begin{eqnarray}
\frac{1}{2}m v_{\mu}v^{\mu}-Q_{\rm dB}-m V'_{\rm
tot}(\rho)=0.
\label{db14}
\end{eqnarray}
Taking the gradient of the quantum Hamilton-Jacobi equation (\ref{db14}) we
obtain \cite{chavmatos} 
\begin{eqnarray}
\frac{dv_\nu}{dt}\equiv v^{\mu}D_\mu v_\nu=\frac{1}{m}\partial_\nu
Q_{\rm dB}+\partial_\nu V'(\rho),
\label{db15}
\end{eqnarray}
which can be interpreted as a relativistic quantum Euler equation (with the
limitation mentioned in footnote 18). The first term on the right hand side can
be
interpreted as a quantum force and the second term as a pressure force
$(1/\rho)\partial_\nu P$. The pressure $P(\rho)$ satisfies the relation
$(1/\rho)P'(\rho)=h'(\rho)=V''(\rho)$, where $h(\rho)=V'(\rho)$ is the
pseudo
enthalpy.
Integrating this relation, we get
\begin{equation}
\label{mad11b}
P(\rho)=\rho h(\rho)-V(\rho)=\rho
V'(\rho)-V(\rho)=\rho^2\left\lbrack
\frac{V(\rho)}{\rho}\right\rbrack'.
\end{equation}
This equation determines the pseudo equation of state $P(\rho)$ as a function of
the potential $V(\rho)$. Inversely, the potential is determined by the
pseudo equation
of state according to
\begin{eqnarray}
V(\rho)=\rho\int\frac{P(\rho)}{\rho^2}\, d\rho.
\label{etoilebis}
\end{eqnarray}

Using the de Broglie
transformation (\ref{db1}) the energy-momentum tensor (\ref{em2}) is given, in
the
hydrodynamic representation, by
\begin{eqnarray}
&&T_{\mu\nu}=\rho\partial_{\mu}\theta\partial_{\nu}\theta+\frac{\hbar^2}{
4m^2\rho }
\partial_{\mu}\rho\partial_{\nu}\rho\nonumber\\
&-&g_{\mu\nu}\left\lbrack
\frac{1}{2}g^{\mu\nu}\rho\partial_{\mu}\theta\partial_{\nu}
\theta+\frac{\hbar^2}{8m^2\rho}g^{\mu\nu}\partial_{\mu}\rho\partial_{\nu}
\rho-V_{\rm tot}(\rho)\right\rbrack.\nonumber\\
\label{em3gen}
\end{eqnarray}
Similarly, the quadricurrent (\ref{charge1}) can be written
as
\begin{eqnarray}
J^{\mu}=-\rho\partial^{\mu}\theta=\rho v^{\mu}.
\label{val1}
\end{eqnarray}
Therefore, the continuity equation (\ref{db8}) or (\ref{db13}) is equivalent to
$D_{\mu}J^{\mu}=0$. It expresses the
conservation of the charge $Q$ of the complex SF (or the conservation of the
boson
number $N$)
\begin{eqnarray}
Q=Ne=-\frac{e}{mc}\int \rho\partial^0\theta \sqrt{-g}  \, d^3x.
\label{charge5}
\end{eqnarray}
In the following we take
$e=1$ so that
$Q=N$. Assuming $\partial_\mu\theta
\partial^\mu\theta>0$, we can introduce the fluid
quadrivelocity
\begin{eqnarray}
u_{\mu}=-\frac{\partial_{\mu}\theta}{\sqrt{\partial_\mu\theta
\partial^\mu\theta} }c,
\label{rtf5bw}
\end{eqnarray}
which satisfies  the  identity
\begin{eqnarray}
u_{\mu}u^{\mu}=c^2.
\label{rtf5cw}
\end{eqnarray}
Using Eqs. (\ref{val1}) and (\ref{rtf5bw}), we have
\begin{eqnarray}
J^{\mu}=\frac{\rho}{c} \sqrt{\partial_\mu\theta
\partial^\mu\theta}  u^{\mu},
\label{val2}
\end{eqnarray}
and we
can write the continuity
equation (\ref{db8}) as
\begin{eqnarray}
D_{\mu}\left \lbrack \rho \sqrt{\partial_\mu\theta
\partial^\mu\theta} \, u^{\mu}\right \rbrack=0.
\label{charge9}
\end{eqnarray}
The rest-mass density $\rho_m=nm$ (which is
proportional to the charge density $\rho_e$) is defined by
\begin{eqnarray}
J^{\mu}=\rho_m u^{\mu}.
\label{val3}
\end{eqnarray}
The continuity equation $D_{\mu}J^{\mu}=0$ can be written as
\begin{eqnarray}
D_{\mu}(\rho_m u^{\mu})=0.
\label{charge7}
\end{eqnarray}
Comparing the expressions of $J^{\mu}$ given by Eqs. (\ref{val2}) and
(\ref{val3}), we
find that the
rest-mass density $\rho_m=nm$ of the SF is given
by
\begin{eqnarray}
\rho_m=\frac{\rho}{c}  \sqrt{\partial_\mu\theta
\partial^\mu\theta}.
\label{charge10a}
\end{eqnarray}
We note that $\rho_m\neq J^0/c$  in general. Using the Bernoulli equation
(\ref{db9}), we get
\begin{eqnarray}
\rho_m=\frac{\rho}{c}
\sqrt{\frac{\hbar^2}{m^2}\frac{\square\sqrt{\rho}}{\sqrt{\rho}}+2V_{\rm
tot}'(\rho)}.
\label{charge10b}
\end{eqnarray}

\subsection{TF approximation}
\label{sec_rtf}

In the classical limit ($\hbar\rightarrow 0$) or in the TF approximation 
where the quantum potential can be neglected, the hydrodynamic equations
(\ref{db8})
and (\ref{db9}) reduce to
\begin{eqnarray}
D_{\mu}\left ( \rho \partial^{\mu}\theta\right )=0,
\label{rtf4}
\end{eqnarray}
\begin{eqnarray}
\frac{1}{2}\partial_{\mu}\theta\partial^{\mu}\theta-V'_{\rm
tot}(\rho)=0.
\label{rtf5}
\end{eqnarray}
Equation (\ref{rtf4})
can be interpreted as a continuity equation and Eq.
(\ref{rtf5}) can be interpreted as a classical relativistic Hamilton-Jacobi (or
Bernoulli)
equation.

In order to determine the rest mass density, we can
repeat the same procedure as before. Assuming
$V_{\rm tot}'>0$, and using Eq. (\ref{rtf5}), we introduce the fluid
quadrivelocity
\begin{eqnarray}
u_{\mu}=-\frac{\partial_{\mu}\theta}{\sqrt{2V_{\rm
tot}'(\rho)}}c,
\label{rtf5b}
\end{eqnarray}
which satisfies the identity (\ref{rtf5cw}). Using Eqs. (\ref{val1}) and
(\ref{rtf5b}), we have
\begin{eqnarray}
J^{\mu}=\frac{\rho}{c} \sqrt{2V_{\rm tot}'(\rho)}u^{\mu},
\label{val4}
\end{eqnarray}
and
we can write the the continuity equation (\ref{rtf4}) as 
\begin{eqnarray}
D_{\mu}\left \lbrack \rho \sqrt{2V_{\rm tot}'(\rho)}u^{\mu}\right \rbrack=0.
\label{charge9b}
\end{eqnarray}
Comparing
the expressions of $J^{\mu}$ given by Eqs. (\ref{val3}) and (\ref{val4}), we
find that the
rest-mass density $\rho_m=nm$   is
given in the TF
approximation, 
by
\begin{eqnarray}
\rho_m=\frac{\rho}{c} \sqrt{2V_{\rm tot}'(\rho)}.
\label{charge10c}
\end{eqnarray}
This is the limit form of Eq. (\ref{charge10b}) with $Q_{\rm dB}=0$. We note
that $\rho_m\neq J^0/c$ in
general. When
$V=\rho_{\Lambda}c^2$ is constant
(corresponding to the $\Lambda$CDM
model), the rest-mass density coincides with the pseudo rest-mass density
($\rho_m=\rho$). We also have $\rho_m=\rho$ in the nonrelativistic limit
$c\rightarrow +\infty$.

In the TF approximation, the energy-momentum tensor from Eq.
(\ref{em3gen}) reduces to
\begin{equation}
T_{\mu\nu}=\rho\partial_{\mu}\theta\partial_{\nu}\theta
-g_{\mu\nu}\left\lbrack
\frac{1}{2}g^{\mu\nu}\rho\partial_{\mu}\theta\partial_{\nu}
\theta-V_{\rm
tot}(\rho)\right\rbrack.
\label{em3genz}
\end{equation}
Using Eq. (\ref{rtf5b}), we get
\begin{equation}
T_{\mu\nu}=2\rho V'_{\rm tot}(\rho)
\frac{u_{\mu}u_{\nu}}{c^2}-g_{\mu\nu}\left\lbrack \rho V'_{\rm tot}(\rho)-V_{\rm
tot}(\rho)\right\rbrack.
\label{em3b}
\end{equation}
The   energy-momentum tensor can be written under the perfect fluid
form
\begin{eqnarray}
T_{\mu\nu}=(\epsilon+P)\frac{u_{\mu}u_{\nu}}{c^2}-P g_{\mu\nu},
\label{em4}
\end{eqnarray}
where $\epsilon$ is the energy density and $P$ is the pressure, provided that we
make the identifications
\begin{eqnarray}
\epsilon=\rho V'_{\rm tot}(\rho)+V_{\rm
tot}(\rho)=\rho c^2+\rho V'(\rho)+V(\rho),
\label{rtf6}
\end{eqnarray}
\begin{eqnarray}
P=\rho V'_{\rm tot}(\rho)-V_{\rm
tot}(\rho)=\rho V'(\rho)-V(\rho),
\label{rtf7}
\end{eqnarray}
where we have used Eq. (\ref{db10}) to get the second equalities. Eliminating
$\rho$ between these equations, we obtain the equation of state
$P(\epsilon)$. The squared speed of sound is
\begin{eqnarray}
c_s^2=P'(\epsilon)c^2=\frac{\rho V''(\rho)c^2}{c^2+\rho V''(\rho)+2V'(\rho)}.
\label{rtf7x}
\end{eqnarray}

{\it Remark:} We note that the expression of the pressure is the same as in Eq.
(\ref{mad11b}). It is also the same as the one obtained in the nonrelativistic
limit $c\rightarrow +\infty$ where the KGE equations reduce to the GPP
equations (see below). By contrast, the enthalpy
differs from the pseudo enthalpy except in the nonrelativistic limit.
Using  Eqs. (\ref{mtd4a}), (\ref{rtf5}), (\ref{charge10c}),
(\ref{rtf6}) and (\ref{rtf7}) we
find that the
enthalpy is given by
\begin{equation}
h=\sqrt{2V'_{\rm tot}(\rho)}\, c=c\sqrt{\partial_\mu\theta\partial^\mu\theta}.
\label{en2}
\end{equation}
Substituting Eq. (\ref{db10}) into Eq. (\ref{en2}), subtracting
$c^2$, and taking the nonrelativistic limit $c\rightarrow +\infty$, we obtain
$h=V'(\rho)$.

\subsection{Nonrelativistic limit}

To take the nonrelativistic limit, it is convenient to work with the
conformal Newtonian gauge which is a perturbed form
of the FLRW line element
\begin{equation}
ds^2=c^2\left(1+2\frac{\Phi}{c^2}\right)dt^2-a(t)^2\left(1-2\frac{\Phi}{c^2}
\right)\delta_{ij}dx^idx^j,
\label{kge2}
\end{equation}
where  $\Phi({\bf r},t)/c^2\ll 1$ represents the
gravitational potential of
classical Newtonian gravity. Making the
Klein
transformation
\begin{eqnarray}
\varphi({\bf r},t)=\frac{\hbar}{m}e^{-i m c^2 t/\hbar}\psi({\bf r},t),
\label{kgp12}
\end{eqnarray}
in the KGE equations (\ref{csf4}) and (\ref{ak21}), using the relation
\begin{equation}
\label{csf3b}
\rho=|\psi|^2=\frac{m^2}{\hbar^2}|\varphi|^2,
\end{equation}
and taking the nonrelativistic limit $c\rightarrow
+\infty$, we get the generalized Gross-Pitaevskii-Poisson (GPP) equations in an
expanding universe (see, e.g., \cite{abrilph,playa,chavmatos} for details)
\begin{equation}
i\hbar\frac{\partial\psi}{\partial
t}+\frac{3}{2}i\hbar
H\psi=-\frac{\hbar^2}{2 m a^2}\Delta\psi+m\Phi \psi
+m\frac{dV}{d|\psi|^2}\psi,
\label{nr1}
\end{equation}
\begin{eqnarray}
\frac{\Delta\Phi}{4\pi G a^2}=|\psi|^2 -\frac{3H^2}{8\pi G},
\label{nr2}
\end{eqnarray}
where ${3H^2}/{8\pi G}=\rho_b$ is the background
density [see Eq. (\ref{fe1})].  

Making the Madelung \cite{madelung} transformation
\begin{eqnarray}
\psi({\bf r},t)=\sqrt{\rho({\bf r},t)} e^{iS({\bf r},t)/\hbar},\label{kgp15}
\end{eqnarray}
\begin{eqnarray}
\rho=|\psi|^2,\qquad {\bf v}({\bf r},t)=\frac{\nabla S}{ma},
\label{kgp16}
\end{eqnarray}
where $\rho$ is the mass density, $S$ is the action and ${\bf v}$ is the
velocity field, we obtain the system of hydrodynamic
equations \cite{prd1,aacosmo,playa}:
\begin{eqnarray}
\frac{\partial\rho}{\partial t}+3H\rho+\frac{1}{a}\nabla\cdot (\rho
{\bf v})=0,
\label{nr3}
\end{eqnarray}
\begin{eqnarray}
\frac{\partial S}{\partial t}+\frac{(\nabla S)^2}{2 m
a^2}=\frac{\hbar^2}{2
m a^2}\frac{\Delta\sqrt{\rho}}{\sqrt{\rho}}
-m\Phi-mh(\rho),
\label{nr6}
\end{eqnarray}
\begin{equation}
\frac{\partial {\bf v}}{\partial t}+H{\bf v}+\frac{1}{a}({\bf v}\cdot
\nabla){\bf v}=
\frac{\hbar^2}{2m^2a^3}\nabla\left( \frac{\Delta\sqrt{\rho}}{\sqrt{
\rho}} \right )-\frac{1}{a}
\nabla\Phi-\frac{1}{\rho a}\nabla P,
\label{nr4}
\end{equation}
\begin{eqnarray}
\frac{\Delta\Phi}{4\pi G a^2}=\rho-\frac{3H^2}{8\pi
G},
\label{nr5}
\end{eqnarray}
where $h(\rho)=V'(\rho)$ is the enthalpy and $P(\rho)$ is the pressure
defined by the relation $h'(\rho)=P'(\rho)/\rho$. It is
explicitly given by $P(\rho)=\rho h(\rho)-\int h(\rho)\, d\rho$, i.e.,
\begin{eqnarray}
P(\rho)=\rho V'(\rho)-V(\rho).
\label{kgp20g}
\end{eqnarray}
The squared speed of sound is $c_s^2=P'(\rho)=\rho V''(\rho)$. 
The hydrodynamic equations  (\ref{nr3})-(\ref{nr5}) have a clear physical
interpretation. Equation (\ref{nr3}), corresponding to the imaginary
part of the GP equation, is the continuity equation.  Equation
(\ref{nr6}),
corresponding to the real part of the GP equation,
is the Bernoulli or Hamilton-Jacobi equation. Equation (\ref{nr4}), obtained by
taking the gradient of  Eq. (\ref{nr6}), is the Euler (momentum) equation.
Equation
(\ref{nr5}) is the Poisson equation. It can be written as $\Delta\Phi=4\pi
Ga^2(\rho-\rho_b)$ where $\rho_b$ is the background
density of the expanding universe.

{\it Remark:} For a complex SF, we note that the potential $V(|\psi|^2)$ that
appears in the GP equation (\ref{nr1}) is the same as the potential
$V(|\varphi|^2)$ that appears in the KG equation (\ref{csf4}) (see, e.g.,
\cite{abrilph,playa,chavmatos}). For a real SF, the two potentials generally
differ (see, e.g., \cite{phi6,tunnel}).

\section{Homogeneous complex SF in an expanding universe}
\label{sec_mtt}

In this Appendix, we apply the results of Appendix \ref{sec_ra} to
a cosmological context, namely for a homogeneous complex SF in an expanding
background, and we recover the results of Sec. \ref{sec_csf}.

\subsection{General results}

The cosmological evolution of a spatially homogeneous complex SF in an
expanding universe is governed by the
KGF equations
\begin{eqnarray}
\frac{1}{c^2}\frac{d^2\varphi}{dt^2}+\frac{3H}{c^2}\frac{d\varphi}{dt}
+2\frac{dV_{\rm tot}}{d|\varphi|^2}\varphi=0,
\label{hj1}
\end{eqnarray}
\begin{equation}
\label{hj2}
H^2=\frac{8\pi
G}{3c^2}\epsilon,
\end{equation}
which can be deduced from the KGE equations (\ref{csf4}) and (\ref{ak21}). The
energy density
$\epsilon(t)$ and the pressure $P(t)$ of the SF are given by
\begin{equation}
\epsilon=\frac{1}{2c^2}\left |\frac{d\varphi}{d
t}\right|^2+V_{\rm tot}(|\varphi|^2),
\label{hsf2kk}
\end{equation}
\begin{equation}
P=\frac{1}{2c^2}\left |\frac{d\varphi}{d
t}\right|^2-V_{\rm tot}(|\varphi|^2).
\label{hsf3kk}
\end{equation}
In the following, we use the hydrodynamic representation of the SF (see
Appendix \ref{sec_db}). Making the de Broglie transformation (\ref{db4}), the
energy density and the pressure of the SF can be written as
\begin{equation}
\epsilon=\frac{1}{2c^2}\rho\dot\theta^2+\frac{\hbar^2}{8m^2\rho
c^2}\dot\rho^2+V_{\rm tot}(\rho),
\label{hhj3}
\end{equation}
\begin{equation}
P=\frac{1}{2c^2}\rho\dot\theta^2+\frac{\hbar^2}{8m^2\rho
c^2}\dot\rho^2-V_{\rm tot}(\rho).
\label{hhj4}
\end{equation}

The equation $D_\nu T^{\mu\nu}=0$ leads to the energy conservation equation 
\begin{eqnarray}
\frac{d\epsilon}{dt}+3H(\epsilon+P)=0.
\label{hj5}
\end{eqnarray}
This equation can also be obtained from the KG equation (\ref{hj1}) with Eqs.
(\ref{hsf2kk}) and (\ref{hsf3kk}).

The equation $D_\mu J^\mu=0$, which is equivalent to the continuity
equation (\ref{db8}), can be written as
\begin{eqnarray}
\label{hj6}
\frac{d}{dt}\left (E_{\rm tot} \rho a^3\right )=0,
\end{eqnarray}
where 
\begin{eqnarray}
\label{hj7}
E_{\rm tot}=-\dot S_{\rm
tot}=-m\dot\theta=\hbar\omega
\end{eqnarray}
is the energy of the SF. Eq. (\ref{hj6}) expresses the
conservation of the charge of the complex  SF (or equivalently the
conservation
of the boson number). It can be written as
\begin{eqnarray}
\label{hj8}
\rho E_{\rm tot}=\frac{Q m^2c^2}{a^3},
\end{eqnarray}
where $Q=Ne$ is a constant of integration representing the charge
 of the SF (proportional to the boson number $N$) 
\cite{arbeycosmo,gh,shapiro,abrilph,abrilphas,kasuya,spintessence}. This
equation can
also be directly obtained from Eq. (\ref{charge5}).

The quantum Hamilton-Jacobi (or Bernoulli) equation (\ref{db9}) takes
the form
\begin{equation}
\label{hj11}
E_{\rm tot}^2=\hbar^2\frac{1}{\sqrt{\rho}}
\frac {
d^2\sqrt{\rho}}{dt^2}+3H\hbar^2\frac{1}{\sqrt{\rho}}\frac{
d\sqrt{\rho}}{dt}+2m^2c^2V'_{ \rm tot}(\rho).
\end{equation}

Finally, we have established in the general case 
that the
rest-mass
density of the SF is given by Eq. (\ref{charge10a}).
For a spatially homogeneous SF in an expanding background, we get
\begin{equation}
\rho_m=-\frac{\rho}{c}
\partial_0\theta=-\frac{1}{c^2}\rho\dot\theta=\rho\frac{\hbar\omega}{mc^2}=\rho
\frac
{E_ { \rm tot}}
{mc^2}=-\rho\frac{\dot S_{\rm
tot}}{mc^2}.
\label{hj13}
\end{equation}
We note that, in this special case, $\rho_m=J^0/c$, where $J_0=-\rho
\partial_0S_{\rm tot}/m$
is the time component of the quadricurrent of charge, but this relation is not
true
for an inhomogeneous SF. Using Eq. (\ref{hj13}), Eqs. (\ref{hj6}) and
(\ref{hj8}) can be rewritten
as
\begin{eqnarray}
\label{hj14}
\frac{d\rho_m}{dt}+3H\rho_m=0,
\end{eqnarray}
and
\begin{eqnarray}
\label{hj15}
\rho_m=\frac{Q m}{a^3}.
\end{eqnarray}
Equation (\ref{hj14}) can
also be
obtained by combining the the first law of thermodynamics for a
cold fluid with the energy conservation equation (see Sec. \ref{sec_dmde}).
These equations
express the
conservation of
the particle number.

\subsection{TF approximation}

In the TF approximation ($\hbar\rightarrow 0$), the energy density
$\epsilon$ and the pressure $P$ of the SF [see Eqs. (\ref{hhj3}) and
(\ref{hhj4})] reduce to
\begin{equation}
\epsilon=\frac{1}{2c^2}\rho\dot\theta^2+V_{\rm
tot}(\rho)=\frac{\rho E_{\rm tot}^2}{2m^2c^2}+V_{\rm
tot}(\rho),
\label{hhj3tf}
\end{equation}
\begin{equation}
P=\frac{1}{2c^2}\rho\dot\theta^2-V_{\rm tot}(\rho)=\frac{\rho E_{\rm
tot}^2}{2m^2c^2}-V_{\rm tot}(\rho),
\label{hhj4tf}
\end{equation}
where we have used Eq. (\ref{hj7}) to get the second equalities.
On the other hand, the quantum Hamilton-Jacobi (or
Bernoulli) equation (\ref{hj11}) reduces to
\begin{eqnarray}
\label{hj17}
E_{\rm tot}^2=2m^2c^2V'_{\rm tot}(\rho).
\end{eqnarray}
Combining Eqs. (\ref{hj8}) and (\ref{hj17}), we obtain
\begin{eqnarray}
\label{hj18}
\frac{\rho}{c}\sqrt{2 V'_{\rm tot}(\rho)}=\frac{Q m}{a^3}.
\end{eqnarray}
This equation determines the relation between the pseudo 
rest-mass density $\rho$ and the scale factor $a$.

According
to Eqs.
(\ref{hj13}) and
(\ref{hj17}), the rest-mass densitity is given by
\begin{eqnarray}
\label{hj19}
\rho_m=\frac{\rho}{c}\sqrt{2 V'_{\rm tot}(\rho)}.
\end{eqnarray}
This shows that (\ref{hj18}) is equivalent to the conservation
of the rest-mass [see Eq. (\ref{hj15})]. Note that
Eq.
(\ref{hj19}) is always true in the TF
approximation even for inhomogeneous systems (see Appendix \ref{sec_rtf}).

Finally, inserting the Bernoulli
equation (\ref{hj17}) into Eqs. (\ref{hhj3tf}) and (\ref{hhj4tf}), we find
that the 
energy density and the pressure of the SF in the TF approximation are given by
\begin{eqnarray}
\epsilon=\rho V'_{\rm tot}(\rho)+V_{\rm
tot}(\rho),
\label{hj20}
\end{eqnarray}
\begin{eqnarray}
P=\rho V'_{\rm tot}(\rho)-V_{\rm
tot}(\rho).
\label{hj21}
\end{eqnarray}
Note that these relations are always true  in the TF
approximation even for inhomogeneous systems (Appendix \ref{sec_rtf}).

{\it Remark:} Eqs. (\ref{hj20}) and (\ref{hj21}) determine
the equation of state $P=P(\epsilon)$. As a result, we can obtain Eq.
(\ref{hj18}) directly from Eqs.
(\ref{hj20}), (\ref{hj21}) and the energy conservation equation (\ref{hj5}).
Indeed, combining these equations we obtain
\begin{eqnarray}
\left\lbrack 2V'_{\rm tot}(\rho)+\rho V''_{\rm tot}(\rho)\right\rbrack
\frac{d\rho}{dt}=-6H\rho V'_{\rm tot}(\rho).
\label{hj24}
\end{eqnarray}
leading to
\begin{eqnarray}
\int \frac{2V'_{\rm tot}(\rho)+\rho V''_{\rm
tot}(\rho)}{\rho
V'_{\rm tot}(\rho)}=-6\ln a.
\label{hj25}
\end{eqnarray}
Equation (\ref{hj25}) integrates to give Eq. (\ref{hj18}).

\subsection{Nonrelativistic limit}

Combining Eqs. (\ref{kgp12}) and (\ref{kgp15}) and comparing the resulting
expression with Eq. (\ref{db1}), we see that $S_{\rm tot}=-mc^2 t+S$,
leading to $E_{\rm tot}=mc^2+E$ with $E=-dS/dt$. We can then rewrite the
previous equations in terms of $S$ and $E$ instead of $S_{\rm tot}=m\theta$ and
$E_{\rm tot}$. Then, taking the nonrelativistic limit
$c\rightarrow +\infty$, we obtain 
\begin{eqnarray}
\frac{d\rho}{dt}+3H\rho=0,\quad E=mV'(\rho),\quad
\frac{3H^2}{8\pi G}=\rho.
\label{sun1}
\end{eqnarray}
These equations can also be obtained from Eqs.
(\ref{nr3})-(\ref{nr5}) by considering a homogeneous SF. They are easily
solved to give  $\rho\propto a^{-3}$, $a\propto t^{2/3}$ and $\rho=1/(6\pi
Gt^2)$. This is the EdS solution. Therefore, in the nonrelativistic limit,  the
homogeneous SF/BEC behaves as CDM. For the power-law potential (\ref{plsf1}), we
get $E=Km\gamma/[(\gamma-1)(6\pi
Gt^2)^{\gamma-1}]$. In particular, for the Chaplygin gas
($\gamma=-1$), we get $E=18\pi^2 Km G^2t^4$; for the standard BEC
($\gamma=2$), we get $E=2a_s\hbar^2/(3Gm^2t^2)$; for the $\Lambda$CDM model
($\gamma=0$), we get $E=0$; for the superfluid ($\gamma=3$), we get  
$E=Km/(24\pi^2
G^2t^4)$. On the other hand, for the potential (\ref{iplsf1}), we get $E=-k_B
T\ln(6\pi G\rho_* t^2)$.

\section{Examples of potentials of self-interaction}
\label{sec_ex}

Let us consider a complex SF $\varphi$ governed by the nonlinear KG
equation
\begin{equation}
\label{s1}
\square\varphi+\frac{m^2c^2}{\hbar^2}\varphi+2\frac{dV}{d|\varphi|^2}\varphi=0
\end{equation}
with a self-interaction potential $V(|\varphi|^2)$. In the TF approximation,
this potential is associated with a
barotropic equation of state $P(\rho)=\rho V'(\rho)-V(\rho)$ where
$\rho=|\psi|^2=(m^2/\hbar^2)|\varphi|^2$ is the pseudo rest-mass
density (see Appendix \ref{sec_ra}).  In the nonrelativistic
limit, the nonlinear KG equation reduces to the generalized GP equation
\begin{equation}
i\hbar\frac{\partial\psi}{\partial
t}+\frac{3}{2}i\hbar
H\psi=-\frac{\hbar^2}{2 m a^2}\Delta\psi+m\Phi \psi
+m\frac{dV}{d|\psi|^2}\psi,
\label{s2}
\end{equation}
which involves the same self-interaction potential $V(|\psi|^2)$. Let us give
some examples of self-interaction potentials and their corresponding equations
of state  $P(\rho)$.

The power-law potential (see Sec. \ref{sec_pwlnon}) 
\begin{eqnarray}
V(|\varphi|^2)=\frac{K}{\gamma-1}\left (\frac{m}{\hbar}\right
)^{2\gamma}|\varphi|^{2\gamma}\qquad (\gamma\neq 1)
\label{s3}
\end{eqnarray}
is associated with the polytropic equation of state $P=K\rho^{\gamma}$. The KG
equation becomes
\begin{equation}
\label{s4}
\square\varphi+\frac{m^2c^2}{\hbar^2}\varphi+\frac{2K\gamma}{\gamma-1}\left
(\frac{m}{\hbar}\right
)^{2\gamma}|\varphi|^{2(\gamma-1)}\varphi=0.
\end{equation}
In the nonrelativistic limit, using
\begin{eqnarray}
V(|\psi|^2)=\frac{K}{\gamma-1}|\psi|^{2\gamma}\qquad (\gamma\neq 1),
\label{s5}
\end{eqnarray}
we obtain the generalized GP equation
\begin{equation}
i\hbar\frac{\partial\psi}{\partial
t}+\frac{3}{2}i\hbar
H\psi=-\frac{\hbar^2}{2 m a^2}\Delta\psi+m\Phi \psi
+\frac{Km\gamma}{\gamma-1}|\psi|^{2(\gamma-1)}\psi.
\label{s6}
\end{equation}

(i) The index $\gamma=-1$ corresponds to the Chaplygin gas
\begin{eqnarray}
V(|\varphi|^2)=-\frac{K}{2}\left (\frac{\hbar}{m}\right
)^{2}\frac{1}{|\varphi|^{2}}, \qquad P=\frac{K}{\rho},
\label{s3chap}
\end{eqnarray}
\begin{equation}
\label{s4chap}
\square\varphi+\frac{m^2c^2}{\hbar^2}\varphi+K\left
(\frac{\hbar}{m}\right
)^{2}\frac{1}{|\varphi|^{4}}\varphi=0,
\end{equation}
\begin{eqnarray}
V(|\psi|^2)=-\frac{K}{2}\frac{1}{|\psi|^{2}},
\label{s5chap}
\end{eqnarray}
\begin{equation}
i\hbar\frac{\partial\psi}{\partial
t}+\frac{3}{2}i\hbar
H\psi=-\frac{\hbar^2}{2 m a^2}\Delta\psi+m\Phi \psi
+\frac{Km}{2}\frac{1}{|\psi|^{4}}\psi.
\label{s6chap}
\end{equation}

(ii) The index $\gamma=2$
corresponds to the standard BEC taking into account two-body interactions
\begin{eqnarray}
V(|\varphi|^2)=K\left (\frac{m}{\hbar}\right
)^{4}|\varphi|^{4},\qquad P=K\rho^2,
\label{s3s}
\end{eqnarray}
\begin{equation}
\label{s4s}
\square\varphi+\frac{m^2c^2}{\hbar^2}\varphi+4K\left
(\frac{m}{\hbar}\right
)^{4}|\varphi|^{2}\varphi=0,
\end{equation}
\begin{eqnarray}
V(|\psi|^2)=K|\psi|^{4},
\label{s5s}
\end{eqnarray}
\begin{equation}
i\hbar\frac{\partial\psi}{\partial
t}+\frac{3}{2}i\hbar
H\psi=-\frac{\hbar^2}{2 m a^2}\Delta\psi+m\Phi \psi
+2Km|\psi|^{2}\psi.
\label{s6s}
\end{equation}

(iii) The
index $\gamma=0$ corresponds to the $\Lambda$CDM model interpreted as an UDM
model
\begin{eqnarray}
V(|\varphi|^2)=-K,\qquad P=K,
\label{s3l}
\end{eqnarray}
\begin{equation}
\label{s4l}
\square\varphi+\frac{m^2c^2}{\hbar^2}\varphi=0,
\end{equation}
\begin{eqnarray}
V(|\psi|^2)=-K,
\label{s5l}
\end{eqnarray}
\begin{equation}
i\hbar\frac{\partial\psi}{\partial
t}+\frac{3}{2}i\hbar
H\psi=-\frac{\hbar^2}{2 m a^2}\Delta\psi+m\Phi \psi.
\label{s6l}
\end{equation}

(iv) The index $\gamma=3$
corresponds to a superfluid taking into account three-body interactions
\begin{eqnarray}
V(|\varphi|^2)=\frac{K}{2}\left (\frac{m}{\hbar}\right
)^{6}|\varphi|^{6},\qquad  P=K\rho^3,
\label{s3su}
\end{eqnarray}
\begin{equation}
\label{s4su}
\square\varphi+\frac{m^2c^2}{\hbar^2}\varphi+3K\left
(\frac{m}{\hbar}\right
)^{6}|\varphi|^{4}\varphi=0,
\end{equation}
\begin{eqnarray}
V(|\psi|^2)=\frac{K}{2}|\psi|^{6},
\label{s5su}
\end{eqnarray}
\begin{equation}
i\hbar\frac{\partial\psi}{\partial
t}+\frac{3}{2}i\hbar
H\psi=-\frac{\hbar^2}{2 m a^2}\Delta\psi+m\Phi \psi
+\frac{3Km}{2}|\psi|^{4}\psi.
\label{s6su}
\end{equation}

(v) For $\gamma=1/2$ we get 
\begin{eqnarray}
V(|\varphi|^2)=-2K \frac{m}{\hbar}|\varphi|,\qquad P=K\rho^{1/2},
\label{s3h}
\end{eqnarray}
\begin{equation}
\label{s4h}
\square\varphi+\frac{m^2c^2}{\hbar^2}\varphi-2K\frac{m}{\hbar}\frac{1}{
|\varphi|}\varphi=0,
\end{equation}
\begin{eqnarray}
V(|\psi|^2)=-2K|\psi|,
\label{s5h}
\end{eqnarray}
\begin{equation}
i\hbar\frac{\partial\psi}{\partial
t}+\frac{3}{2}i\hbar
H\psi=-\frac{\hbar^2}{2 m a^2}\Delta\psi+m\Phi \psi
-Km\frac{1}{|\psi|}\psi.
\label{s6h}
\end{equation}

The potential (see Sec. \ref{sec_pwloui})
\begin{eqnarray}
V(|\varphi|^2)=\frac{k_B T m}{\hbar^2}|\varphi|^2\left\lbrack \ln\left
(\frac{m^2|\varphi|^2}{\rho_*\hbar^2}\right )-1\right\rbrack
\label{s7}
\end{eqnarray}
is associated with the isothermal equation of state $P=\rho k_B T/m$.
It can take into account finite temperature effects in
DM. The KG equation becomes
\begin{equation}
\label{s8}
\square\varphi+\frac{m^2c^2}{\hbar^2}\varphi+\frac{2k_B T
m}{\hbar^2}\ln\left
(\frac{m^2|\varphi|^2}{\rho_*\hbar^2}\right )\varphi=0.
\end{equation}
In the nonrelativistic limit, using
\begin{eqnarray}
V(|\psi|^2)=\frac{k_B T}{m}|\psi|^2\left\lbrack \ln\left
(\frac{|\psi|^2}{\rho_*}\right )-1\right\rbrack,
\label{s9}
\end{eqnarray}
we obtain the generalized GP equation
\begin{equation}
i\hbar\frac{\partial\psi}{\partial
t}+\frac{3}{2}i\hbar
H\psi=-\frac{\hbar^2}{2 m a^2}\Delta\psi+m\Phi \psi
+k_B T  \ln\left
(\frac{|\psi|^2}{\rho_*}\right )\psi.
\label{s10}
\end{equation}

The logarithmic potential \cite{graal}
\begin{eqnarray}
V(|\varphi|^2)=-A\left\lbrack \ln\left
(\frac{m^2|\varphi|^2}{\hbar^2\rho_P}\right )+1\right\rbrack
\label{s11}
\end{eqnarray}
is associated with the logotropic equation of state $P=A\ln\left
({\rho}/{\rho_P}\right )$. The KG equation
becomes
\begin{equation}
\label{s12}
\square\varphi+\frac{m^2c^2}{\hbar^2}\varphi-\frac{2A}{|\varphi|^2}\varphi=0.
\end{equation}
In the nonrelativistic limit, using
\begin{eqnarray}
V(|\psi|^2)=-A\left\lbrack \ln\left
(\frac{|\psi|^2}{\rho_P}\right )+1\right\rbrack,
\label{s13}
\end{eqnarray}
we obtain the generalized GP equation
\begin{equation}
i\hbar\frac{\partial\psi}{\partial
t}+\frac{3}{2}i\hbar
H\psi=-\frac{\hbar^2}{2 m a^2}\Delta\psi+m\Phi \psi
-\frac{mA}{|\psi|^2}\psi.
\label{s14}
\end{equation}
This model is studied in detail in \cite{graal}.

Instead of the logarithmic potential we can consider the
constant potential
\begin{eqnarray}
V(|\varphi|^2)=V_0=\epsilon_\Lambda,
\label{s11n}
\end{eqnarray}
mimicking a
cosmological constant like in Appendix E of \cite{graal}. It is associated with
the constant equation of state $P=-\epsilon_\Lambda$. It leads to
the usual KG and GP equations (\ref{s4l}) and (\ref{s6l}).
Note that a constant potential has no effect on the  KG and GP equations but it
adds a constant term $\epsilon_\Lambda$ in the energy density interpreted as DE.

Of course, we can consider a
multitude of more
general
SF models by summing the potentials [like, e.g., Eqs. (\ref{s3}),
(\ref{s7}), (\ref{s11}), and (\ref{s11n})] or, equivalently,
by summing the corresponding pressures (polytropic, isothermal, logotropic and
constant). This is a form of Dalton's law. 
The mixed equation of state generically writes
\begin{equation}
\label{s15}
P=K\rho^{\gamma}+\rho\frac{k_B T}{m}-\rho_{\Lambda}c^2+A\ln\left
(\frac{\rho}{\rho_P}\right ),
\end{equation}
where we can add several polytropic terms with different index $\gamma$. This
mixed equation of state is
associated with a complex SF
potential of the form
\begin{eqnarray}
V(|\varphi|^2)=\frac{K}{\gamma-1}\left (\frac{m}{\hbar}\right
)^{2\gamma}|\varphi|^{2\gamma}\nonumber\\
+\frac{k_B T
m}{\hbar^2}|\varphi|^2\left\lbrack \ln\left
(\frac{m^2|\varphi|^2}{\rho_*\hbar^2}\right
)-1\right\rbrack\nonumber\\
+\rho_{\Lambda}c^2-A\left\lbrack \ln\left
(\frac{m^2|\varphi|^2}{\hbar^2\rho_P}\right )+1\right\rbrack.
\label{s16}
\end{eqnarray}

\end{document}